\documentclass[11pt]{article}

\usepackage[dvipsnames]{xcolor}
\usepackage{graphicx}

\usepackage{fullpage, parskip}
\usepackage{enumitem, multicol, multirow}
\usepackage[symbol]{footmisc}
\usepackage{subcaption, float}
\usepackage{authblk}

\usepackage{amsmath, amsfonts, amsthm, amssymb, nicefrac, thmtools}
\allowdisplaybreaks
\newtheorem{theorem}{Theorem}

\newtheorem{lemma}{Lemma}

\newtheorem{claim}{Claim}

\newtheorem{result}{Result}
\newtheorem{fact}{Fact}

\theoremstyle{definition}
\newtheorem{definition}{Definition}

\usepackage{tikz}
\usetikzlibrary{decorations.pathreplacing,calc,arrows}

\usepackage[ruled,linesnumbered,noend]{algorithm2e}
\let\oldnl\nl
\newcommand{\nonl}{\renewcommand{\nl}{\let\nl\oldnl}}

\usepackage{url}
\usepackage[colorlinks=true, linkcolor=blue, citecolor=blue, urlcolor=blue, breaklinks=true]{hyperref}

\usepackage{bold-extra}

\newcommand{\poly}{\mathrm{poly}}

\newcommand{\OPT}{\mathrm{OPT}}
\newcommand{\ALG}{\mathrm{ALG}}

\newcommand{\R}{\mathbb{R}}

\newcommand{\Z}{\mathbb{Z}}

\newcommand{\ov}{\overline}

\newcommand{\half}{\mathrm{half}}

\title{Which $L_p$ Norm is the Fairest? Approximations for Fair Facility Location Across All ``$p$''}

\author[1]{Swati Gupta\footnote{Part of this work was done while the author was at Georgia Institute of Technology.}}
\affil[1]{Massachusetts Institute of Technology}
\affil{\texttt{swatig@mit.edu}}

\author[2]{Jai Moondra}
\author[2]{Mohit Singh}
\affil[2]{Georgia Institute of Technology}
\affil{\texttt{jmoondra3@gatech.edu}, \texttt{mohit.singh@isye.gatech.edu}}

\date{}

\begin{document}

    \maketitle

    \begin{abstract}
        Fair facility location problems try to balance access costs to open facilities borne by different groups of people by minimizing the $L_p$ norm of these group distances. However, there is no clear choice of ``$p$'' in the current literature. We present a novel approach to address the challenge of choosing the right notion of fairness. We introduce the concept of portfolios, a set of solutions that contains an approximately optimal solution for each objective in a given class of objectives, such as $L_p$ norms. This concept opens up new possibilities for getting around the “right” notion of fairness for many problems. For $r$ client groups, we demonstrate portfolios of size $\Theta(\log r)$  for the facility location and $k$-clustering problems, with an $O(1)$-approximate solution for each $L_p$ norm. Further, motivated by the Justice40 Initiative that provides rolling budget investments, we impose a refinement-like structure on the portfolio. We develop novel approximation algorithms for these structured portfolios and show experimental evidence of their performance in two US counties. We also present a planning tool that provides potential ways to expand access to US healthcare facilities, which might be of independent interest to policymakers.


    \end{abstract}

    \section{Introduction}\label{sec: introduction}

    \begin{figure}
        \centering
        \frame{\includegraphics[width=\textwidth]{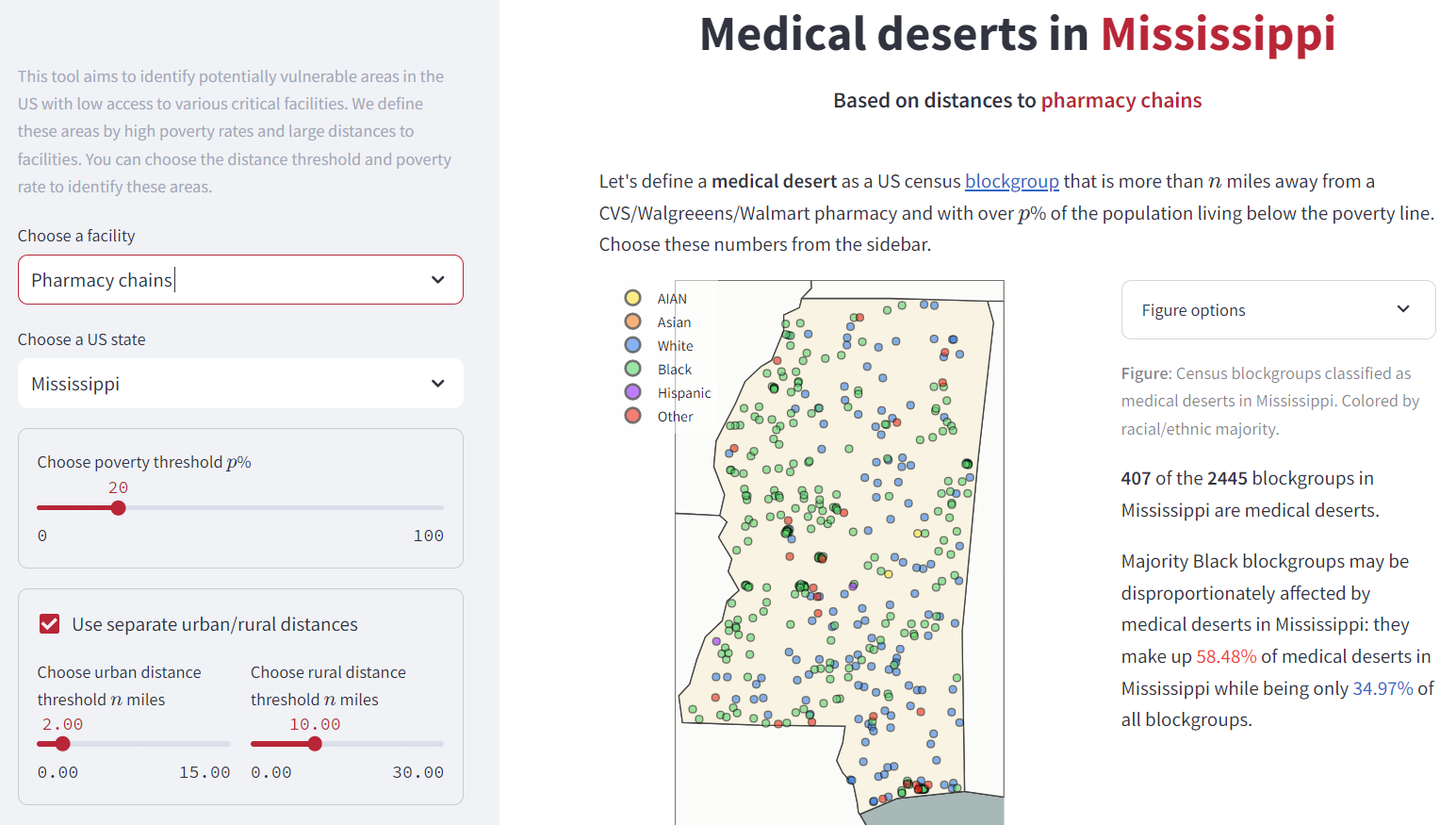}}
        \caption{A screenshot from our online \href{https://usa-medical-deserts.streamlit.app/}{tool} indicating potential medical deserts in Mississippi, USA. The tool can be found at \url{https://usa-medical-deserts.streamlit.app/}}
        \label{fig: online-tool}
    \end{figure}

    Inequity in the placement of critical facilities is a well-documented problem \cite{multidimensional_poverty_2023, DPF12, calderon2004effects, GJRYZ20}.
    For instance, it is largely suspected that profit maximization by grocery chains has led to the formation of food deserts spread widely across the US, which are defined as regions with low-income populations and low access to fresh food (e.g., people with no cars, and no grocery stores within a mile of their house) \cite{DPF12, balancing2022john}. An estimated 40 million Americans live in these food deserts that disproportionately affect racial minorities \cite{PSMBC07}, thus leading to even greater structural disadvantages. Recent studies show similar disparities in the locations of COVID testing sites \cite{tao_examining_2020, rader_spatial_2022}.
    We highlight how severe this concern is {in healthcare} by developing an online tool to indicate the presence of medical deserts due to the absence of critical medical facilities (e.g., three major pharmacy chains namely CVS, Walgreens, and Walmart\footnote{Large pharmacy chains were chosen as facilities in our online tool to display the effect of opening a large number of facilities at scale.}), within a user-specified distance of low-income census blockgroups\footnote{A census blockgroup is the smallest administrative region in the US with public census data.} (see Figure \ref{fig: online-tool}).
    Using a similar definition to USDA, our tool shows a persistent racial disparity in the makeup of medical deserts across the US.
    Although one solution to fix-all remains elusive, it is clear that the pursuit of maximizing a single objective such as the number of customers served (reasonably) through these facilities can lead to exacerbated social inequity, especially for socioeconomic groups at the intersection of multiple disadvantaged identities (e.g., poor, uninsured and people of color). \footnote{For instance, the proposed closure of the Alta Bates medical facility in  Berkeley, California is estimated to disproportionately impact vulnerable minorities \cite{ho2018report}.}

    As our society is becoming more reliant on automated decision-making, much effort has been directed at making the underlying algorithms fairer for disadvantaged socioeconomic groups \cite{EO2023AI}. For applications modeled as optimization problems such as the facility location problem, a common approach in the fairness literature is to suitably tweak the optimization problem to incorporate fairness, e.g., by proposing a fairer objective \cite{MS94}.

\begin{figure}[t]
        \begin{minipage}{0.54\textwidth}
            \centering
            \includegraphics[width=0.9\textwidth]{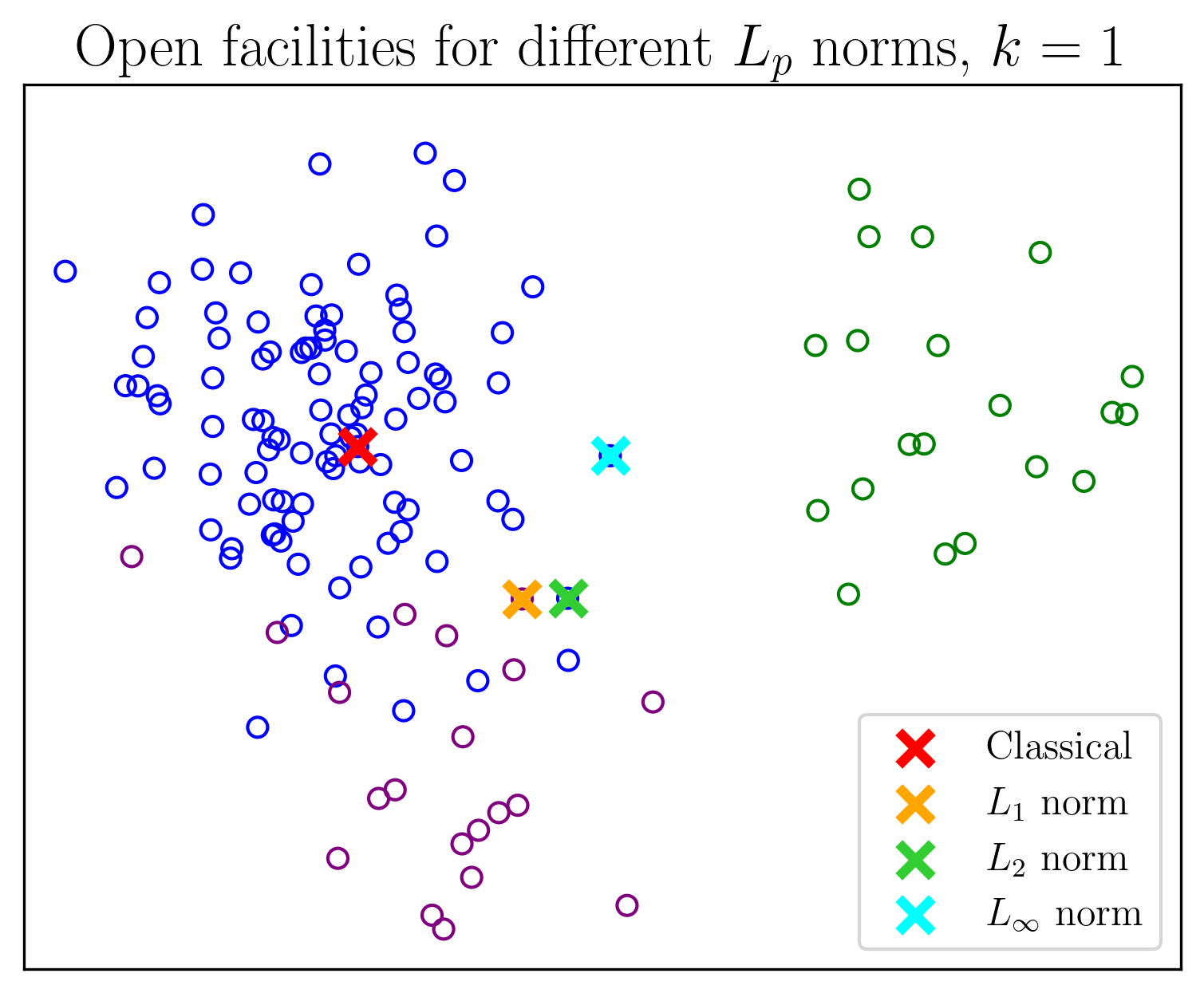}
            \footnotesize
            \begin{tabular}{|c|c|c|c|c|}
                \hline
                Group & Classical & $L_1$ norm & $L_2$ norm & $L_\infty$ norm \\ \hline
                1 (blue) & 0.21 & 0.42 & 0.47 & 0.46 \\ \hline
                2 (green) & 0.89 & 0.71 & 0.66 & 0.54 \\ \hline
                3 (purple) & 0.51 & 0.32 & 0.36 & 0.55 \\ \hline
                all clients & 0.35 & 0.45 & 0.48 & 0.48 \\ \hline
            \end{tabular}
        \end{minipage}
        \hfill
        \begin{minipage}{0.44\textwidth}
            \caption{An illustrative example for $k$-clustering with three client groups $X_1, X_2, X_3$ (in blue, green, and purple respectively) that partition client set $X$. We seek to open one facility anywhere in $X$. The optimal solution for classical objective $\sum_{j \in X} d(j, f)$ opens facility $f$ near the center of the blue group. If we minimize the $L_p$ norm of vector $\left(\frac{1}{|X_s|} \sum_{j \in X_s} d(j, f)\right)_{s = 1, 2, 3}$ of average group distances, then $f$ moves closer to the center of all groups as $p$ increases from $1$ to $\infty$. The table below shows average group distances for optimal solutions to different objectives:}
            \label{fig: three-groups-example}


        \end{minipage}
    \end{figure}

    However, this idea encounters two limitations for practical decision-making. First, fairness criteria, represented as different objective functions in optimization problems, are subjective and it is unclear how the choice of the `right' fairness criteria must be made. For example, \cite{KMR17} and \cite{chouldechova2018frontiers} have shown that it is impossible to satisfy various statistical notions of fairness simultaneously when classifying imperfectly across groups. In a similar vein, a large number of fairness criteria have been proposed for facility location problems (e.g., \cite{MS94, GSV22, ABV21, GJRYZ20} for clustering), and these can themselves be incompatible as well. For example, Figure \ref{fig: three-groups-example} considers a facility location problem with $3$ groups and four different objectives, each of which opens a facility at different locations. These solutions induce different average group distances for the three groups (blue, green, and purple). Note that the classical objective, $L_1$, and $L_2$ norms make the green group travel (on average) more than 1.375 times compared to the overall population. The $L_{\infty}$ norm on the other hand is fairer across groups but causes the overall population to travel 1.37 times more than the classical objective. Second, such optimization models treat fairness interventions as one-time solutions when in the real world these interventions are often spread across time.
    For example, the Justice40 initiative by the Biden administration has reserved 40\% of the resources for infrastructure development for historically disadvantaged communities. These resources will be released on a \emph{rolling} basis, making more budget available gradually over time. This paper aims for the modest goal of theoretically understanding what it means to have `fair' placement of necessary facilities and attempts to address the aforementioned challenges.

    \subsection{Portfolios}

    To address the first challenge of handling a large number of fairness objectives, we propose the notion of \emph{portfolios}. Given an optimization problem (e.g. facility location), a class $\mathbf{C}$ of objective functions (e.g., different fairness metrics) and a desired approximation $\alpha \ge 1$, a \emph{portfolio} $P$ is a \emph{small} set of feasible solutions to the optimization problem with the following guarantee: each objective function in the given class $\mathbf{C}$ must have an $\alpha$-approximate solution to it in the portfolio. Thus, portfolios are sets of solutions that are equitable under various fairness criteria.
    A decision maker can simply weigh the properties of the solutions in the portfolio, potentially including external factors as well, and make an informed choice. A similar concept has been used in other critical resource allocation scenarios, such as for organ transplants: e.g., the United Network for Organ Sharing (UNOS) chooses a policy for kidney transplants after carefully considering a portfolio of candidate policies and evaluating them on multiple axes of interest \cite{unos_2021}.
    Portfolios also generalize the well-studied notion of \emph{simultaneous approximations} \cite{KK00, GM06}, which are single solutions with approximation guarantees for all fairness metrics, i.e., portfolios of size $1$. Importantly, note that portfolios help translate the choice of the notion of fairness, to a choice in the space of solutions, thereby bridging theory with practice.

    There is a natural trade-off between portfolio size $|P|$ and the approximation factor $\alpha$. It is clear that if the approximation factor is relaxed, then the size of the portfolio required can only decrease. This paper attempts to study this trade-off for facility location problems while keeping the context of rolling budgets in mind:
    \begin{center}
    {\it When do small portfolios with good approximation guarantees exist for various fairness objectives in facility location problems? How do we obtain these portfolios efficiently? Can we further impose structural properties on solutions in a portfolio to accommodate for rolling budget settings?}
    \end{center}

    We study portfolios for fair versions of the \emph{(uncapacitated) facility location} and \emph{$k$-clustering} problems: two very well-studied problems in computer science and operations research \cite{hochbaum1982heuristics, STA97, CG99, CFS03, Li13, williamShmoysBook}.
    Both problems consider a set of clients $X$ (e.g. locations of individuals in a city) with distances $d$ (e.g. physical distance or travel time between locations) and seek to \emph{open} a subset $F \subseteq X$ of \emph{facilities} (e.g. hospitals) and assign each client $j \in X$ to an open facility $\Pi(j) \in F$. In (uncapacitated) facility location, we are also given \emph{facility opening costs} $c: X \to \R_{\ge 0}$ and the goal is to choose $(F, \Pi)$ that minimizes the sum of facility opening cost $c(F) := \sum_{f \in F} c(f)$ and total client distance $\sum_{j \in X} d(j, \Pi(j))$. In the $k$-clustering problem, exactly $k$ facilities must be opened to minimize the total client distance $\sum_{j \in X} d(j, \Pi(j))$. The latter problem explicitly avoids the trade-off between the cost of opening facilities and the distances, which will make it better suited to the rolling budget setting. In particular, solutions to the fair versions of the facility location problem sometimes tend to worsen the access costs for all groups to achieve a better trade-off between equity (i.e., balance of access costs) and efficiency (i.e., cost of opening facilities) (see an example in Appendix \ref{app: facility-access-cost-tradeoff}), especially when the number of groups is large. Instead, fair $k$-clustering lets us explicitly control the number of open facilities.

    \begin{table}[t]
        \centering
        \caption{A summary of approximations in this paper for arbitrary approximation factor $\alpha > 1$. $n = |X|$ is the number of clients and $r$ is the number of client groups. Given a budget $k$ for open facilities and real numbers $\alpha, \beta \ge 1$, a solution is called $(\beta, \alpha)$-bicriteria approximation if (1) its objective value is within factor $\alpha$ of the optimum and (2) it opens at most $\beta k$ facilities. Results marked with * are restricted to $L_p$ norms and do not generalize to other classes of norms considered in this work.}
        \label{tab: results-summary}
        \hspace*{-1.8em}
        \footnotesize
        \begin{tabular}
        {|ccc|c|c|c|c|}
            \hline
            \multicolumn{3}{|c|}{Problem} &
            Reference &
            \begin{tabular}[c]{@{}c@{}}Portfolio\\ approximation ratio\end{tabular} &
            \begin{tabular}[c]{@{}c@{}}Portfolio\\ size\end{tabular} &
            Complexity
            \\ \hline
            \multicolumn{1}{|c|}{\multirow{4}{*}{\begin{tabular}[c]{@{}c@{}}Fair Facility\\ Location\end{tabular}}} &
            \multicolumn{2}{c|}{\multirow{3}{*}{\begin{tabular}[c]{@{}c@{}}Upper  bound\end{tabular}}} &
            \textbf{This work} &
            \multirow{2}{*}{$\alpha$} &
            \multirow{2}{*}{$O\left(\log_{\alpha} r\right)$} &
            \multirow{2}{*}{Existence} \\
            \cline{4-4}
            \multicolumn{1}{|c|}{} &
            \multicolumn{1}{c}{} &
            &
            Golovin et al.* &
            &
            &
            \\ \cline{4-7}
            \multicolumn{1}{|c|}{} &
            \multicolumn{2}{c|}{} &
            \textbf{This work} &
            $4 \alpha$ &
            $O\left(\log_{\alpha} r\right)$ &
            Polynomial-time
            \\ \cline{2-7}
            \multicolumn{1}{|c|}{} &
            \multicolumn{2}{c|}{Lower bound} &
            \textbf{This work*} &
            $\alpha$ &
            $\Omega\left(\log_{2\alpha} r\right)$ &
            Existence
            \\ \hline\hline
            \multicolumn{1}{|c|}{\multirow{5}{*}{\begin{tabular}[c]{@{}c@{}}Fair\\ $k$-Clustering\end{tabular}}} &
            \multicolumn{2}{c|}{\multirow{3}{*}{\begin{tabular}[c]{@{}c@{}}Upper  bound\end{tabular}}} &
            \textbf{This work} &
            \multirow{2}{*}{$(1, \alpha)$} &
            \multirow{2}{*}{$O\left(\log_{\alpha} r\right)$} &
            \multirow{2}{*}{Existence}
            \\ \cline{4-4}
            \multicolumn{1}{|c|}{} &
            \multicolumn{1}{c}{} &
            &
            Golovin et al.* &
            &
            &
            \\ \cline{4-7}
            \multicolumn{1}{|c|}{} &
            \multicolumn{2}{c|}{} &
            \textbf{This work} &
                {$(4, 4\alpha)$} &
            $O\left(\log_{\alpha} r\right)$ &
            Polynomial-time
            \\ \cline{2-7}
            \multicolumn{1}{|c|}{} &
            \multicolumn{2}{c|}{\multirow{2}{*}{Lower bound}} &
            \textbf{This work*} &
            $(1, \alpha)$ &
            $\Omega\left(\log_{2\alpha} r\right)$ &
            \multirow{2}{*}{Existence}
            \\ \cline{4-6}
            \multicolumn{1}{|c|}{} &
            \multicolumn{2}{c|}{} &
            Golovin et al.* &
            $(1, \alpha)$ &
            $\Omega\left(\left(\log_{\alpha} r\right)^{1/3}\right)$
            &
            \\ \hline\hline
            \multicolumn{1}{|c|}{\multirow{4}{*}{\begin{tabular}[c]{@{}c@{}}Clustering\\ Refinement\end{tabular}}} &
            \multicolumn{1}{c|}{\multirow{2}{*}{\begin{tabular}[c]{@{}c@{}}Upper\\ bound\end{tabular}}} &
            Arbitrary metrics &
            \textbf{This work} &
            $(O(1), \poly(n^{\frac{1}{\sqrt{\log n}}}))$ &
            $O(\log r \cdot \log n)$ &
            \multirow{2}{*}{Polynomial-time}
            \\ \cline{3-6}
            \multicolumn{1}{|c|}{} &
            \multicolumn{1}{c|}{} &
            \begin{tabular}[c]{@{}c@{}}Line and tree\\ metrics\end{tabular} &
            \textbf{This work} &
            $(O(1), O(\log n))$ &
            $O(\log r \cdot \log n)$ &
            \\ \cline{2-7}
            \multicolumn{1}{|c|}{}  &
            \multicolumn{2}{c|}{Lower bound} &
            \multicolumn{4}{c|}{Open}
            \\ \hline
        \end{tabular}
    \end{table}

    Next, to define the fair versions of the above-mentioned two problems, we consider the social theory of \emph{intersectionality}, which deals with understanding how various aspects of a person's social and political identities (like race, gender, class, etc.) overlap and intersect \cite{benet2014oxford, foulds2020intersectional}. We allow each person or client in the model to have a fractional group membership in various racial, economic, and gender groups. Mathematically, we represent the fractional membership with $\mu_{j, s} \ge 0$ for each client $j \in X$ and each of $r$ groups $s = 1, \ldots, r$.
    Fractional group memberships have not received much attention in the existing literature at the intersection of optimization, ML, and algorithmic fairness, which usually assumes that clients divide into non-intersecting groups, or considers the finest partition of intersections as unique individual groups \cite{DHPRZ12, CKLV17, celis2020interventions}.

    Given these groups, we would like the fair facility placement problems to balance the total distances traveled by the $r$ different groups, defined as $d_\Pi(s) := \sum_{j \in X} \mu_{j, s} d(j, \Pi(j))$ for each group $s \in [r]$.
    To balance these group distances, we consider the $L_p$ norms $\|d_\Pi\|_p := \left(\sum_{s \in [r]} d_\Pi(s)^p\right)^{1/p}$, i.e, the $L_p$ norm objective is $\sum_{f \in F} c(f) + \|d_\Pi\|_p$ for \emph{fair} facility location and $\|d_\Pi\|_p$ for \emph{fair} $k$-clustering. Fair facility location with $L_1$ norm (i.e. sum of group distances) and individual fairness is the classical (uncapacitated) facility location problem. Fair $k$-clustering with individual fairness is the classical $k$-median problem for the $L_1$ norm and the classical $k$-center problem for the $L_\infty$ norm.
    Parameter $p$ provides a \emph{smooth, monotonic interpolation} between $L_1$ and $L_\infty$ norms; see Figure \ref{fig: three-groups-example} for an illustration. We next discuss our results for the first part of the paper on portfolios, and will subsequently introduce the rolling budget setting.

    \textbf{Results.} An $\alpha$-approximate portfolio for $L_p$ norms is a set of solutions such that for any $p \in [1, \infty]$, this set has an $\alpha$-approximate solution for the $L_p$ norm objective.
    Since there are an exponential number of feasible solutions, it is unclear whether small (i.e., constant or logarithmic in the size of $r$) portfolios for $L_p$ norms even exist. Further, as we note in Section \ref{sec: portfolio-upper-bound}, there are instances where a single solution cannot be an $o(\sqrt{r})$-approximation for all $L_p$ norms.
    It is easy to show that logarithmic-sized portfolios for $L_p$ norms always exist for this problem, by generalizing the key argument for top-$l$ norms \cite{GM06} to $L_p$ norms:

    \begin{result}[Upper Bounds on the Size of Portfolios, Theorem \ref{thm: portfolio-upper-bound}]
        For both fair facility location and fair $k$-clustering with $r$ groups, for each $\alpha > 1$, there exists an $O\left(\log_\alpha r\right)$-size portfolio, admitting an $\alpha$-approximation for $L_p$ norms. Further, one can compute in polynomial time using Algorithm \textsc{StepPortfolio} an $O(\log_\alpha r)$-size portfolio for fair facility location and fair $k$-clustering attaining slightly weaker approximations of $4\alpha$ and $(4, 4\alpha)$ respectively, where the latter bicriteria approximation relaxes the requirement of exactly $k$ facilities to guarantee that at most 4$k$ will be open.
    \end{result}

    The proof is constructive and relies on two properties: (1) the optimal value for both fair facility location and fair $k$-clustering  is monotone as a function of $p$, and (2) the optimal values for the $L_1$ and $L_\infty$ norms are within factor $r$ of each other. One can then select a subset $S =\{1=p_0 < p_1 < \hdots < p_{O(\log_\alpha r)}=\infty\}$ of $p$ values so that the value of corresponding optimal solution to $p_i$ is a factor $\alpha$ away from the optimal solution to $p_{i+1}$. We call this algorithm \textsc{StepPortfolio}, and show that it works for various classes of norms, including convex combinations of $L_1$ and $L_{\infty}$ norms.

    Further, note that the above result highlights a crucial difference between the existence and polynomial time computable portfolios for given approximation factors. This difference stems from the fact that both fair facility location and $k$-clustering are APX-complete even for $p = 1$ and $p = \infty$, so portfolios with arbitrarily good approximation factors cannot be found in polynomial time unless P = NP.

    In terms of lower bounds of achievable sizes of portfolios given an approximation factor, there is much less that can be inferred from existing results. For example, \cite{GGKT08} showed that to attain a size-1 portfolio for $k$-clustering that achieves simultaneous approximation for arbitrary $\alpha>1$, one needs to open at least $\Omega((\log_{\alpha k}r)^{1/3})$ facilities. There was no such lower bound known for fair facility location, to the best of our knowledge. We improve on both these settings and show complementing lower bounds compared to Result 1 (Theorem \ref{thm: portfolio-upper-bound}). These bounds do not exploit the structure of groups, and in fact, work in the case where every individual is their own group (i.e., $r=|X|$).

    \begin{result}[Lower Bounds on the Size of Portfolios, Theorem \ref{thm: portfolio-lower-bound}]
        For each $\alpha > 1$ and for all $r$ large enough, there exist instances of fair facility location and fair k-clustering with $r$ groups ($r=|X|$), where any $\alpha$-approximate portfolio for $L_p$ norms must have size $\Omega(\log_{2\alpha}r)$.
    \end{result}

    Note that Results 1 and 2 imply that our constructions are order-optimal. So far, we have modeled fairness in facility placement problems as a static problem: given the locations of clients and their group memberships, we wanted to find a portfolio of approximate solutions. We next introduce portfolios that accommodate rolling budgets, so that the number of facilities grows over time with additional structure.

    \begin{figure}
        \centering
        \begin{minipage}{0.30\textwidth}
            \centering
            \includegraphics[width=0.8\textwidth]{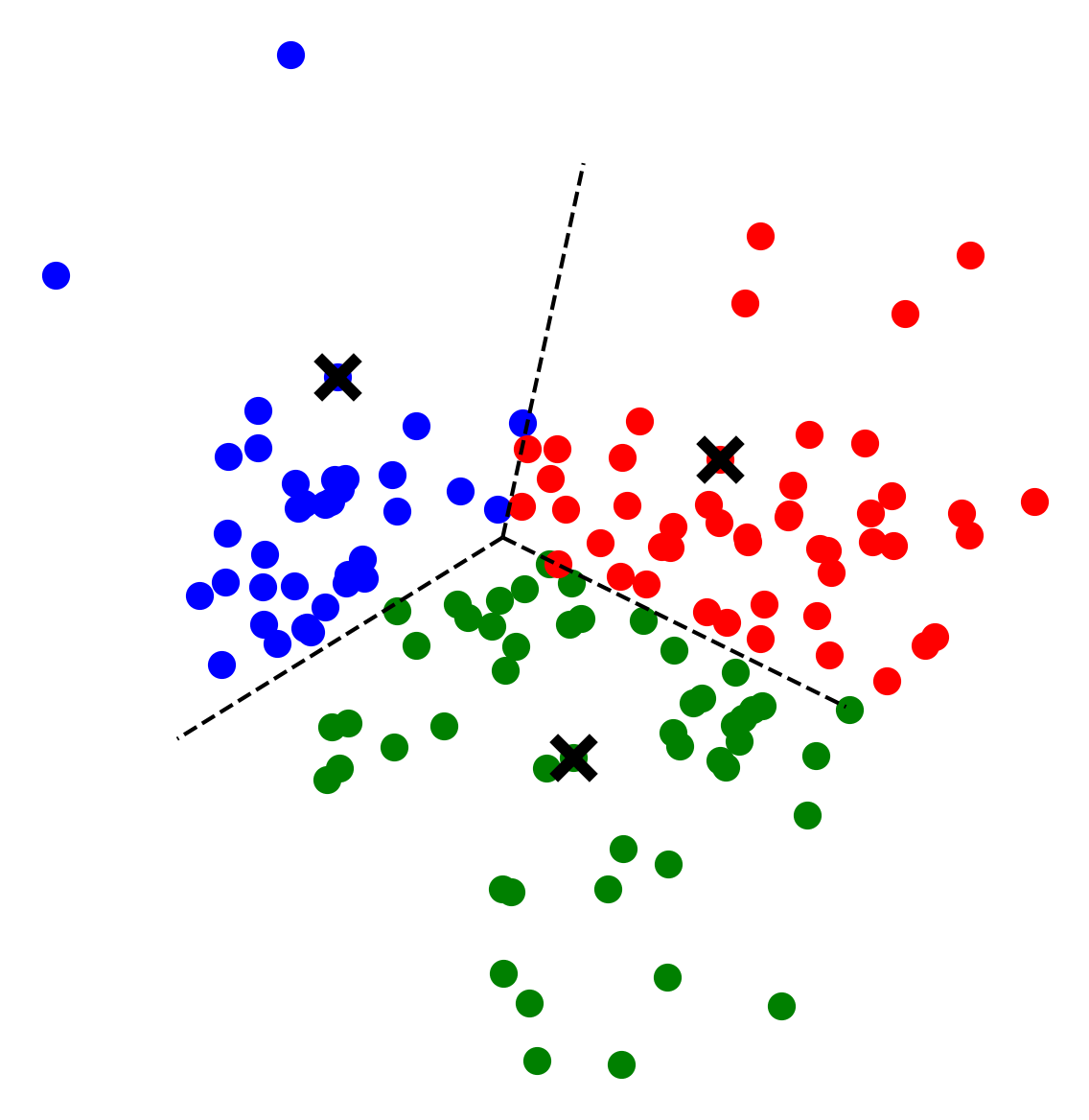}
        \end{minipage}
        \hfill
        \begin{minipage}{0.30\textwidth}
            \centering
            \includegraphics[width=0.8\textwidth]{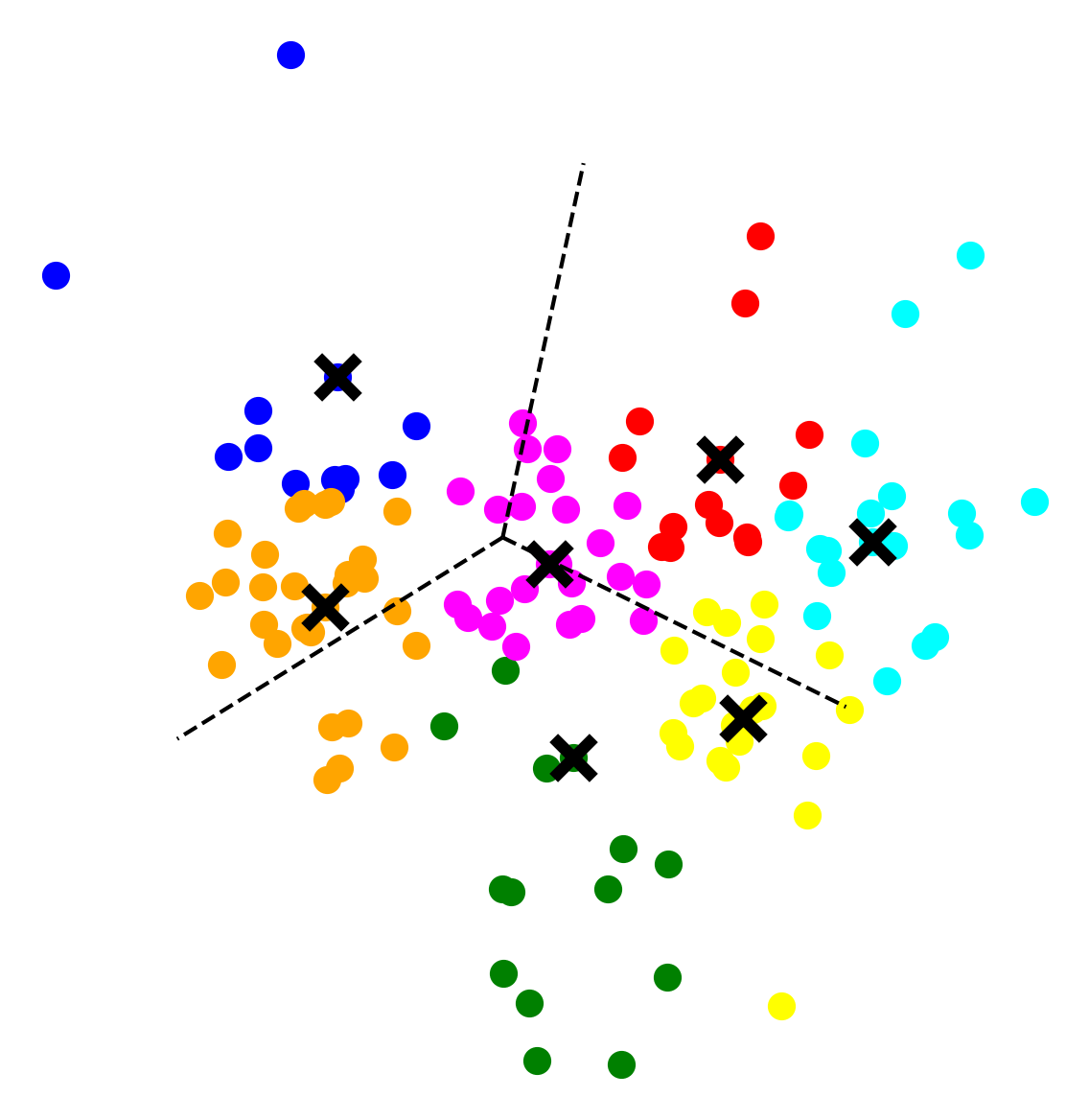}
        \end{minipage}
        \hfill
        \begin{minipage}{0.30\textwidth}
            \centering
            \includegraphics[width=0.8\textwidth]{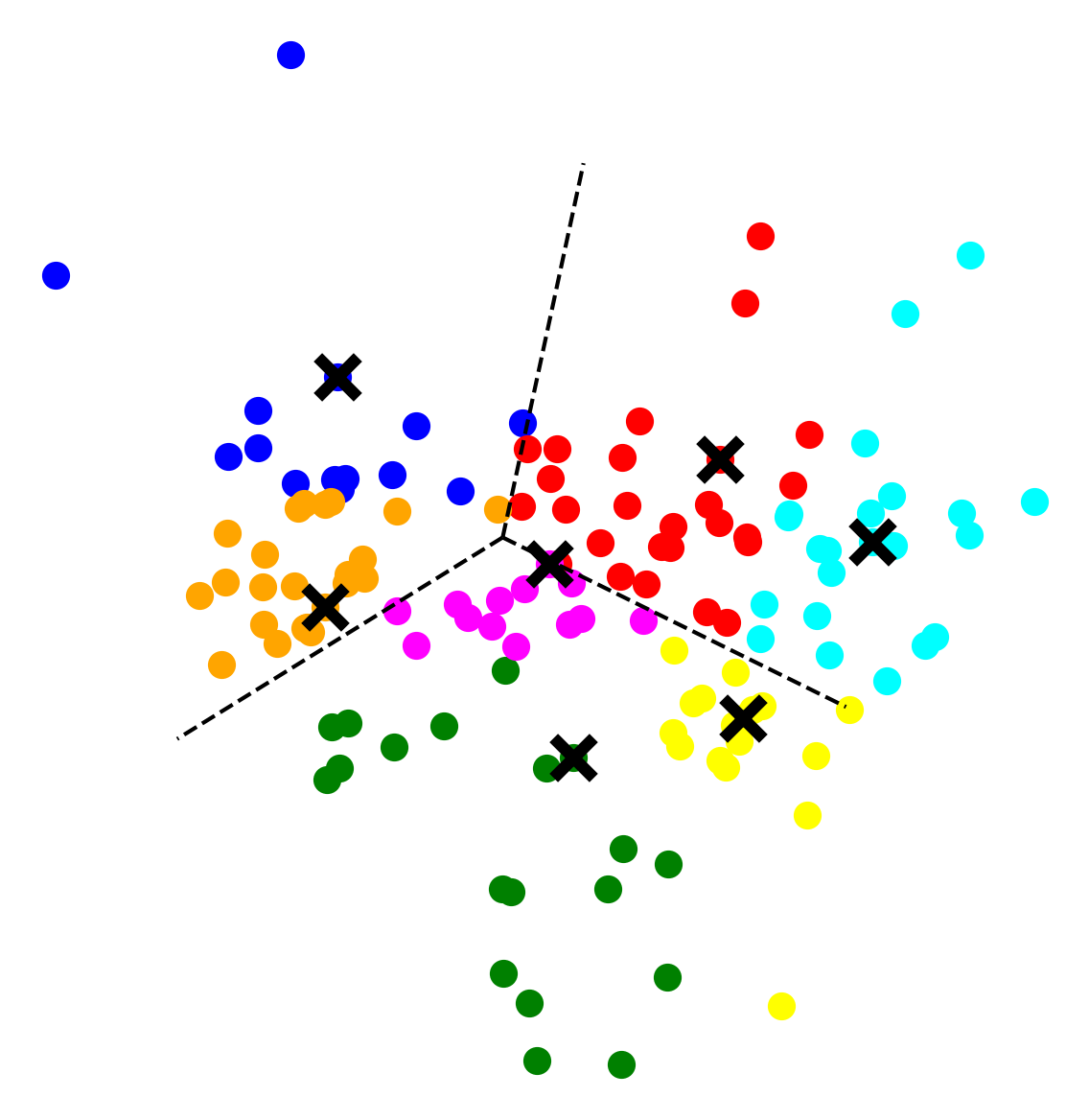}
        \end{minipage}
        \caption{An illustration for refinements for $k$-clustering. Consider budgets $k_1 = 3, k_2 = 7$. Each circle is a client and each cross is an open facility. (left) A solution with $k_1 = 3$ facilities with colors representing assignment to facilities.  (center) A solution with $k_2 = 7$ open facilities, obtained by opening $4$ more facilities. However, this does not form a refinement for $k$-clustering since the set of clients colored orange, pink, and yellow intersect the dotted boundaries. (right) A different assignment for the same open facilities that forms a refinement.}
        \label{fig: refinement-example}
    \end{figure}

    \subsection{Structured Portfolios with Refinements}

    As alluded to before, infrastructure investments often spread over time, e.g., the Justice40 policy plans to provide a series of investments to ``improve how the Federal government ensures equitable distribution of the benefits of many programs'' \cite{Justice40}. In such a setting, for each new facility opened with a rolling budget, it is useful to refine the partition of clients served by the current set of facilities. This eases the administrative burden of managing clients, as well as improves accessibility in such a system. For example, for existing medical centers, new hospitals or even new local clinics can be opened with a rolling investment, so that local clinics can serve a set of patients assigned to a particular medical center. This is a practically motivated constraint and has been considered in the literature in the context of hierarchical facilities \cite{farahani2014hierarchical}. We do not consider facilities to be hierarchically placed, but instead ask that for any facility $f_t$ that is open at time $t$, all its clients should come from those served by a single other open facility $f_{t - 1}$ at time $t - 1$. Figure \ref{fig: refinement-example} gives an example of a refined partition (right) of the original facility locations (left), and a comparative partition of the clients that does not satisfy refinements (middle).

    To construct portfolios of solutions with refinement guarantees, we develop an algorithm \hspace{0pt}\textsc{DiscountedLookAhead} that given any sets $F_1 \subseteq \ldots \subseteq F_l \subseteq X$ of open facilities, constructs assignments of clients to open facilities for each time so that the assignments satisfy (1) refinements as explained above, and (2) distances of clients to assigned facilities are only some factor $\alpha$ more than their distances to the nearest open facility.
    The latter property allows us to prove approximation guarantees on the refined portfolios.

    \paragraph{Refinements for Facility Location:} Recall that the portfolios from Section 1.2 did not enforce any structural properties on the various solutions. In particular, one of the solutions might open facilities $f_1, f_2, f_3$, and the other solution might open $f_2, f_5$. We convert this portfolio to have a chain structure as discussed above, and then using {\sc DiscountedLookAhead} we immediately get the following result:

    \begin{result}[Facility Location Portfolios that Satisfy Refinements, Theorem \ref{thm: facility-location-refinement}]
        Given an instance of fair facility location on metric space $(X, d)$ and $r$ client groups, using Algorithm \textsc{DiscountedLookahead}, we construct a portfolio $P = \left\{F_1 \subseteq \ldots \subseteq F_l\right\}$ with corresponding assignments that form a refinement, so that for each $p \in [1, \infty]$, there exists an $F_t$ that is $\poly(2^{\sqrt{\log r}})$-approximation for the $L_p$ norm objective.
    \end{result}

    Note that, without the refinement property, it is easy to construct such portfolios with a $O(1)$-approximation. The hard part is to not increase the client-facility distances by a lot (in fact, a greedy assignment can increase the distances polynomial in the number of groups, Section \ref{sec: discounted-lookahead}), by modifying assignments so that they satisfy the refinements condition. Result 3 shows an approximation of $2^{\sqrt{\log_2 r}} = r^{1/\sqrt{\log_2 r}}$, which is \emph{subpolynomial} in $r$, i.e., it is asymptotically smaller than $r^\delta$ for any constant $\delta > 0$. Result 4 will show better approximations when the structure in the underlying metric can be exploited.

    \paragraph{Refinements for $k$-Clustering:} Next, to explicitly consider the number of facilities placed, and not trade-off the facility opening costs with distances traveled by clients implicitly, we first consider refinements for the $k$-clustering problem for a given $L_p$-norm. Given budgets $k_1 < k_2 < \ldots < k_l$ for the numbers of open facilities at different times $t = 1, \ldots, l$, we would like to open subsets of facilities $F_1 \subseteq \ldots \subseteq F_l$ so that these satisfy the refinements structure, and for each $t$, $|F_t| = O(k_t)$ in size, as well as $\alpha$-approximate for the given $L_p$ norm. Since this definition requires $\alpha$-approximate cost guarantees at all times $t \in [l]$, it is unclear whether there even exists such a refinement for reasonable $\alpha$.
    We use the ideas in {\sc DiscountedLookAhead} to show analogous results to Result 3, depending on the distance metric assumed.

    \begin{result}[$k$-clustering Refinement for given $L_p$-norm, Theorem \ref{thm: refinement-general-metric}]
        Given a metric space $(X,d)$ on $n$ points, norm $p \in [1, \infty]$, and budgets $k_1 < \ldots < k_l$ for opening facilities at times $1, \ldots, l$, using \textsc{DiscountedLookahead}
        we can obtain a sequence $\{F_1 \subseteq \ldots \subseteq F_l\}$ of facilities with assignments that satisfy refinement, so that
        \begin{enumerate}
            \item for each time $t \in [l]$, at most $O(k_t)$ facilities are open (i.e., $|F_t| = O(k_t)$), and
            \item $F_t$ is a $\poly(2^{\sqrt{\log n}})$-approximation for the $L_p$ norm objective for $k_t$ facilities.
        \end{enumerate}
    \end{result}

    Next, we consider all $L_p$ norms to circumvent the issue of the choice of fairness. We can similarly design a portfolio of solutions that is a refinement so that for each $L_p$-norm there exists a chain in the portfolio that provides a good approximation for each given budget of the open facilities.

    \begin{result}[$k$-clustering Portfolio with Refinement, Theorem \ref{thm: refinement-general-metric}]
        Given a metric space $(X,d)$ on $n$ points and budgets $k_1 < \ldots < k_l$ for opening facilities at times $1, \ldots, l$, using \textsc{DiscountedLookahead} and \textsc{StepPortfolio} we can construct a portfolio $\mathbf{P}$ of size $O(\log r \cdot \log n)$, with each solution $\mathbf{S}_i$ of the form $\mathbf{S}_i = \{F_1(i) \subseteq \ldots \subseteq F_l(i)\}$ with assignments that satisfy refinement, so that for any $p \in [1, \infty]$, there exists an $\mathbf{S}_i\in P$, so that:
        \begin{enumerate}
            \item it opens at most $O(k_t)$ facilities for each time $t \in [l]$, and
            \item $F_t(i)$ is a $\poly(2^{\sqrt{\log n}})$-approximation for the $L_p$ norm objective for $k_t$ facilities for each $t \in [l]$.
        \end{enumerate}
    \end{result}

    The approximation factor we obtain for refinements of $k$-clustering is not one that one commonly encounters in the literature. Though it is open whether an improved approximation can exist in general, a natural question is if we could do better for more structured metrics. We answer affirmatively when distances $d$ correspond to shortest path distances on a tree. We can improve this approximation factor significantly, using our novel algorithm \textsc{ExpandIntervals} for line metric and \textsc{BranchAndLinearize} that extends it to tree metrics. These algorithms can be adapted to all of the above-mentioned settings, but we state the result only for $k$-clustering portfolios:

    \begin{result}[$k$-Clustering Portfolio Refinement for Tree Metrics, Theorem \ref{thm: refinement-general-metric}]
        Given a metric space $(X,d)$ on $n$ points and budgets $k_1 < \ldots < k_l$ for opening facilities at times $1, \ldots, l$, using \textsc{ExpandIntervals} and \textsc{BranchAndLinearlize} we can construct a portfolio $\mathbf{P}$ of size $O(\log r \cdot \log n)$, with each solution $\mathbf{S}_i$ of the form $\mathbf{S}_i = \{F_1(i) \subseteq \ldots \subseteq F_l(i)\}$ with assignments that satisfy refinement, so that for any $p \in [1, \infty]$, there exists an $\mathbf{S}_i\in P$, so that:
        \begin{enumerate}
            \item it opens at most $O(k_t)$ facilities for each time $t \in [l]$, and
            \item $F_t(i)$ is an $O(\log n)$-approximation for the $L_p$ norm objective for $k_t$ facilities for each $t \in [l]$.
        \end{enumerate}
    \end{result}

    We remark that the natural approach of embedding any metric into a distribution of tree metrics \cite{alon1995graph, bartal1998approximating, fakcharoenphol2004approximating} with bounded distortion does not give any guarantee for our problem. This is due to the fact that our objective is non-linear while such an approach utilizes the linearity of the objective function crucially.

    \subsection{Experiments}

    We complement our theoretical results with two sets of experiments on US census data.
    The first builds on our online tool for identifying medical deserts (see Figure \ref{fig: online-tool}) and proposes potential sites for new pharmacies alongside existing CVS, Walgreens, and Walmart pharmacies in the state of Mississippi, USA.
    We divide the population into $18$ demographic groups and give a portfolio of three solutions based on different $L_p$ norm objectives with these groups.
    Each solution proposes opening up to 75 new facilities and recommends the sequence to open them, thus incorporating rolling budgets (i.e., each solution forms a $k$-clustering refinement but without specific assignments).
    Although each solution in the portfolio recommends different facilities, they all reduce the number of medical deserts identified by our tool from $407$ to less than $190$, and \emph{opening just the first 15 of the 75 proposed facilities reduces the number of medical deserts by about $70$.} Further, most of these $\sim 70$ blockgroups are majority Black or African American, thus \emph{mitigating some of the disproportionate impact of medical deserts on the Black population.}
    These results are presented with further details in Section \ref{sec: experiment-pharmacies}. Our online tool, combined with these insights, could be a valuable resource for policymakers evaluating the locations of medical deserts and potential new pharmacy sites.

    Our second experiment on Fulton County and Chatham County in the state of Georgia, USA validates our theoretical results for refinements with additional structure on assignments. We consider opening a sequence of $l$ facilities (e.g. hospitals) in both counties (where $l = 16$ for Fulton County and $l = 12$ for Chatham County). We create 24 client groups based on racial demographics, health insurance level, and poverty level.
    In the first stage, we seek to open $l$ facilities iteratively, i.e., seek refinements without assignments. In this stage, each client is assigned to its closest open facility at each time. We compare our algorithm against the natural greedy algorithm that iteratively opens the best facility and show that \emph{our algorithm has a better approximation ratio than the greedy algorithm at most times.}
    In the second stage, we also seek assignments that form a refinement and show that using \textsc{DiscountedLookahead} to impose this additional structure on assignments does not significantly change the approximation. That is, \emph{additional refinement structure on assignments can be imposed with little additional cost in practice}.

    \paragraph{Outline.} We present related work next in Section \ref{sec: related-work} and formal problem statements and preliminaries in Section \ref{sec: preliminaries}.
    In Sections \ref{sec: portfolio-upper-bound} and \ref{sec: portfolio-lower-bound}, we deal respectively with upper and lower bounds on portfolio sizes (Results 1, 2).
    In Section \ref{sec: refinements}, we reduce refinements for both fair facility location and $k$-clustering refinements to the \textsc{DiscountedLookahead} algorithm.
    Next, in Section \ref{sec: discounted-lookahead}, we discuss \textsc{DiscountedLookahead} to get refinements for arbitrary metrics. In Section \ref{sec: line-metric}, we give algorithm \textsc{ExpandIntervals} that improves upon the guarantee of \textsc{DiscountedLookahead} for line metrics, while the algorithm for tree metric is included in Appendix \ref{app: tree-metric}.
    We discuss the first experiment in Section \ref{sec: experiment-pharmacies} and conclude with open questions in Section \ref{sec: conclusion}.
    The second experiment is deferred to Appendix \ref{app: second-experiment}.

    \section{Related Work}\label{sec: related-work}

    Variants of both uncapacitated facility location \cite{hochbaum1982heuristics, CNW83, STA97, Sviridenko97, GK98, KGPR98, CG99, KMS02, JMMSV03, CFS03, Byrka07, Li13} and $k$-clustering \cite{LV92, arora1998approximation, charikar1999constant, SO13, cohen2022improved, chakrabarty2022approximation} are very well-studied in the computer science and operations research literature, with current best approximation ratios of $1.488$ \cite{Li13} for uncapacitated facility location, $2.675$ \cite{BPRST17} for $k$-median, and $2$ \cite{HS86} for $k$-center problem respectively. Each of the three problems is APX-hard, and the $2$-approximation for $k$-center in particular is tight unless P = NP \cite{hsu1979easy}.
    Most of these approximation algorithms crucially use the linearity of the objective and therefore do not generalize to our setting.

    $L_p$ norm objectives are widely considered in approximation algorithms literature as a model for fairness and as interesting theoretical questions \cite{AYZ04, GGKT08, FTT21, MSV21}. In particular, \cite{KK00, GGKT08, CMV21} study various fixed norm objectives including $L_p$ norm objectives for $k$-clustering.

    To the best of our knowledge, we are the first to explicitly study portfolios for fair optimization, although similar notions are implicit in the works of \cite{KK00, GM06, GGKT08}. \cite{GGKT08} in particular study $L_p$ norms and effectively give $O(\log r)$-sized $O(1)$-approximate portfolios for $k$-clustering (see Result 1) and other problems. They generate $O(\log r)$ $L_p$ norms such that any $L_q$ norm is $2$-approximated by one of these $O(\log r)$ norms; their technique however crucially uses the structure of $L_p$ norms.
    Our technique for portfolio upper bounds closely resembles the technique of \cite{GM06} who use it to get portfolios for top-$l$ norms. \cite{GGKT08} also give a lower bound of $\Omega\left(\left(\log_\alpha r\right)^{1/3}\right)$ for $\alpha$-approximate portfolios for $L_p$ norms for individually fair $k$-clustering; we improve this bound to $\Omega(\log_\alpha r)$.

    The special case of size-$1$ portfolios, also called simultaneous approximations, has been more extensively studied (e.g. \cite{KK00, AERW04, GM06}). \cite{GM06} in particular study simultaneous approximations for various problems and establish a fundamental result that shows that simultaneous $\alpha$-approximations for top-$l$ norms are in fact simultaneous $\alpha$-approximations for many other classes of norms, including for $L_p$ norms; this turns out to be false when portfolio size is $> 1$ (see Section \ref{sec: portfolio-upper-bound}).

    Many other kinds of fairness criteria in facility location problems are well-studied from both theoretical and applied perspectives \cite{Truelove93, Talen98}. A rich variety of variants and fairness objectives has developed over the years for settings including for individual fairness \cite{jung2019center, MC21, bajpai2022revisiting}, group fairness \cite{ABV21, MSV21, GSV22}, games \cite{goemans2004cooperative, HT09}, and proportional fairness \cite{CKLV17, BCFN19, ahmadi2020fair, harb2020kfc, SZ22}.  Most of these models either fix a fairness criterion or do not discuss how a choice among a suite of fairness criteria must be made. Portfolios can generally be defined for any model with multiple fairness criteria with corresponding questions on the trade-off between portfolio size and approximation.

    Facility location problems with multiple sets of open facilities have been widely studied under the umbrella term `hierarchical facility location' \cite{Bumb01, Ageev02, AYZ04, Zhang04, KP09, GJ10, WDGX10}. All of these problems specify different `levels' similar to the notion of `time' for refinements; each level has its own set of open facilities.
    Unlike refinements that require cost guarantees at each `level', these problems are all unicriteria (i.e. the objective is to minimize total cost across all levels).
    Additionally, most of these problems do not impose a chain structure over sets of open facilities.

    \section{Preliminaries}\label{sec: preliminaries}

    In this section, we introduce some notation and give formal problem definitions. We also show that the algorithm of \cite{STA97} for classical uncapacitated facility location gives constant-factor approximations for fair facility location and fair $k$-clustering with any fixed norm objective. This result will be useful later to obtain portfolios and solutions that satisfy refinements.

    Throughout, we work with a given metric space $(X, d)$ on $|X| = n$ points or \emph{clients} and $r$ \emph{client groups} given to us through group memberships $\mu_{j, s} \ge 0$ for each client $j \in X$ and group $s \in [r]$. The special case $r = n$ and $\mu_{j, s} = 0$ for all $j \neq s$ (i.e., each client is in their own group) is called the \emph{individually fair} problem. Facilities can be opened anywhere in $X$; given a set of \emph{open facilities} $F \subseteq X$ and a client assignment $\Pi: X \to F$, the group distance $d_\Pi(s)$ for group $s \in [r]$ is defined as $\sum_{j \in X} \mu_{j, s} d(j, \Pi(j))$.

    \begin{definition}[Fair Facility Location (FFL) and Fair $k$-Clustering (\textsc{FC})]
        Given a metric space $(X, d)$, a set $X$ of clients with group memberships $\mu$ and facility opening costs $c: X \to \R_{\ge 0}$, the fair facility location (FFL) problem seeks to find a set of open facilities $F^*$ and an assignment $\Pi^*: X \rightarrow F^*$ of clients to open facilities that minimize the following objective
        $$
        (F^*, \Pi^*) \in {\arg\min}_{\substack{F \subseteq X, \\ \Pi: X \to F}}
        \underbrace{\sum_{f \in F} c(f)}_{\text{Facility cost }  c(F)} + \underbrace{g\left(d_{\Pi}(1), \ldots, d_{\Pi}(r) \right)}_{\text{Access cost } g(d_\Pi)}, \quad\quad  (\text{FFL}_{g})
        $$
        where $g$ is a given norm on $\R^r$. In particular, when $g$ is the $L_p$ norm for some $p \in [1, \infty]$, we denote this problem as FFL$_p$. An instance of fair $k$-clustering (FC) is similarly defined, where instead of facility opening costs, a budget $k$ is given on the number of open facilities. The FC$^{(k)}_g$ problem seeks to find $(F^*, \Pi^*)$ so that
        $$
        (F^*, \Pi^*) \in {\arg\min}_{\substack{F \subseteq X: |F| \le k, \\ \Pi: X \to F}} g\big(d_{\Pi}(1), \ldots, d_{\Pi}(r) \big). \quad\quad (\text{FC}_{g}^{(k)})
        $$
    \end{definition}

    A solution $(F, \Pi)$ to FFL$_g$ (or FC$^{(k)}_g$) is $\alpha$-approximate if the objective value of $(F, \Pi)$ is within factor $\alpha$ of the optimal. For FC$_g^{(k)}$, we also study the weaker notion of \emph{bicriteria solutions}, where a solution to $(F, \Pi)$ is bicriteria $(\beta, \alpha)$-approximation if (1) it opens at most $\beta k$ facilities (i.e. $|F| \le \beta k$) and (2) the objective value of $(F, \Pi)$ for FC$_g^{(k)}$ is within factor $\alpha$ of the optimal for FC$_g^{(k)}$. Bicriteria $(1, \alpha)$-approximations are the same as $\alpha$-approximations.

    Next, we define portfolios for fair facility location and fair $k$-clustering as a way of providing representative solutions for a given class $\mathbf{C}$ of equity objectives:

    \begin{definition}[Portfolios for Fair Facility Location and Fair $k$-Clustering]\label{def: portfolios}
    Given an instance of fair facility location with $r$ client groups and a class $\mathbf{C}$ of norms on $\R^r$, a set $P$ of solutions to the instance is called an $\alpha$-approximate portfolio for $\mathbf{C}$ if for each norm $g \in \mathbf{C}$, $P$ contains a solution that is an $\alpha$-approximation to FFL$_g$.
    Bicriteria approximate portfolios are similarly defined for fair $k$-clustering.
    \end{definition}

    All our upper bounds for portfolios and refinements generalize beyond $L_p$ norms to any class of norms that \emph{interpolate} between the $L_1$ and $L_{\infty}$ norms, which are typically the extremes of efficiency and equity objectives respectively. We make this notion of `interpolation' precise in the next definition:

    \begin{definition}[Monotonically Interpolating Norm Class]\label{def: monotonically-interpolating-norm-class}
    Consider a class $\mathbf{C} = \{g_\lambda: \R^r \to \R; \lambda \in [a, b]\}$ of monotone norms $g_\lambda$ on $\R^r$ parameterized by $\lambda$ that varies between fixed real numbers $a, b$. We say that $\mathbf{C}$ \emph{interpolates monotonically} between $L_1, L_\infty$ norms if (1) $g_a$ is the $L_1$ norm on $\R^r$, (2) $g_b$ is the $L_\infty$ norm, and (3) $g_\lambda$ decreases with $\lambda$, i.e., for $\lambda < \mu$, for all $x \in \R^r$, $g_\lambda(x) \ge g_{\mu}(x)$.
    \end{definition}

    Geometrically, monotonic interpolations correspond to continuously expanding the $L_1$ norm ball to the $L_\infty$ ball. Many norm classes including $L_p$ norms fall in this category; these are discussed in Section \ref{sec: portfolio-upper-bound}.

    Next, we define refinements for fair facility location and fair $k$-clustering as a way to incorporate rolling budgets in portfolios.

    \begin{definition}[Refinements]\label{def: refinements}
    Given metric space $(X, d)$, group memberships $\mu$, and a class of norms $\mathbf{C}$, a sequence $(\mathbf{F}, \mathbf{\Pi}) = (F_1, \Pi_1), \ldots, (F_l, \Pi_l)$ of solutions is called a refinement if
    \begin{enumerate}
        \item (facility condition) the facility sets form a chain, i.e., $F_1 \subseteq F_2 \subseteq \ldots \subseteq F_l$, and
        \item (assignment condition) for each $t \ge 2$ and for each facility $f_t \in F_t$, there is some facility $f_{t - 1} \in F_{t - 1}$ such that all clients assigned to $f_{t}$ by $\Pi_t$ are assigned to $f_{t - 1}$ by $\Pi_{t - 1}$, i.e.,
    \end{enumerate}
    \begin{equation}\label{eqn: assignment-condition}
    \{j \in X: \Pi_t(j) = f_t\} \subseteq \{j \in X: \Pi_{t - 1}(j) = f_{t - 1}\}.
    \end{equation}

    Additionally, given facility costs $c: X \to \R_{\ge 0}$, refinement $(\mathbf{F}, \mathbf{\Pi})$ is called $\alpha$-approximate for fair facility location if it is also an $\alpha$-approximate portfolio for fair facility location for $\mathbf{C}$, i.e., for each norm $g \in \mathbf{C}$, there is some $(F_t, \Pi_t)$ in the refinement that is an $\alpha$-approximation to FFL$_g$.

    To define this concept for clustering, we consider $l$ budgets $k_1 < \ldots < k_l$ and a norm $g \in \mathbf{C}$. Then, a refinement $(\mathbf{F}, \mathbf{\Pi})$ is called $(\beta, \alpha)$-approximate for clustering with norm $g$ if for each $t \in [l]$, $(F_t, \Pi_t)$ is a $(\beta, \alpha)$-approximation to FC$_g^{(k_t)}$. A set $\mathbf{P}$ of refinements is a $(\beta, \alpha)$-approximate \emph{portfolio of refinements} for clustering if for each norm $g \in \mathbf{C}$, there is some refinement $(\mathbf{F}, \mathbf{\Pi}) \in \mathbf{P}$ that is $(\beta, \alpha)$-approximate for clustering with norm $g$.
    \end{definition}

    We also define $\alpha$-\emph{reassignments} that we will use as a subroutine for refinements. Unlike refinements where both facility sets and assignments are sought, facility sets are a part of the \emph{input} for the $\alpha$-reassignment problem.

    \begin{definition}[$\alpha$-Reassignments]\label{def: reassignment-problem}
    Given a metric space $(X, d)$ and facility sets $F_1 \subseteq \ldots \subseteq F_l \subseteq X$, the $\alpha$-reassignment problem seeks assignments $\Pi_t: X \to F_t$ for $t \in [l]$ such that (1) $(F_1, \Pi_1), \ldots, (F_l, \Pi_l)$ is a refinement and (2) the following cost upper bound is satisfied for each $t \in [l]$ and each client $j \in X$:
    \begin{equation}\label{eqn: cost-upper-bound}
    d\left(j, \Pi_t(j)\right) \le \alpha \cdot \min_{f \in F_t} d(j, f).
    \end{equation}
    \end{definition}

    \textbf{Approximations for FFL and FC for fixed norm.} Before we prove our results on portfolios for fair facility location and fair $k$-clustering, we establish constant-factor approximations for any fixed norm by extending the relaxation and rounding algorithm of \cite{STA97} for uncapacitated facility location to more general objectives. Specifically, we obtain a $4$-approximation for FFL$_g$ and bicriteria $(4, 4)$-approximation for FC$_g$ for arbitrary norm $g$, and in particular for any fixed $L_p$ norm.
    We assume that norm $g$ is given to us through a polynomial time value oracle to compute $g(x)$ for any vector $x$. \cite{STA97}'s algorithm considers a linear programming (LP) relaxation and then uses a filtering technique introduced by \cite{LV92} to round the optimal (fractional) solution to the LP to an integral solution. We observe that this algorithm individually rounds client distances, getting constant-factor approximations to FFL$_g$ and FC$^{(k)}_g$; for completeness, we include the algorithm and the proof in Appendix \ref{app: relaxation-and-rounding}.

    \begin{theorem}\label{thm: facility-location-solvability-theorem}
    There exists a polynomial time algorithm that given any norm $g$, obtains (1) a $4$-approximation to FFL$_g$ and (2) bicriteria $(4, 4)$-approximation to FC$_g^{(k)}$ for any budget $k$.
    \end{theorem}

    \section{Upper Bounds on Portfolio Sizes}\label{sec: portfolio-upper-bound}

    In this section, we obtain upper bounds on portfolio sizes (Theorem~\ref{thm: portfolio-upper-bound}) using Algorithm \textsc{StepPortfolio} for any class $\mathbf{C} = \{g_\lambda: \lambda \in [a, b]\}$ of $r$-dimensional norms $g_\lambda$ that monotonically interpolates between $L_1, L_\infty$ norms. Recall (Definition \ref{def: monotonically-interpolating-norm-class}) that (1) $g_a$ is the $L_1$ norm and $g_b$ is the $L_\infty$ norm, (2) $g_\lambda(x)$ is non-increasing with $\lambda$ for each $x \in \R^r$, i.e., if $\lambda < \mu$ then $g_\lambda(x) \ge g_{\mu}(x)$. Geometrically, such a class corresponds to a continuous expansion of the $L_1$ norm unit ball into the $L_\infty$ norm unit ball. $L_p$ norms form one such class with $\lambda = p$, $a = 1$ and $b = \infty$. Other examples include:
    \begin{enumerate}
        \item Top-$l$ norms \cite{GM06, CS17}: for $l \in [r]$, the top-$l$ norm of $x \in \R^r$ is the sum of $l$ highest coordinates of $x$ by absolute value. Parameter $\lambda = 1/l$ varies from $a = 1/r$ (top-$r$ or $L_1$ norm) to $b = 1$ (top-$1$ or $L_\infty$ norm).
        \item For parameter $\lambda \in [0, 1]$ and $x \in \R^r$, define the convex combination norm
        $$
        \|x\|^{(\lambda)} = (1 - \lambda) \|x\|_1 + \lambda \|x\|_{\infty}.
        $$
        Then $\|x\|^{(\lambda)}$ decreases as $\lambda$ increases, $\|x\|^{(0)} = \|x\|_1$ and $\|x\|^{(1)} = \|x\|_\infty$. Similar norms are considered, for example, in \cite{GJRYZ20}, in the context of facility location.
    \end{enumerate}

    One might naturally ask if portfolios for one class of monotonically interpolating norms (e.g., top-$l$ norms) are also portfolios for another such class (e.g., $L_p$ norms); indeed, a portfolio of size $1$ that is $\alpha$-approximate for top-$l$ norms is $\alpha$-approximate for a much larger class of norms called symmetric monotonic norms; this class includes both top-$l$ and $L_p$ norms.
    At the end of this section, we show that this is false for portfolios of size $> 1$ for fair facility location; Appendix \ref{app: portfolios-are-not-tranferable} gives an example for more general problems.

    We prove the following theorem that guarantees logarithmic portfolio sizes for a given monotonically interpolating norm class:

    \begin{theorem}\label{thm: portfolio-upper-bound}
    Given a class $\mathbf{C}$ of monotonically interpolating norms on $\R^r$, approximation $\alpha > 1$, and an instance of fair facility location or fair $k$-clustering with $r$ client groups, there exists an $\alpha$-approximate portfolio of size $O(\log_{\alpha} r)$ for $\mathbf{C}$. Further, there is a polynomial-time algorithm to find a portfolio of size $O(\log_{\alpha} r)$ for $\mathbf{C}$ that is $4\alpha$-approximate for the given instance of fair facility location and $(4, 4\alpha)$-approximate for the given instance of fair $k$-clustering.
    \end{theorem}

    \begin{proof}
        Denote the class $\mathbf{C} = \{g_\lambda: \lambda \in [a, b]\}$.
        We prove the theorem for fair facility location; the proof for fair $k$-clustering is similar and omitted. We will prove the following: given an oracle to obtain $\theta$-approximations for FFL$_{g_\lambda}$ for any given $\lambda \in [a, b]$, we can find in a polynomial number of oracle calls an $(\alpha \theta)$-approximate portfolio for $\mathbf{C}$. The portfolios size is $N + 1$ where $N = \log_{\alpha} (\theta r)$.
        Choosing $\theta = 1$ gives the existence result. Choosing $\theta = 4$ and using Theorem \ref{thm: facility-location-solvability-theorem} gives the polynomial-time result.

        We construct a sequence of $\lambda$ values $a = \lambda(0) < \lambda(1) < \ldots < \lambda(N) = b$ such that the set $P$ of $\theta$-approximate solutions for these $N + 1$ norms forms the desired portfolio. For $\lambda \in [a, b]$, let $\OPT_\lambda$ be the optimal value of FFL$_{g_\lambda}$, let $(F_\lambda, \Pi_\lambda)$ be the $\theta$-approximate solution obtained by the oracle. For brevity, denote the access cost $g_\lambda(d_{\Pi_{\lambda}})$ by $g_\lambda(\Pi_\lambda)$, so that the objective value of this solution for $g_\lambda$ is $\ALG_\lambda = c(F_\lambda) + g_\lambda(\Pi_{\lambda})$.
        By definition, $\ALG_{\lambda} \le \theta \cdot \OPT_{\lambda}$.

        We can assume without loss of generality that $\ALG_\lambda$ is non-increasing with $\lambda$: indeed, given $\mu > \lambda$, the cost of $(F_\lambda, \Pi_\lambda)$ for $g_\mu$ is $c(F_\lambda) + g_\mu(\Pi_\lambda)$, which is at most $c(F_\lambda) + g_\lambda(\Pi_\lambda) = \ALG_\lambda$ since $g_\mu(x) \le g_\lambda(x)$ for all vectors $x$.

        Next, for $\lambda \in [a, b]$, denote by $\lambda_{\alpha}$ the minimum value of $\mu \in [a, b]$ such that $\ALG_{\mu} \le \ALG_{\lambda}/\alpha$. Intuitively, we take a `step' of size $\alpha$ in the objective value.
        If no such $\mu$ exists (i.e.,  when ${\ALG_{b}} > \ALG_{\lambda}/\alpha$), define $\lambda_\alpha = b$. Construct portfolio $P$ as follows: initially set $\lambda(0) = a$ and $i = 0$. While $\lambda(i)_\alpha < b$, keep taking $\alpha$-steps, i.e., setting $\lambda(i + 1) = \lambda(i)_{\alpha}$ and increasing the counter $i$. Suppose $\lambda(0), \ldots, \lambda(N)$ is the sequence of $\lambda$ values generated by this algorithm. Algorithm \textsc{StepPortfolio} outputs the corresponding $\theta$-approximations, i.e., $P = \left\{(F_{\lambda(i)}, \Pi_{\lambda(i)}): i \in [0, N]\right\}$.

        We claim that for each $i \in [0, N - 1]$ and norm $\mu \in [\lambda(i), \lambda(i + 1))$, solution $(F_{\lambda(i)}, \Pi_{\lambda(i)})$ is a $(\theta \alpha)$-approximation to FFL$_{g_{\mu}}$. This is sufficient to prove the approximation guarantee of the portfolio. Since $g_\mu \le g_{\lambda(i)}$, the cost of $(F_{\lambda(i)}, \Pi_{\lambda(i)})$ for FFL$_{g_{\mu}}$ is  $c(F_{\lambda(i)}) + g_\mu(\Pi_{\lambda(i)}) \le c(F_{\lambda(i)}) + g_{\lambda(i)}(\Pi_{\lambda(i)}) = \ALG_{\lambda(i)}$.
        Now, by definition of $\lambda(i + 1)$, we have $\ALG_{\mu} \ge \frac{\ALG_{\lambda(i)}}{\alpha}$, so that this cost is at most $\alpha \ALG_{\mu} \le \alpha \theta \cdot \OPT_\mu$. This proves the approximation guarantee for $P$.

        We now prove that $|P| = O(\log_{\alpha} (\theta r))$. By construction, $\ALG_{\lambda(i + 1)} \le \ALG_{\lambda(i)}/\alpha$, i.e., the value of $\ALG_{\lambda(i)}$ decreases by factor $\ge \alpha$ in each step (except possibly the last). Therefore, the number of steps $N$ is bounded by $\log_\alpha \frac{\ALG_{a}}{\ALG_b}$. It is now sufficient to prove that $\ALG_a \le r\theta \cdot \ALG_b$. To see this claim, since $(F_a, \Pi_a)$ is a $\theta$-approximation to FFL$_a$ and $g_a$ is the $L_1$ norm, we have $\ALG_a = c(F_a) + \|\Pi_a\|_1 \le \theta \cdot \left(c(F_b) + \|\Pi_b\|_1\right)$. Next, since $\|\Pi_b\|_1 \le r \|\Pi_b\|_\infty$, we get $\ALG_a \le \theta \cdot \left(c(F_b) + r \cdot \|\Pi_b\|_\infty\right) \le \theta r \cdot \left(c(F_b) + \|\Pi_b\|_\infty\right) = \theta r \cdot \ALG_b$ since $g_b$ is the $L_\infty$ norm.
    \end{proof}

    \textbf{Portfolios for top-$l$ norms vs $L_p$ norms.} Consider an instance of fair facility location with $10$ clients, each located at $x = 0$ or $x = 1$ on the real line. There are three client groups $X_1, X_2, X_3$ that partition clients $X$, and group memberships $\mu_{j, s} = 1/|X_s|$ each client $j \in X_s$ and $0$ otherwise. $X_1$ and $X_2$ each consist of a client at $x = 0$ and two clients at $x = 1$. $X_3$ consists of $3$ clients at $x = 0$ and $1$ client at $x = 1$.
    A facility can be opened at $x = 1/2, 2/3$ or $1$ with facility cost $c = 2$. Table \ref{tab: top-l-counterexample-objective-values} gives the objective values for various norms for opening facilities at these three locations. The optimal portfolio for top-$l$ norms has size $2$ but the optimal portfolio for $L_p$ norms has size $3$.

    \vspace{0.5em}

    \begin{minipage}{0.49\textwidth}
        \footnotesize
        \begin{tabular}{|@{}c@{}|c|c|c|c|}
            \hline
            \begin{tabular}{c} Open \\ facility \end{tabular} & \begin{tabular}{c} $L_1$, i.e. \\ top-$3$ \end{tabular}  & \begin{tabular}{c} $L_\infty$, i.e. \\ top-$1$ \end{tabular} & $L_2$ & top-$2$ \\ \hline
            $x = 1/2$ & $3.500$ & \textbf{2.500}  & $2.866$  & \textbf{3.000}  \\ \hline
            $x = 2/3$ & $3.472$ & $2.583$ & \textbf{2.858}  & $3.028$  \\ \hline
            $x = 1$ &  \textbf{3.417} & $3.750$ & $2.886$ & $3.083$  \\ \hline
        \end{tabular}
    \end{minipage}
    \begin{minipage}{0.50\textwidth}
        \captionsetup{type=table}
        \captionof{table}{
            Objective values for various norms for various locations of open facility. The two solutions corresponding to open facilities at $x = 1/2$ and $x = 1$ respectively form an optimal portfolio for top-$l$ norms but the optima for $L_1, L_2, L_\infty$ norms open facilities at different locations.}
        \label{tab: top-l-counterexample-objective-values}
    \end{minipage}

    \section{Lower Bounds on Portfolio Sizes}\label{sec: portfolio-lower-bound}

    In this section, we give lower bounds (Theorem~\ref{thm: portfolio-lower-bound}) on portfolio sizes for $L_p$ norms for both fair facility location and fair $k$-clustering, matching the upper bound for constant-factor approximations. Our examples are all in the individually fair case so that each client is in their unique group and therefore $r = n$. To illustrate why portfolios with multiple solutions are necessary, we note an instance of fair facility location where no single solution achieves $o(\sqrt{r})$-approximation for all $L_p$ norms (a similar example was noted for fair $k$-clustering by \cite{GM06}; their example crucially uses the bound $k$ while ours crucially uses facility opening costs.)

    Consider the real line; we place $1$ client at the origin $x=0$, and $n - 1$ clients at a unit distance from them (i.e., $x=1$). The cost for opening the facility at the origin is $c(0) = 1$, and at $x=1$ is $c(1) = \sqrt{n}$. There are only three feasible solutions: (i) open facility only at $x=0$, which is optimal for $L_{\infty}$ with cost 2, or (ii) only at $x=1$, which is optimal for $L_1$ norm with cost $1 + \sqrt{n}$, or (iii) open both facilities. It can be seen that none of the three solutions is better than $o(\sqrt{n})$-approximation for both $L_1$ and $L_\infty$ norms.
    More generally, we have the following theorem:

    \begin{theorem}\label{thm: portfolio-lower-bound}
    For $\alpha > 1$ and for all large enough $r$, there exist instances of fair facility location and fair $k$-clustering with $r$ groups where any $\alpha$-approximate portfolio for $L_p$ norms must have size $\Omega\left(\log_{2\alpha} r\right)$.
    \end{theorem}

    \begin{proof}[Proof for fair facility location.]
        We will construct an instance where any $\alpha$-approximate portfolio for a set of $S \subseteq [1, \infty]$ of $\Omega(\log_{2\alpha} r)$ norms must be of the same size $|S|$, i.e., no single solution will be $\alpha$-approximate for two norms in this set. Denote $\gamma = 2\alpha$. Our metric space is a star graph with $L + 1$ leaf vertices $x_0, \ldots, x_{L}$ and $n = r = \gamma^{2L}$ clients all located at the central vertex $v$ of the star. Each client $j \in [n]$ is in their unique group with $\mu_{j, j} = 1$.

        A facility can be opened at one of the leaf vertices but not at $v$ (i.e. $c(v) = \infty$).
        The cost of opening a facility at $x_i$ is $c(x_i) = \gamma^{i}$ for each $i \in [0, L]$, so that $c(x_0) < c(x_1) < \ldots < c(x_L)$. For each $i \in [0, L]$, distance $d(v, x_i) = \gamma^{-i}$ so that $d(v, x_0) > d(v, x_1) > \ldots > d(v, x_L)$.

        Fix some $p \in [1, \infty]$ and $i \in [0, L]$.
        If none of the facilities at $x_i, x_{i + 1}, \ldots, x_L$ is opened, then the total cost (facility opening cost + norm of distance vector) for FFL$_p$ is strictly greater than  $\|\underbrace{d(v, x_{i - 1}), \ldots, d(v, x_{i - 1})}_{\gamma^{2L}}\|_p
        = d(v, x_{i - 1}) \gamma^{\frac{2L}{p}} = \gamma^{-(i - 1) + \frac{2L}{p}}$.
        Specifically, for $p_i = \frac{L}{i}$, we get that this cost is at least $\gamma^{i + 1}$, so that the total cost is strictly greater than $\gamma^{i + 1}$.

        Next, if any of the facilities at $x_{i + 1}, \ldots, x_{L}$ is opened, then the total cost is strictly greater than the cost of opening the facility at $x_{i + 1}$, which is $\gamma^{i + 1}$.
        When only the facility at $x_i$ is open, the total cost is exactly
        $$
        c(x_i) + \|d(v, x_i), \ldots, d(v, x_i)\|_{p_i} = \gamma^{i} + \gamma^{\frac{2L}{p_i} - i} = 2 \gamma^i.
        $$
        Since $\gamma^{i + 1} = \alpha \times (2 \gamma^i)$, this implies that the optimal solution for FFL$_{p_i}$ is to open only $x_i$.
        Further, any $\alpha$-approximate solution for FFL$_{p_i}$ must open the facility at $x_i$ but none of the facilities at $x_{i + 1}, \ldots, x_L$. Therefore, an $\alpha$-approximate portfolio for norms $\left\{\frac{L}{L}, \frac{L}{L - 1}, \ldots, \frac{L}{1}, \frac{L}{0} \overset{\text{def}}{=} \infty\right\}$ must contain at least $L + 1$ distinct solutions. This completes the proof since $L = \frac{1}{2}\log_{\gamma} r$.
    \end{proof}

    \begin{proof}[Proof for fair $k$-clustering.] We employ a similar proof strategy, but this time, we choose $L$ such that it satisfies $(1 + L)(\gamma^{2L} + 1) = n = r$ where $\gamma = 2\alpha$. Since $\alpha > 1$, it can be verified that for large enough $r$, $\frac{1}{4} \log_{(2\alpha)} r \le L \le \frac{1}{2} \log_{(2\alpha)} r$. Our metric space is the complete graph $K_{L + 1}$ on $L + 1$ vertices $x_0, \ldots, x_L$ with distance $1$ between any two vertices. At each vertex $x_i$, there are $1 + \gamma^{2L}$ clients denoted $(i, 0), (i, 1), \ldots, (i, \gamma^{2L})$.
    Each client is in their unique group and fractional group memberships are defined as follows:
    $$
    \mu_{(i, s), (i, s)} = \begin{cases}
                               \gamma^{i} & \text{if} \; s = 0,  \\
                               \gamma^{-i} & \text{if} \; s > 0.
    \end{cases}
    $$
    $k = L$ facilities are allowed to open, i.e., a facility is open at all but one vertex $x_i$. For the solution that does not open the facility at $x_i$, each client at $x_i$ must be connected to some other open facility with $L_p$ norm cost
    $$
    \|(\gamma^{i}, \underbrace{\gamma^{-i}, \ldots, \gamma^{-i}}_{\gamma^{2L}}, \underbrace{0, \ldots, 0}_{\text{remaining}})\|_p = \left(\gamma^{ip} + \gamma^{2L - ip}\right)^{1/p}.
    $$
    For $p = L/i$, this cost is $2^{i/L} \gamma^{i} < 2 \gamma^i$. For $i' \neq i$, the cost of the solution that does not open the facility at $x_{i'}$ is $\Big(\gamma^{L\frac{i'}{i}} + \gamma^{L\left(2 - \frac{i'}{i}\right)} \Big)^{i/L}$.
    For $i' > i$, this is dominated by the first term and strictly greater than $\gamma^{i'} \ge \alpha \times (2\gamma^i)$. Similarly, for $i' < i$, the second term dominates and this cost is strictly greater than $\alpha \times (2\gamma^i)$. Together, this implies that any $\alpha$-approximate portfolio for norms $\left\{\frac{L}{L}, \frac{L}{L-1}, \ldots, \frac{L}{1}, \frac{L}{0}\right\}$ must contain all the $L + 1$ possible solutions. Since $L = \Theta\left(\log_{2\alpha} r\right)$, this finishes the proof.
    \end{proof}

    \section{Refinements}\label{sec: refinements}

    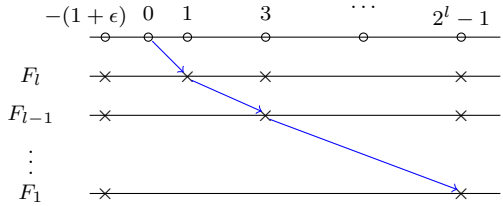
\begin{figure}[h]
        \begin{minipage}{0.45\textwidth}
            \begin{tikzpicture}[scale=0.52]
                \scriptsize
                \filldraw[black] (-1.6,1) circle (0pt) node[anchor=north]{$-(1 + \epsilon)$};
                \filldraw[black] (0,1) circle (0pt) node[anchor=north]{$0$};
                \filldraw[black] (1,1) circle (0pt) node[anchor=north]{$1$};
                \filldraw[black] (3,1) circle (0pt) node[anchor=north]{$3$};
                \filldraw[black] (5.5,1) circle (0pt) node[anchor=north]{$\hdots$};
                \filldraw[black] (8,1) circle (0pt) node[anchor=north]{$2^l - 1$};

                \draw[black] (-1.1,0) circle (3pt);
                \draw[black] (0,0) circle (3pt);
                \draw[black] (1,0) circle (3pt);
                \draw[black] (3,0) circle (3pt);
                \draw[black] (5.5,0) circle (3pt);
                \draw[black] (8,0) circle (3pt);

                \filldraw[black] (-3,-1) circle (0pt) node[anchor=center]{$F_l$};
                \filldraw[black] (-3,-2) circle (0pt) node[anchor=center]{$F_{l - 1}$};
                \filldraw[black] (-3,-3) circle (0pt) node[anchor=center]{$\vdots$};
                \filldraw[black] (-3,-4) circle (0pt) node[anchor=center]{$F_1$};

                \filldraw[black] (-1.1,-1) circle (0pt) node[anchor=center]{$\times$};
                \filldraw[black] (1,-1) circle (0pt) node[anchor=center]{$\times$};
                \filldraw[black] (3,-1) circle (0pt) node[anchor=center]{$\times$};
                \filldraw[black] (8,-1) circle (0pt) node[anchor=center]{$\times$};

                \filldraw[black] (-1.1,-2) circle (0pt) node[anchor=center]{$\times$};
                \filldraw[black] (3,-2) circle (0pt) node[anchor=center]{$\times$};
                \filldraw[black] (8,-2) circle (0pt) node[anchor=center]{$\times$};
                \filldraw[black] (-1.1,-4) circle (0pt) node[anchor=center]{$\times$};
                \filldraw[black] (8,-4) circle (0pt) node[anchor=center]{$\times$};

                \draw[] (-1.5,0) -- (9,0);
                \draw[] (-1.5,-1) -- (9,-1);
                \draw[] (-1.5,-2) -- (9,-2);
                \draw[] (-1.5,-4) -- (9,-4);

                \draw[->, blue] (0.1,-0.1) -- (0.9,-0.9);
                \draw[->, blue] (1.1,-1.1) -- (2.9,-1.9);
                \draw[->, blue] (3.1,-2.1) -- (7.9,-3.9);
            \end{tikzpicture}
        \end{minipage}
        \hfill
        \begin{minipage}{0.54\textwidth}
            \caption{An example to show that the greedy algorithm is $\Omega(2^l)$-approximate for the reassignment problem with $l$ facility sets. Client locations are drawn as circles and facility locations are drawn as a cross. A copy of the line is drawn for each $s \in [l]$ for clarity. The greedy algorithm assigns $\Pi_l(0) = 1$ since the closest facility in $F_l$ to the client at $x = 0$ is at $x = 1$. The closest facility to $x = 1$ in $F_{l - 1}$ is at $x = 3$, so the algorithm assigns $\Pi_{l - 1}(0) = 3$, and so on until $\Pi_1(0) = 2^{l} - 1$. However, the closest facility to $x = 0$ in $F_1$ is at $-(1 + \epsilon)$, so that $\frac{d(0, \Pi_1(0))}{\min_{f \in F_1} d(0, f)} \simeq 2^l - 1$.}
            \label{fig: greedy-assignment-example}
        \end{minipage}
    \end{figure}

    Next, we reduce the problem of finding refinements for both fair facility location and fair $k$-clustering (Definition \ref{def: refinements}) to $\alpha$-reassignments (Definition \ref{def: reassignment-problem}). Recall that an $\alpha$-reassignment of a chain of facilities $F_1 \subseteq \ldots \subseteq F_l$ is a set of assignments $\Pi_t: t \in [l]$ that satisfies the assignment condition (eqn. (\ref{eqn: assignment-condition})) with an added cost upper bound: each client $j \in X$ must be assigned to a facility that is within a distance $\alpha$ of its closest facility for each facility set $F_t$, i.e., $d(j, \Pi_t(j)) \le \alpha \cdot \min_{f \in F_t} d(j, f)$ for all $t \in [l]$.

    To see why this problem is non-trivial, we first observe that the natural greedy algorithm only achieves $\alpha = \Omega(2^l)$ in the cost upper bound. Consider the following greedy assignment for each client $j \in X$: assign $j$ first to its closest facility in $F_l$, say $f_l$, i.e., set $\Pi_l(j) = {\arg\min}_{f \in F_l} d(j, f) := f_{l}$. Next, find facility $f_{l - 1} \in F_{l -1}$ that is closest to $f_l$ and assign $\Pi_{l - 1}(j) = f_{l - 1}$. This ensures that each client mapped to $f_l$ under $\Pi_l$ is mapped to $f_{l - 1}$ under $\Pi_{l - 1}$, i.e., the assignment condition is satisfied. Extend this to assignments $\Pi_1, \ldots, \Pi_l$.
    In the example in Figure \ref{fig: greedy-assignment-example}, the greedy algorithm's assignment $\Pi_1$ assigns client at $x = 0$ to the facility at $x = 2^l - 1$ in $F_1$, even though the closest facility in $F_1$ is at $-(1 + \epsilon)$ (for arbitrarily small $\epsilon > 0$), and therefore $\alpha \ge \frac{2^l - 1}{1 + \epsilon}$ is \emph{exponential} in $l$.

    We next state our results for the $\alpha$-reassignment problem for arbitrary metric, line metric, and tree metric in Theorem \ref{thm: reassignment-problem}, with \emph{subexponential} in $l$ approximations. Assuming these, we prove corresponding results for refinements for fair facility location in Section \ref{sec: refinements-fl} (Theorem \ref{thm: facility-location-refinement}) and for fair $k$-clustering in Section \ref{sec: refinements-clustering} (Theorem \ref{thm: refinement-general-metric}).
    The proof of Theorem \ref{thm: reassignment-problem} is deferred to Section \ref{sec: discounted-lookahead} for arbitrary metric, to Section \ref{sec: line-metric} for line metric, and to Appendix \ref{app: tree-metric} for tree metric.

    \begin{theorem}\label{thm: reassignment-problem}
    There is a polynomial-time algorithm that given metric space $(X, d)$ and non-empty facility sets $F_1 \subseteq \ldots \subseteq F_l \subseteq X$, obtains
    \begin{enumerate}
        \item (Algorithm \textsc{DiscountedLookahead}) $\alpha_l$-reassignments with $\alpha_l = O(e^{3 \sqrt{l}})$ for any metric $d$,
        \item (Algorithm \textsc{ExpandIntervals}) $O(l)$-reassignments when $d$ is a line metric,
        \item (Algorithm \textsc{BranchAndLinearize}) $O(l)$-reassignments for modified facility sets $F_1' \subseteq F_2' \subseteq \ldots \subseteq F_l' \subseteq X$ such that each $F_t \subseteq F_t'$ and $|F_t'| \le 2 |F_t|$ when $d$ is a tree metric. That is, it obtains $O(l)$-reassignments on tree metrics with facility sets at most double the size of the input facility sets.
    \end{enumerate}
    \end{theorem}

    The rest of this section is devoted to showing how guarantees for refinements for fair facility location (Section \ref{sec: refinements-fl}) and fair $k$-clustering (Section \ref{sec: refinements-clustering}) can be obtained by using the reassignment guarantees in Theorem \ref{thm: reassignment-problem}. The strategy for both problems involves first obtaining appropriate facility sets $F_1 \subseteq \ldots \subseteq F_l$ without necessarily satisfying the assignment condition (eqn. (\ref{eqn: assignment-condition})). Then, corresponding assignments that also satisfy the assignment condition are obtained using Theorem \ref{thm: reassignment-problem}.

    \textbf{Notation}. Given any instance of fair facility location or fair $k$-clustering, a norm $g$ on $\R^r$, and a solution $(F, \Pi)$, we denote the access cost $g(d_{\Pi})$ simply as $g(\Pi)$ for brevity. Given facilities $F \subseteq X$, by `solution $F$' we mean the induced solution $(F, \Theta(F))$ where $\Theta(F): X \to F$ assigns each client to its nearest facility in $F$. The access cost of this solution is denoted variously as $g(d_{\Theta(F)}) = g(\Theta(F)) = g(F)$ for brevity. We will throughout work with a fixed monotonically interpolating (Definition \ref{def: monotonically-interpolating-norm-class}) norm class $\mathbf{C} = \{g_\lambda: \lambda \in [a, b]\}$.

    \subsection{Refinements for Facility Location}\label{sec: refinements-fl}

    In this section, we give approximation guarantees on refinements for fair facility location using Theorem \ref{thm: portfolio-upper-bound} on portfolio sizes and Theorem \ref{thm: reassignment-problem} on $\alpha$-reassignments.

    \begin{theorem}\label{thm: facility-location-refinement}
    There is a polynomial-time algorithm that given an instance of fair facility location with $r$ groups and a family $\mathbf{C}$ of monotonically interpolating norms on $\R^r$, gives an $\alpha$-approximate refinement where $\alpha = \poly(e^{\sqrt{\log r}})$.
    \end{theorem}

    \begin{proof}
        We invoke Theorem \ref{thm: portfolio-upper-bound} for $\alpha = 8$ and obtain $l = O(\log r)$ solutions $(G_1, \Theta_1), \ldots, (G_l, \Theta_l)$ such that for each $\lambda \in [a, b]$ some $(G_t, \Theta_t)$ is an $8$-approximation to FFL$_{g_\lambda}$. Relabel the indices so that $c(G_1) \le \ldots \le c(G_l)$, where $c(G_t) := \sum_{f \in G_t} c(f)$ is the facility opening cost. Define `prefix' facility sets $F_t = \bigcup_{s \in [t]} G_s$ for all $t \in [l]$ so that facility sets $F_1 \subseteq F_2 \subseteq \ldots \subseteq F_l$ form a chain. Also, $c(F_t) \le t \cdot c(G_t) \le l \cdot c(G_t)$.

        Set $\alpha = O(e^{3 \sqrt{ \log r}})$. By Theorem \ref{thm: reassignment-problem}, we can find assignments $\Pi_1, \ldots, \Pi_l$ that satisfy (1) the assignment condition (eqn. (\ref{eqn: assignment-condition})) for refinement, and (2) the cost upper bound (eqn. (\ref{eqn: cost-upper-bound})) $d(j, \Pi(j)) \le \alpha \cdot \min_{f \in F_t} d(j, f)$ for each client $j \in X$ and each $t \in [l]$. Since $G_t \subseteq F_t$, we get $\min_{f \in F_t} d(j, f) \le \min_{f \in G_t} d(j, f) \le d(j, \Theta_t(j))$. Therefore, group access cost vectors satisfy $d_{\Pi_t} \le \alpha \cdot d_{\Theta_t}$ for each $t \in [l]$. In particular $g(\Pi_t) \le \alpha g(\Theta_t)$ for any norm $g$.

        Pick $\lambda \in [a, b]$. By the portfolio guarantee, there is some $(G_t, \Theta_t)$ that is $8$-approximation to FFL$_{g_\lambda}$. The cost of $(F_t, \Pi_t)$ for $g_\lambda$ is $c(F_t) + g_\lambda(\Pi_t) \le l c(G_t) + \alpha g_\lambda(\Theta_t)
        \le \alpha \left(c(G_t) + g_\lambda(\Theta_t)\right).$
        That is, $(F_t, \Pi_t)$ is an $O(\alpha)$-approximation for FFL$_{g_\lambda}$.
    \end{proof}

    \subsection{Refinements for $k$-Clustering}\label{sec: refinements-clustering}

    In this section, we reduce refinements for fair $k$-clustering to the $\alpha$-reassignment problem. As with facility location, we first obtain appropriate facility sets $F_1 \subseteq \ldots \subseteq F_l$ and then find assignments using Theorem \ref{thm: reassignment-problem}. The difference in approximation guarantees for various metrics in Theorem \ref{thm: reassignment-problem} is reflected in approximation guarantees for refinements:

    \begin{theorem}\label{thm: refinement-general-metric}
    There are polynomial-time algorithms that given a metric space $(X, d)$ on $n$ clients, $l$ instances of fair $k$-clustering with budgets $k_1 < \ldots < k_l$, and a class $\mathbf{C}$ of monotonically interpolating norms on $\R^r$, give
    \begin{enumerate}
        \item A bicriteria $\left(O(1), \poly\big(n^{\frac{1}{\sqrt{\log n}}}\big)\right)$-approximate refinement for any fixed norm $g \in \mathbf{C}$. This can be improved to $\left(O(1), O(\log n)\right)$-approximation when $d$ is a tree metric.
        \item A bicriteria $\left(O(1), \poly\big(n^{\frac{1}{\sqrt{\log n}}}\big)\right)$-approximate portfolio $\mathbf{P}$ of refinements of size $O(\log n \cdot \log r)$. This can be improved to $\left(O(1), O(\log n)\right)$-approximation when $d$ is a tree metric.
    \end{enumerate}
    \end{theorem}

    We first tackle a fixed norm $g$ and later extend it to a portfolio of refinements. Recall that we are looking for a refinement $(\mathbf{F}, \mathbf{\Pi}) = (F^{(1)}, \Pi^{(1)}), \ldots, (F^{(l)}, \Pi^{(l)})$ such that for each $t \in [l]$, $(F^{(t)}, \Pi^{(t)})$ is an $\left(O(1), \alpha\right)$-bicriteria approximation to FC$_g^{(k_t)}$ for appropriate $\alpha$. Our strategy is similar to refinements for fair facility location: we first obtain just the facility sets $F^{(t)}$ and then apply Theorem \ref{thm: reassignment-problem} to obtain assignments.
    Our first lemma allows us to assume without loss of generality that $l = \log_2 n$ and $k_t = 2^t$ for each $t \in [l]$:

    \begin{lemma}
        Given any fixed norm $g$, refinement $(\tilde{\mathbf{F}}, \tilde{\mathbf{\Pi}})$ that is $(\beta, \alpha)$-approximate for budgets $\tilde{k}_s = 2^s, s \in [\log_2 n]$ can be converted to a $(2\beta, \alpha)$-approximate refinement $({\mathbf{F}}, {\mathbf{\Pi}})$ for arbitrary budgets $k_1 < \ldots < k_l$ in polynomial time.
    \end{lemma}

    \begin{proof}
        Suppose $(\tilde{\mathbf{F}}, \tilde{\mathbf{\Pi}}) = (\tilde{F}^{(1)}, \tilde{\Pi}^{(1)}), \ldots, (\tilde{F}^{(\log_2 n)}, \tilde{\Pi}^{(\log_2 n)})$.
        For each budget $k_, t \in [l]$, there is some $s \in [\log_2 n]$ such that $\frac{\tilde{k}_s}{2} < k_t \le \tilde{k}_{s}$.
        Define $(F^{(t)}, \Pi^{(t)}) = (\tilde{F}^{(s)}, \tilde{\Pi}^{(s)})$.
        Then $|F^{(t)}| = |\tilde{F}^{(s)}| \le \beta \tilde{k}_s \le 2 \beta k_t$.
        Further, since $(\tilde{F}^{(s)}, \tilde{\Pi}^{(s)})$ has objective value within $\alpha$ of the optimal for FC$_g^{(\tilde{k}_s)}$ and $k_t \le \tilde{k}_s$, $(F^{(t)}, \Pi^{(t)})$ has objective value within $\alpha$ of the optimal for FC$_g^{(k_t)}$.
        Thus,  $({F}^{(s)}, {\Pi}^{(s)}), t \in [l]$ forms a $(2\beta, \alpha)$-approximate refinement for budgets $k_1, \ldots, k_l$.
    \end{proof}

    We assume for the rest of this section that $k_t = 2^t$ and $l = \log_2 n$.
    Using Theorem \ref{thm: facility-location-solvability-theorem}, obtain facility sets $G^{(t)}$ that are $4$-approximations to FC$_g^{(k_t)}$ for each $t \in [l]$, with $|G^{(t)}| \le 4k_t = 2^{t + 2}$. Define prefix sets $F^{(t)} := G^{(1)} \cup G^{(2)} \cup \ldots \cup G^{(t)}$; then $|F^{(t)}| \le 4(2 + 4 + \ldots + 2^t) \le 8 k_t$. Further, $F^{(t)} \supseteq G^{(t)}$ implies that $F^{(t)}$ is also a $4$-approximation to FC$_g^{(k_t)}$.

    Using Theorem \ref{thm: reassignment-problem}, obtain assignments $\Pi^{(1)}, \ldots, \Pi^{(l)}$ that form a refinement and such that for each $t \in [l]$, each client's distance to the assigned facility in $F^{(t)}$ increases by factor at most $O(e^{3\sqrt{l}})$. Since $l = \log_2 n$, this implies that each $(F^{(t)}, \Pi^{(t)})$ is a bicritera $(O(1), O(e^{3\sqrt{\log_2 n}}))$-approximation to FC$_g^{(k_t)}$.

    The result for line and tree metrics follows from the corresponding result in Theorem \ref{thm: reassignment-problem}, with the size of facility sets at most doubling for tree metrics. This completes the result for a fixed norm.

    For each $\lambda \in [a, b]$, denote by $(\mathbf{F}_\lambda, \mathbf{\Pi}_\lambda) = (F^{(1)}_\lambda, \Pi^{(1)}_\lambda), (F^{(2)}_\lambda, \Pi^{(2)}_\lambda), \ldots, (F^{(l)}_\lambda, \Pi^{(l)}_\lambda)$ the $(O(1), O(\alpha))$-approximate refinement for norm $g_\lambda$ for appropriate $\alpha$. To construct a portfolio $\mathbf{P}$ of refinements with approximation guarantees for each norm $g_\lambda \in \mathbf{C}$, we will construct a sequence $\lambda(0), \lambda(1), \ldots, \lambda(N) \in [a, b]$ of $\lambda$ values with corresponding refinements $(\mathbf{F}_{\lambda(i)}, \mathbf{\Pi}_{\lambda(i)})$ such that
    \begin{enumerate}
        \item $\lambda(0) = a$, $\lambda(N) = b$, $N = O(\log n \cdot \log r)$, and
        \item for each $i \in [0, N - 1]$ and for each $\lambda \in [\lambda(i), \lambda(i + 1))$, refinement $(\mathbf{F}_{\lambda(i)}, \mathbf{\Pi}_{\lambda(i)})$ will be an $(O(1), O(\alpha))$-approximation for norm ${g_{\lambda}}$ for appropriate $\alpha$.
    \end{enumerate}

    For each $\lambda$, denote the cost of $F_\lambda^{(t)}$ (without the assignment $\Pi_\lambda^{(t)}$) for FC$_{g_\lambda}^{(k_t)}$ as $\ALG^{(k_t)}_{\lambda}$, then, as we noted, $\ALG^{(k_t)}_{\lambda} \le 4 \cdot \OPT_{\lambda}^{(k_t)}$ and $|F_\lambda^{(t)}| \le 8 k_t$.
    For $\lambda \in [a, b]$ and $t \in [l]$, define \emph{the halving parameter} $\lambda^{(t)}_{\mathrm{half}}$ to be the minimum value of $\mu \in [a, b]$ such that $\ALG^{(k_t)}_\mu \le \frac{1}{2} \ALG^{(k_t)}_{\lambda}$. If such a norm does not exist, then define $\lambda^{(t)}_{\mathrm{half}} = b$. Further, define the \emph{global halving parameter} $\half{\lambda} = \min_{t \in [l]} \lambda^{(t)}_{\mathrm{half}}$.

    Our algorithm is analogous to that for Theorem \ref{thm: portfolio-upper-bound}: initially $\lambda(0) = a$ and $i = 0$, and while $\half{\lambda(i)} < b$, keep setting $\lambda(i + 1) = \half{\lambda(i)}$ and increasing the counter $i \gets i + 1$.
    Let $\lambda(0), \ldots, \lambda(N)$ denote the sequence of parameter values output by this algorithm; then $\lambda(0) = a$ and $\lambda(N) = b$.

    We prove the size bound $N = O(\log r \cdot \log n)$ first. As in the proof of Theorem \ref{thm: portfolio-upper-bound}, since $g_a$ is the $L_1$ norm and $g_b$ is the $L_\infty$ norm, we have for each $t$ that $\ALG^{(k_t)}_{a} \le 4r \cdot \ALG^{(k_t)}_b$. Since $\lambda(i + 1) = \half{\lambda(i)}$ for each $i \in [N - 1]$, by definition, there is some $t \in [l]$ such that $\ALG^{(k_t)}_{\lambda(i + 1)} \le \frac{1}{2} \ALG^{(k_t)}_{\lambda(i)}$, i.e., the value of $\ALG^{(k_t)}$ reduces by factor $\ge 2$ in the $i$th step. Since $\ALG^{(k_t)}_{a} \le 4r \cdot \ALG^{(k_t)}_b$, this can happen at most $\log_2 (4r) = O(\log r)$ times for each $t \in [l]$. This implies that $N = O(\log r \cdot \log n)$.

    Finally, for each $i \in [0, N - 1]$, we prove the approximation guarantee of the sequence $(\mathbf{F}_{\lambda(i)}, \mathbf{\Pi}_{\lambda(i)})$ for each $\lambda \in [\lambda(i), \lambda(i + 1))$. For each $t \in [l]$, the cost of $F^{(t)}_{\lambda(i)}$ for FC$_{g_\lambda}^{(k_t)}$ is
    $$
    g_{\lambda(i)}(F^{(t)}_{\lambda}) \le g_\lambda(F^{(t)}_{\lambda(i)}) = \ALG^{(k_t)}_{\lambda(i)} \le 2 \ALG^{(k_t)}_{\lambda} \le 2 \big(4 \cdot \OPT_{\lambda}^{(k_t)}\big).
    $$
    The first inequality follows since $\lambda(i) \le \lambda$ and therefore $g_\lambda(x) \ge g_{\lambda(i)}(x)$ for each vector $x$ and the second inequality follows since $\lambda \in [\lambda(i), \lambda(i)_{\text{half}})$. Thus, $F^{(t)}_{\lambda}$ is an $(O(1), 8)$-approximation to FC$_{g_\lambda}^{(k_t)}$ and therefore $(F^{(t)}_{\lambda}, \Pi^{(t)}_\lambda)$ is an $(O(1), O(\alpha))$-approximation to FC$_{g_\lambda}^{(k_t)}$ for appropriate $\alpha$ using Theorem \ref{thm: reassignment-problem}.

    \section{\textsc{DiscountedLookahead} Algorithm}\label{sec: discounted-lookahead}

    We discuss algorithm \textsc{DiscountedLookahead} for the reassignment problem (Definition \ref{def: reassignment-problem}) and prove Theorem \ref{thm: reassignment-problem} for arbitrary metric $d$. Recall that given facility sets $F_1 \subseteq \ldots \subseteq F_l \subseteq X$, we seek assignments $\Pi_t: X \to F_t, t \in [l]$ that satisfy (1) the assignment condition (eqn. (\ref{eqn: assignment-condition}) and (2) the cost upper bound (eqn. (\ref{eqn: cost-upper-bound})) $d(j, \Pi_t(j)) \le \alpha \cdot \min_{f \in F_t} d(j, f)$ for each client $j \in X$ and $t \in [l]$ with $\alpha = O(e^{3\sqrt{l}})$.

    \subsection{Algorithm Idea}\label{sec: general-algorithm-idea}

    Recall the example in Figure \ref{fig: greedy-assignment-example}, where the greedy algorithm's assignment $\Pi_1$ assigns client at $x = 0$ to the facility at $x = 2^l - 1$ in $F_1$, even though the closest facility in $F_1$ is at $-(1 + \epsilon)$ (for arbitrarily small $\epsilon > 0$), and therefore $\alpha \ge \frac{2^l - 1}{1 + \epsilon}$ is \emph{exponential} in $l$.

    A better solution is to assign client $j$ at $x = 0$ to the facility at $-(1 + \epsilon)$ under each assignment $\Pi_1, \ldots, \Pi_l$. The greedy algorithm misses this solution because it only looks at assignments one `level' ahead at a time. The key idea is to look at closest facilities $h_t := {\arg\min}_{f \in F_t} d(j, f)$ at each `level'. Instead of assigning $\Pi_l(j) = h_l$, consider distances $d(j, h_1), \ldots, d(j, h_l)$, fix a parameter $\gamma \ge 1$ and choose
    $$
    s^* = {\arg\min}_{t \in [l]}  \left(\gamma^t \cdot d(j, h_t) \right).
    $$
    Since $F_t \subseteq F_{s^*}$ for each $t \ge s^*$, we can assign $\Pi_t(j) = h_{s^*}$ for all $t \ge s^*$. By definition of $s^*$, we are guaranteed that $\frac{d(j, \Pi_t(j))}{d(j, h_t)} = \frac{d(j, h_{s^*})}{d(j, h_t)} \le  \gamma^{t - s^*} \le \gamma^{l - 1}$. For $t < s^*$, we can assign $\Pi_t(j)$ recursively by treating $h_{s^*}$ as a client in the restricted instance with facility sets $F_1 \subseteq \ldots \subseteq F_{s^*}$. We show that assignments $\Pi_1, \ldots, \Pi_l$ satisfy the assignment condition and therefore form a refinement, although it takes some work to show this.

    \begin{algorithm}[t]
        \caption{\textsc{DiscountedLookahead}($F_1, \ldots, F_l, \gamma$)}\label{alg: metric-partition-refinement}
        \nonl \textbf{input}: Facility sets $F_1 \subseteq \ldots \subseteq F_l \subseteq X$, with $X$ denoted as $F_{l + 1}$, parameter $\gamma \ge 1$ \\
        \nonl \textbf{output}: Assignments $\Pi_t: X \to F_t$ for all $t \in [l]$ \\
        initialize $\Pi_{t}^{(k)}: F_k \to F_t$ to be empty for all $1 \le t < k \le l + 1$ \\
        for any $t \in [l]$ and any $f \in X$, denote $h_t(f) = {\arg\min}_{h \in F_t} d(f, h)$ \\
        \For{$k = 1$ to $l + 1$}
        {
            \For{each $f \in F_k$}
            {
                $s^* = {\arg\min}_{t \in [k - 1]} \left( \gamma^{t} \cdot d\left(f, h_t(f)\right) \right)$,
                where $s^*$ is the highest such number in case of ties
            \label{step: metric-partition-index-selection} \\
                $\text{set } \Pi^{(k)}_t(f) = \begin{cases}
                                                  h_{s^*}(f) & \text{if} \: t \in [s^*, k - 1]; \phantom{text}\\
                                                  \Pi^{(s^*)}_t\left(h_{s^*}(f)\right) & \text{if} \: t \in [s^* - 1] \\
                \end{cases}
                $\label{step: metric-partition-assignment}
            }
        }
        \textbf{return} assignments $\Pi_1^{(l + 1)}, \ldots, \Pi_l^{(l + 1)}$
    \end{algorithm}

    For $\gamma = 1$, this reduces to the greedy algorithm since $F_1 \subseteq \ldots \subseteq F_l$ and therefore $d(j, h_1) \ge \ldots \ge d(j, h_l)$. In choosing $s^*$, we are \emph{discounting} the larger distance $d(j, h_t)$ by a factor $\gamma^{l - t}$ as compared to $d(j, h_l)$. We call this algorithm \textsc{DiscountedLookahead} (Algorithm \ref{alg: metric-partition-refinement}).

    For brevity, denote $X = F_{l + 1}$. Formally, for each $k \in [l + 1]$ and each $t \in [k - 1]$, the algorithm constructs an assignment $\Pi^{(k)}_t: F_k \to F_t$ from facilities at a `higher level' $k$ to facilities at a `lower level' $t$. The final assignments on clients $F_{l + 1} = X$ are $\Pi_k^{(l + 1)}: F_{l + 1} \to F_k$ for $k \in [l]$.

    \subsection{Algorithm Analysis}\label{sec: general-algorithm-analysis}

    First, we prove that the assignments form a refinement (Claim \ref{claim: metric-refinement-claim-2}) and then prove the cost upper bound.
    Omitted proofs can be found in Appendix \ref{app: missing-refinements-proofs}.

    For convenience, define the identity assignments $\Pi_{k}^{(k)}: F_k \to F_k$ where $\Pi_{k}^{(k)}(f) = f$ for all $f \in F_k$, $k \in [l + 1]$.
    Our first claim proves a crucial \emph{convolution property} for the assignments.

    \begin{restatable}{claim}{ConvolutionProperty}
        \label{claim: metric-refinement-claim-2}
        For all $t, s, k \in [l + 1]$ such that $t \le s \le k$ and for all $f \in F_k$,
        $\Pi_{t}^{(k)}(f) = \Pi_{t}^{(s)}\left(\Pi_{s}^{(k)}\left( f\right)\right).$
    \end{restatable}

    With this claim in hand, proving that the assignments form a refinement is straightforward algebra:

    \begin{lemma}\label{lem: algorithm-for-general-metric-forms-partition-refinement}
    The assignments $\Pi_1^{(l + 1)}, \ldots, \Pi_l^{(l + 1)}$ output by Algorithm \ref{alg: metric-partition-refinement} satisfy the assignment condition and therefore form a refinement.
    \end{lemma}

    \begin{proof}
        Consider facility $f_k \in F_k$ for some $k \in [2, l + 1]$ and denote $f_{k - 1} = \Pi^{(k)}_{k - 1}(f_k) \in F_{k - 1}$. Let client $j \in X = F_{l + 1}$ be such that $\Pi_{k}^{(l + 1)}(j) = f_k$. By Claim \ref{claim: metric-refinement-claim-2}, we have $\Pi_{k - 1}^{(l + 1)}(j) = \Pi_{k - 1}^{(k)}\left(\Pi_{k}^{(l + 1)}(j)\right).$ But $\Pi_{k - 1}^{(k)}\left(\Pi_{k}^{(l + 1)}(j)\right) = \Pi_{k - 1}^{(k)}\left(f_k\right) = f_{k - 1}$. That is, all clients assigned to $f_k$ by $\Pi_k^{(l + 1)}$ are assigned to $f_{k - 1}$ by $\Pi_{k - 1}^{(l + 1)}$.
    \end{proof}

    To prove the cost upper bound, we need the following technical claim:

    \begin{restatable}{lemma}{RecurrenceRelation}\label{lem: general-metric-recurrence-relation}
    Given real $\gamma > 1$ and some integer $l \in \Z_+$, consider the set of recursively defined numbers $u_{k, t}$ for $1 \le t, k \le l + 1$, where $u_{k, t} = \gamma^{t-1}$ when $t=k$, and $u_{k,t} = \max \big\{ \gamma^{t}, \underset{s \in [t, k - 1]}{\max} \left(\gamma^{t - s} + u_{s, t} (1 + \gamma^{t - s})\right) \big\}$ otherwise.
    Then, for any $k\geq 1$ and $t\geq 1$, we have that $u_{k, t} \le e^\frac{2 \gamma}{\gamma - 1} \gamma^{l}$. Further, $\min_{\gamma > 1} e^\frac{2 \gamma}{\gamma - 1} \gamma^{l} \le e^{3\sqrt{l}+2}$.
    \end{restatable}

    We are ready to prove the approximation guarantee and therefore our theorem:

    \begin{proof}[Proof of Theorem \ref{thm: reassignment-problem} for arbitrary metric.]
        We seek $\gamma > 1$ such that the assignments $\Pi_1^{(l + 1)}, \ldots,\Pi_l^{(l + 1)}$ satisfy the cost upper bound. We prove the following stronger statement where $u$ is the sequence defined in Lemma \ref{lem: general-metric-recurrence-relation}:
        $$
        d\left(f, \Pi^{(k)}_t(f) \right) \le u_{k, t} \: d(f, h_t(f)), \\ \quad \forall \: f \in F_k, \: \forall \: k, t \in [l + 1], t \le k.
        $$
        Choosing $k = l + 1$ and using Lemma \ref{lem: general-metric-recurrence-relation} to pick the appropriate $\gamma$ then implies the cost upper bound since $\max_{k, t} u_{k, t} = O(e^{3\sqrt{l}})$ and $d(f, h_t(f)) = \min_{h \in F_t} d(f, h)$.

        We induct on $k - t$. When $k - t = 0$, $\Pi_{t}^{(k)}(f) = f$ by definition for all $f \in F_k$. Then $d(f, \Pi_{t}^{(k)}(f)) = 0$, and the claim follows since $u_{k, t} \ge 0$ for all $k, t$. Suppose $k - t \ge 1$. Fix $t \in [k - 1]$ and $f \in F_k$. Let $s^*$ be the index in $[k - 1]$ chosen in step \ref{step: metric-partition-index-selection} of the algorithm. As in the algorithm, we denote $h = h_{s^*}(f)$.

        \textbf{Case I}: $t \in [s^*, k - 1]$. From step \ref{step: metric-partition-assignment}, we have $\Pi_t^{(k)}(f) = h_{s^*}(f) := h$. By the choice of $s^*$, we have $\gamma^{t} d(h_t(f), f) \ge \gamma^{s^*} d(h, f)$. This implies that $d(h, f) \le \gamma^{t - s^*} d(h_t(f), f) \le \gamma^{t - 1} d(h_t(f), f) \le u_{k, t} d(h_t(f), f)$ as claimed.

        \textbf{Case II}: $t \in [s^* - 1]$.  Then, we have the following inequalities
        $$
        d\left(f, \Pi_t^{(k)}(f)\right) \le d\left(f, h\right) + d\left(h, \Pi_t^{(k)}(f)\right) \le d\left(f, h\right) + d\left(h, \Pi_t^{(s^*)}(h)\right) \le d\left(f, h\right) + u_{s^*, t} \: d\left(h, h_t(h)\right).
        $$
        The first inequality is the triangle inequality. The second inequality follows from step \ref{step: metric-partition-assignment} of the algorithm that shows $\Pi_t^{(k)}(f) = \Pi_t^{(s^*)}(h)$.
        The third inequality follows from the induction hypothesis since $s^* - t < k - t$.

        By definition, $h_t(f) \in F_t$ and $d(h, h_t(h)) = \min_{f' \in F_t} d(h, f') \le d(h, h_t(f))$. Further, $d(h, h_t(f)) \le d(h, f) + d(f, h_t(f))$ by the triangle inequality. Therefore, we have that $d\left(f, \Pi_t^{(k)}(f)\right) \le d(f, h) + u_{s^*, t} \left(d(f, h) + d(f, h_t(f))\right) = (1 + u_{s^*, t}) d(f, h) + d(f, h_t(f))$. By the choice of $s^*$, we have $\gamma^{s^*} d(f, h) \le \gamma^{t} d(f, h_t(f))$, so that $$
        d\left(f, \Pi_t^{(k)}(f)\right) \le \left( \left(1 + u_{s^*, t}\right) \gamma^{t - s^*} + 1 \right) d(f, h_t(f)) \le u_{k, t} d(f, h_t(f)).
        $$
    \end{proof}

    \section{\textsc{ExpandIntervals} Algorithm for Line Metric}\label{sec: line-metric}

    In this section, we give Algorithm \textsc{ExpandIntervals} (Algorithm \ref{alg: line-partition-refinement}) that outputs an $\alpha$-reassignment for $\alpha = 2l$ when given facility sets $F_1 \subseteq \ldots \subseteq F_l \subseteq X \subseteq \R$ on a line, thus finishing the proof of Theorem \ref{thm: reassignment-problem} for line metric. Recall that we seek assignments $\Pi_1, \ldots, \Pi_l$ that satisfy the assignment condition (eqn. (\ref{eqn: assignment-condition})) and the cost upper bound $d(j, \Pi_k(j)) \le \alpha \cdot \min_{f \in F_k} d(j, f)$ for each client and each `level' $k \in [l]$.

\begin{figure}[t]
        \begin{minipage}{0.32\textwidth}
            \centering
            \includegraphics[width=\columnwidth]{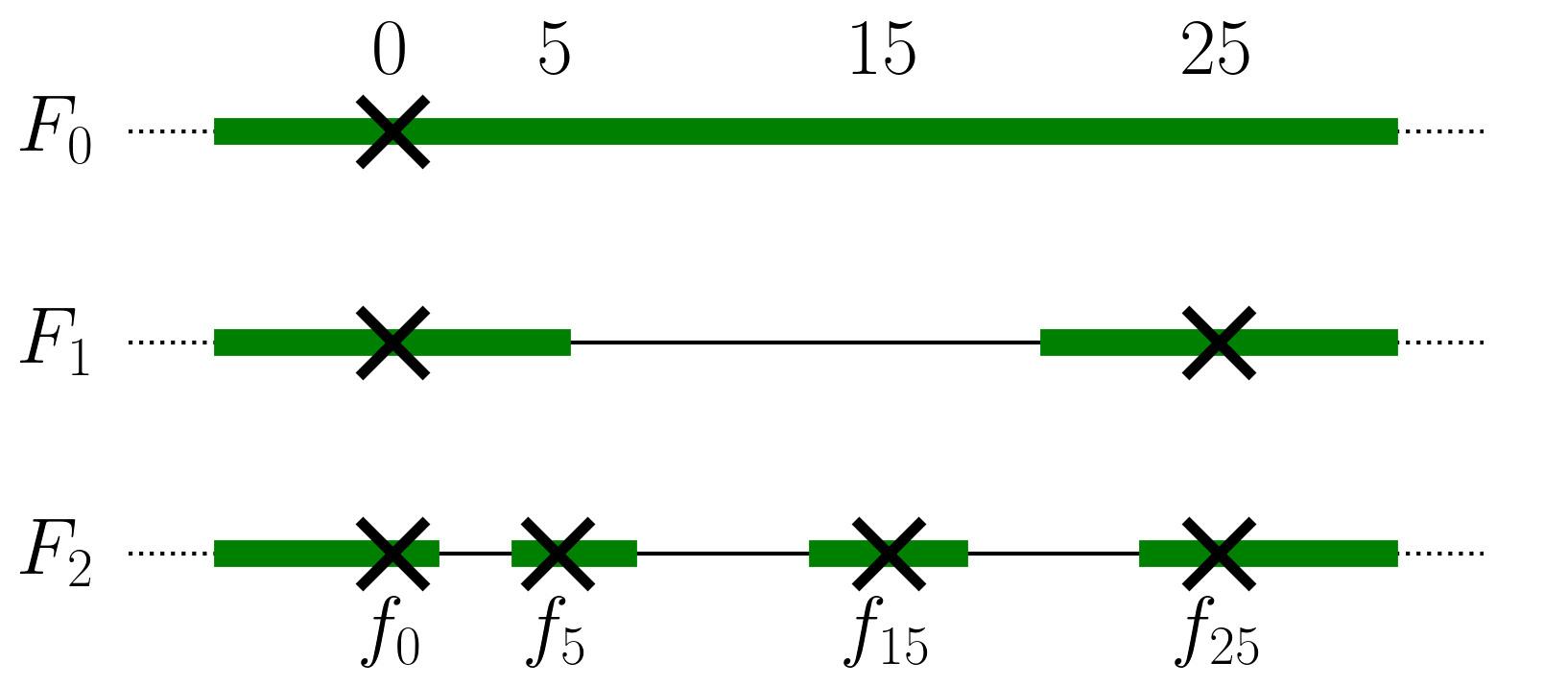}
        \end{minipage}
        \begin{minipage}{0.67\textwidth}
            \caption{An example to illustrate intervals $N_\alpha(f, k)$, with $l = 2$ levels and an additional auxiliary level $F_0 = \{f_0\}$; $\alpha = 2l = 4$. Facilities $f_0, f_5, f_{15}, f_{25} \in F_2$ are at $x = 0, 5, 15, 25$ respectively with $N_\alpha(f_0, 0) = (-\infty, 1]$, $N_\alpha(f_5, 0) = [4, 7]$, $N_\alpha(f_{15}, 0) = [13, 17]$, and $N_\alpha(f_{25}, 0) = [23, \infty)$. For facilities $F_1$, $N_\alpha(f_0, 1) = (-\infty, 5]$ and $N_\alpha(f_{25}, 1) = [20, \infty)$.}
            \label{fig: line-interval-example}
        \end{minipage}
    \end{figure}

    \textsc{ExpandIntervals} begins by identifying clients that must be assigned to specific facilities if the cost upper bound is to be satisfied: for any client $j \in X$ and facility $f \in F_k$, if $\alpha \cdot d(j, f) < d(j, f')$ for all $f' \in F_k \setminus \{f\}$, then we must have $\Pi_k(j) = f$. In particular, consider three consecutive facilities $f'', f, f' \in F_k$ located at $x(f'') < x(f) < x(f')$ respectively. If we only allow clients to be assigned to their immediate left or right facilities, then any client $j$ in the `necessary' interval $N_\alpha(f, k) := \left[x(f) - \frac{x(f) - x(f'')}{\alpha + 1}, x(f) + \frac{x(f') - x(f)}{\alpha + 1}\right]$ is closer to $f$ than its other immediate neighbor in $F_k$ by factor $\alpha$ (see Figure \ref{fig: line-interval-example} for an illustration). It is then necessary to assign $\Pi_k(j) = f$ for all clients $j \in N_\alpha(f, k)$ to satisfy the cost upper bound.

    Further, if $N_\alpha(f, k)$ intersects $N_\alpha(h, t)$ for some $h \in F_{t}, t < k$, then by the assignment condition, we must have all clients assigned to $f$ under $\Pi_k$ assigned to $h$ under $\Pi_{t}$ and in particular $\Pi_{t}(j) = h$ for all clients $j \in N_\alpha(f, k)$. Step 1 of our algorithm \textsc{ExpandIntervals} identifies intervals $A(f, k) \supseteq N_\alpha(f, k)$ for each facility $f \in F_k$ and level $k \in [l]$ such that each client in $A(f, k)$ is assigned to $f$ under $\Pi_k$.

    \begin{figure}
        \centering
        \begin{subfigure}{0.32\textwidth}
            \includegraphics[width=\textwidth]{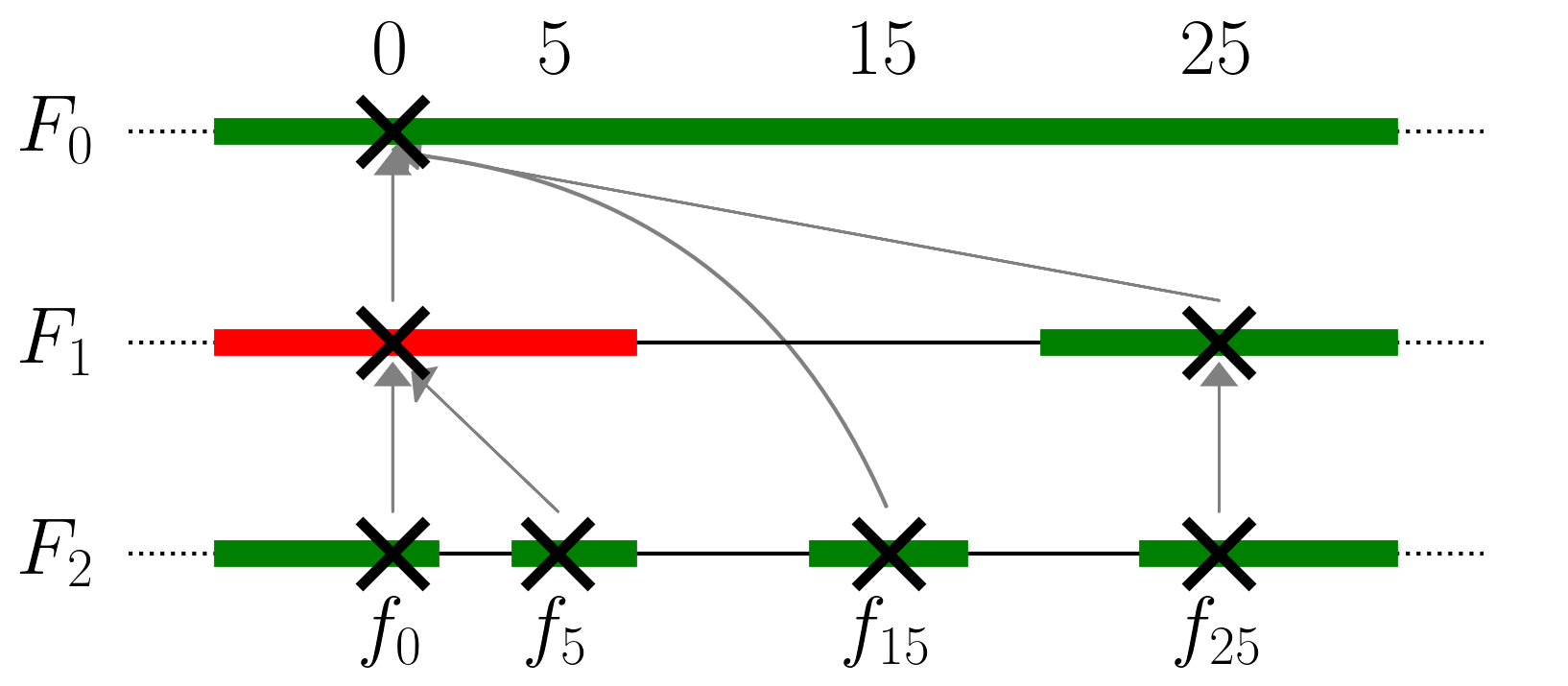}
            \caption*{(a)}
        \end{subfigure}
        \hfill
        \begin{subfigure}{0.32\textwidth}
            \centering
            \includegraphics[width=\textwidth]{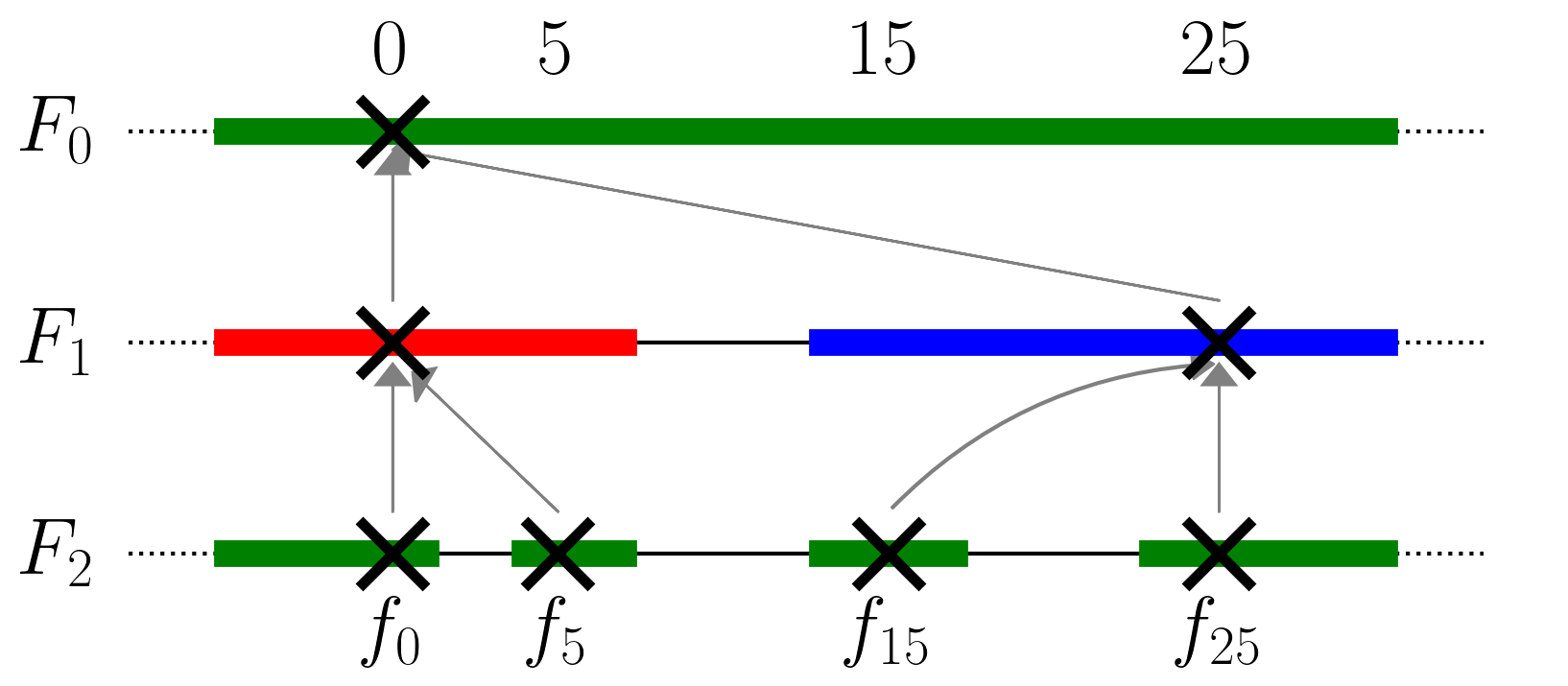}
            \caption*{(b)}
        \end{subfigure}
        \hfill
        \begin{subfigure}{0.32\textwidth}
            \includegraphics[width=\textwidth]{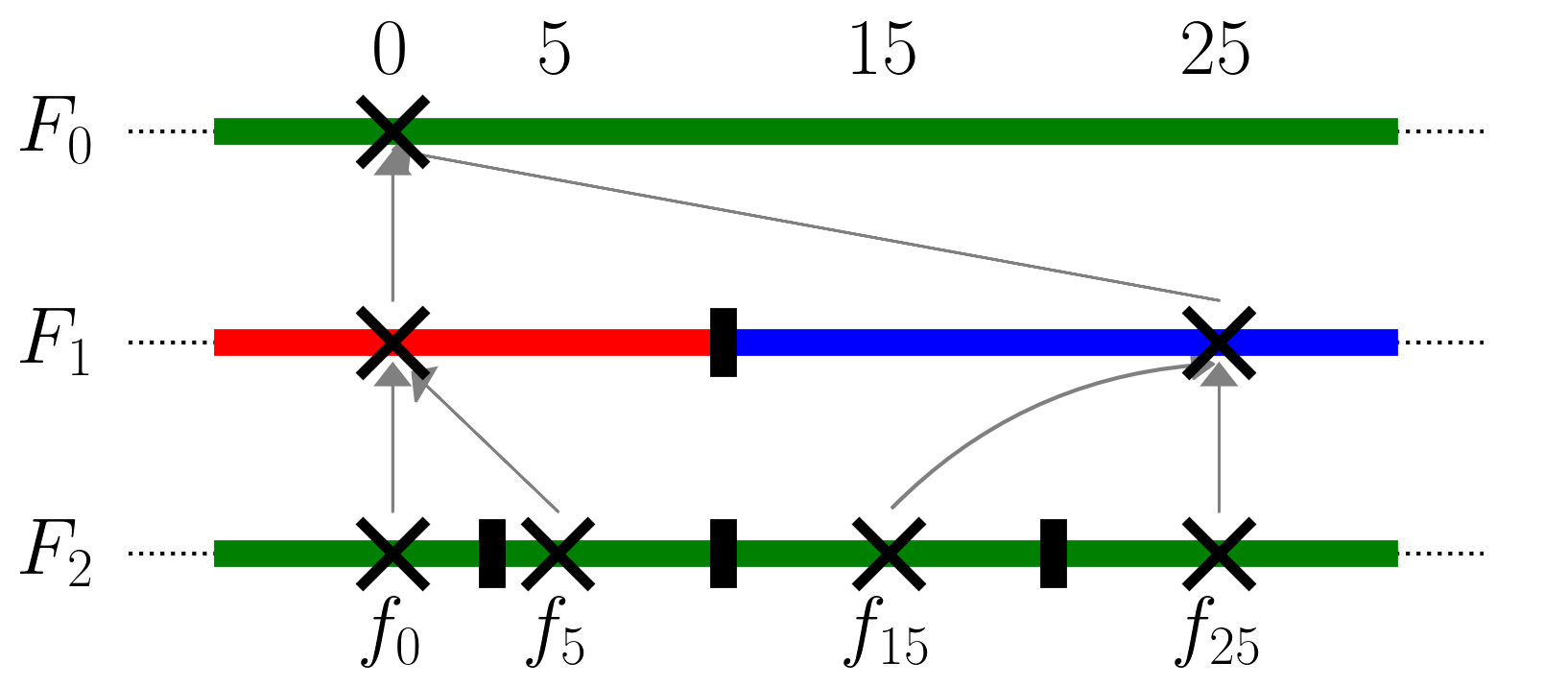}
            \caption*{(c)}
        \end{subfigure}
        \caption{An example to illustrate Algorithm \textsc{ExpandIntervals}. Intervals $A(f, k)$ for facilities in $F_2$, $F_1$, and for auxiliary level $F_0$ are colored and initialized to $N_\alpha(f, k)$, as in Figure \ref{fig: line-interval-example}. The edges of graphs $H$ are drawn in grey. Steps of  \textsc{ExpandIntervals}: (a) Step 1 where the interval $A(f_{0}, 1)$ (in red) expands (b) Step 2, where edge $((f_{15}, 2), (f_0, 0))$ is removed and replaced by edge $((f_{15}, 2), (f_{25}, 1))$ and corresponding interval $(f_{25}, 1)$ (in blue) expands (c) Step 3 where all intervals expand arbitrarily to cover $\R$ at each level.}
        \label{fig: line-interval-partition}
    \end{figure}

    \SetKwComment{Comment}{/* }{ */}

    \begin{algorithm}[t]
        \caption{\textsc{ExpandIntervals}($F_0, F_1, \ldots, F_l$)}
        \label{alg: line-partition-refinement}
        \nonl \textbf{input}: Facility sets $F_0 \subseteq F_1 \subseteq \ldots \subseteq F_l \subseteq X = \R$\\
        \nonl \textbf{output}: Assignments $\Pi_k: X \to F_k$ for all $k \in [l]$\\
        initialize $H = (\ov{F}, \emptyset)$ to be the empty graph on vertex set $\ov{F} = \{(f, k): f \in F_k, k \in [0, l]\}$ \\
        \label{step: A-initialization} initialize $A(f, k) = N_\alpha(f, k)$ for all $(f, k) \in \ov{F}$\\
        \nonl \textbf{invariant}: $A(f, k) \supseteq N_\alpha(f, k)$ for all $(f, k) \in \overline{F}$ throughout the algorithm\\
        \nonl \textbf{step 1}: add edges to $H$ \\
        \For{level $k = l - 1$ to $0$}
        {
            \label{step: A_assignment_loop}
            \For{each facility $f \in F_k$}
            {
                \label{step: A_update} define the set of facilities $S(f, k) = \{(h, s): s > k, A(h, s) \cap N_\alpha(f, k) \neq \emptyset\}$ \\
            \For{each $(h, s) \in S(f, k)$}{
                update the interval $A(f, k) \gets A(f, k) \cup A(h, s)$ \\
                add edge $\big((h, s), (f, k)\big)$ to $H$
            }
            }
        }\label{step: first-for-loop-ends}
        \nonl Lemma \ref{lem: disjoint-interval-trees-A}: $H$ is an interval tree after step 1 \\
        \nonl \textbf{step 2}: rearrange $H$ to satisfy \emph{immediate parent} property\\
        \For{level $k = 0$ to $l - 1$ \label{step: step-2-outer-loop}}
        {
            \For{each facility $f \in F_k$}
            {
                \For{each child $(h, s)$ of $(f, k)$ in $H$ with $s > k+1$}
                {
                    let $(g, k + 1)$ be the child of $(f, k)$ in $F_{k + 1}$ closest to $(h, s)$ \\
                remove edge $\big((h, s), (f, k)\big)$ from $H$ and add edge $\big((h, s), (g, k + 1) \big)$ to $H$ \\
                update $A(g, k + 1) \gets \text{conv}\big(A(g, k + 1) \cup A(h, s)\big)$ \label{step: combined-convex-hull} \\
                }
            }
        }
        \nonl Lemma \ref{lem: immediate-ancestor-property}: $H$ is an interval tree  and satisfies the immediate parent condition after step 2 \\
        \nonl \textbf{step 3}: assign any unassigned intervals to satisfy \emph{completeness} \\
        \label{step: complete-assignment}\textsc{CompleteAssignment}$\big((f_0, 0), H\big)$ \\
        Lemma \ref{lem: line-metric-completeness}: $H$ is a hierarchy tree after step 3 \\
        \Return{assignments $\Pi_1, \ldots, \Pi_l$ induced by intervals $A(f, k)$ for $(f, k) \in \ov{F}$} \\
    \end{algorithm}

    \begin{algorithm}
        \caption{\textsc{CompleteAssignment}($(f, k), H$)}\label{alg: refine-boundary-subroutine}
        \nonl \textbf{input}: Facility, level pair $(f, k)$ and interval tree $H$ \\
        let $(f_1, k + 1), \ldots, (f_t, k + 1)$ be the children of $(f, k)$ in $H$. \\
        expand intervals $A(f_1, k + 1), \ldots, A(f_t, k + 1)$ arbitrarily so that they partition $A(f, k)$ \\
        \For{children $f \in \{f_1, \ldots, f_t\}$}
        {
            \textsc{CompleteAssignment}($(f, k + 1), H$) \\
        }
    \end{algorithm}

    In Figure \ref{fig: line-interval-example}, for example, we have two levels ($l = 2$); a third auxiliary facility set $F_0 = \{f_0\}$ is introduced for convenience where $f_0$ is an arbitrarily chosen facility from $F_1$. Intervals $A(f, k)$ are initialied to $N_\alpha(f, k)$ for $\alpha = 2l = 4$ in Figure \ref{fig: line-interval-example}. Since $A(f_5, 2) \cap A(f_0, 1) \neq \emptyset$, interval $A(f_0, 2)$ is expanded to $A(f_5, 2) \cup A(f_0,1)$. Additionally, some edges (in grey) are added so that intervals $A(f, k)$ form an `interval tree':

    \begin{definition}[Interval Tree and Hierarchy Tree]
        Given intervals $A(f, k)$ for facilities in $\overline{F} := \{(f, k): f \in F_k, k \in [0, l]\}$, a directed graph $H$ with vertex set $\overline{F}$ is called an \emph{interval tree} if
        \begin{enumerate}
            \item It is a tree rooted at $(f_0, 0)$ with edges pointed towards $(f_0, 0)$.
            \item For any edge $((h, s), (f, k))$ in $H$, we have $s > k$ and $A(f, k) \supseteq A(h, s)$.
            \item For any two siblings $(f, k)$ and $(f', k')$ in $H$, we have $A(f, k) \cap A(f', k') = \emptyset$.
        \end{enumerate}

        Further, we call $H$ a hierarchy tree if it also satisfies
        \begin{enumerate}
            \item (Immediate parent property) For all $(f, k) \in \ov{F} \setminus \{(f_0, 0)\}$, the parent of $(f, k)$ in $H$ is $(g, k - 1)$ for some $g \in F_{k - 1}$. That is, the only edge exiting $(f, k)$ is of the form $((f, k), (g, k - 1))$,
            \item (Completeness) For all $k \in [0, l]$, $\bigcup_{f \in F_k} A(f, k) = \R$.
        \end{enumerate}
    \end{definition}

    We formalize this in the following lemma:

    \begin{restatable}{lemma}{LineStepOneIntervalTree}\label{lem: disjoint-interval-trees-A}
    $H$ is an interval tree after step 1 in Algorithm \textsc{ExpandIntervals}.
    \end{restatable}

    We still need to ensure that the algorithm produces well-defined assignments, i.e., for each level $k \in [l]$, each client is assigned once (and exactly once) under each $\Pi_k$; this is equivalent to saying that intervals $\{A(f, k): f \in F_k\}$ partition $\R$. Note that the length $\frac{x(f') - x(f'')}{\alpha + 1}$ of interval $N_\alpha(f, k)$ is inversely proportional to $\alpha$; we show that choosing $\alpha \ge 2l$ ensures that $A(f, k)$ does not intersect with $A(f', k)$ for $f \neq f'$. Steps 2 and 3 of the algorithm ensure that $\bigcup_{f \in F_k} A(f, k) = \R$, i.e., all clients are covered by $\Pi_k$.

    Step 2 ensures that interval tree $H$ satisfies the immediate parent property by rearranging its edges. In Figure \ref{fig: line-interval-partition}(a), the problematic edge is $((f_{15}, 0), (f_0, 2))$. The child of $(f_{0}, 0)$ closest to $(f_{15}, 2)$ is $(f_{25}, 1)$, and so edge $((f_{15}, 2), (f_{25}, 1))$ is added and $A(f_{25}, 1)$ is expanded to the convex hull of $A(f_{25}, 1) \cup A(f_{15}, 2)$ in Figure \ref{fig: line-interval-partition}(b). As shown in Figure \ref{fig: line-interval-partition}(c), step 3 expands intervals $A(f, k)$ to cover $\R$ for each level $k \in [0, l]$. The next two lemmas formalize this:

    \begin{restatable}{lemma}{LineStepTwo}\label{lem: immediate-ancestor-property}
    $H$ is an interval tree rooted at $(f_0, 0)$  and satisfies the immediate parent property after step 2 in Algorithm \ref{alg: line-partition-refinement}.
    \end{restatable}

    \begin{lemma}\label{lem: line-metric-completeness}
    $H$ satisfies the completeness condition at the end of Algorithm \ref{alg: line-partition-refinement}.
    \end{lemma}

    As discussed, the cost upper bound is satisfied since $A(f, k) \supseteq N_\alpha(f, k)$, and the assignment condition is satisfied since we ensure that whenever $A(f, k)$ intersects $A(h, t)$ for some $f \in F_k, h \in F_{t}, t < k$, then $A(f, k) \subseteq A(h, t)$. Our next lemma asserts that this is equivalent to $H$ being a hierarchy tree:

    \begin{restatable}{claim}{LineHierarchyTreesAreRefinements}
        \label{claim: line-interval-tree}
        Suppose we are given intervals $A(f, k)$ for all $(f, k) \in \ov{F}$ and a hierarchy tree $H = (\ov{F}, E)$ such that $A(f, k) \supseteq N_\alpha(f, k)$ for each $(f, k) \in \ov{F}$ some $\alpha \ge 1$. Define assignments $\Pi_k: X \to F_k, k \in [0, l]$ as $\Pi_k(j) = f$ for all clients $j \in A(f, k)$ and $(f, k) \in \overline{F}$. Then,
        \begin{enumerate}
            \item Assignments $\Pi_0, \Pi_1, \ldots, \Pi_l$ satisfy the assignment condition, i.e., form a refinement,
            \item For all clients $j \in X$ and levels $k \in [0, l]$, assignments $\Pi_k$ satisfy the cost upper bound $
            d(j, \Pi_k(j)) \le \alpha \cdot \min_{f \in F_k} d(j, f).$
        \end{enumerate}
    \end{restatable}

    Finally, we note that intervals $A(f, k)$ are initialized to $N_\alpha(f, k)$ and only expanded in the whole algorithm, thus satisfying that $A(f, k) \supseteq N_\alpha(f, k)$ at the end of the algorithm. Together with Lemma \ref{lem: line-metric-completeness} and Claim \ref{claim: line-interval-tree}, this finishes the proof of Theorem \ref{thm: reassignment-problem} for line metric.

    \section{Portfolios in Practice: Recommending New Pharmacies}\label{sec: experiment-pharmacies}

    \begin{figure}[h]
        \begin{minipage}{0.63\textwidth}
            \centering
            \includegraphics[width=\columnwidth]{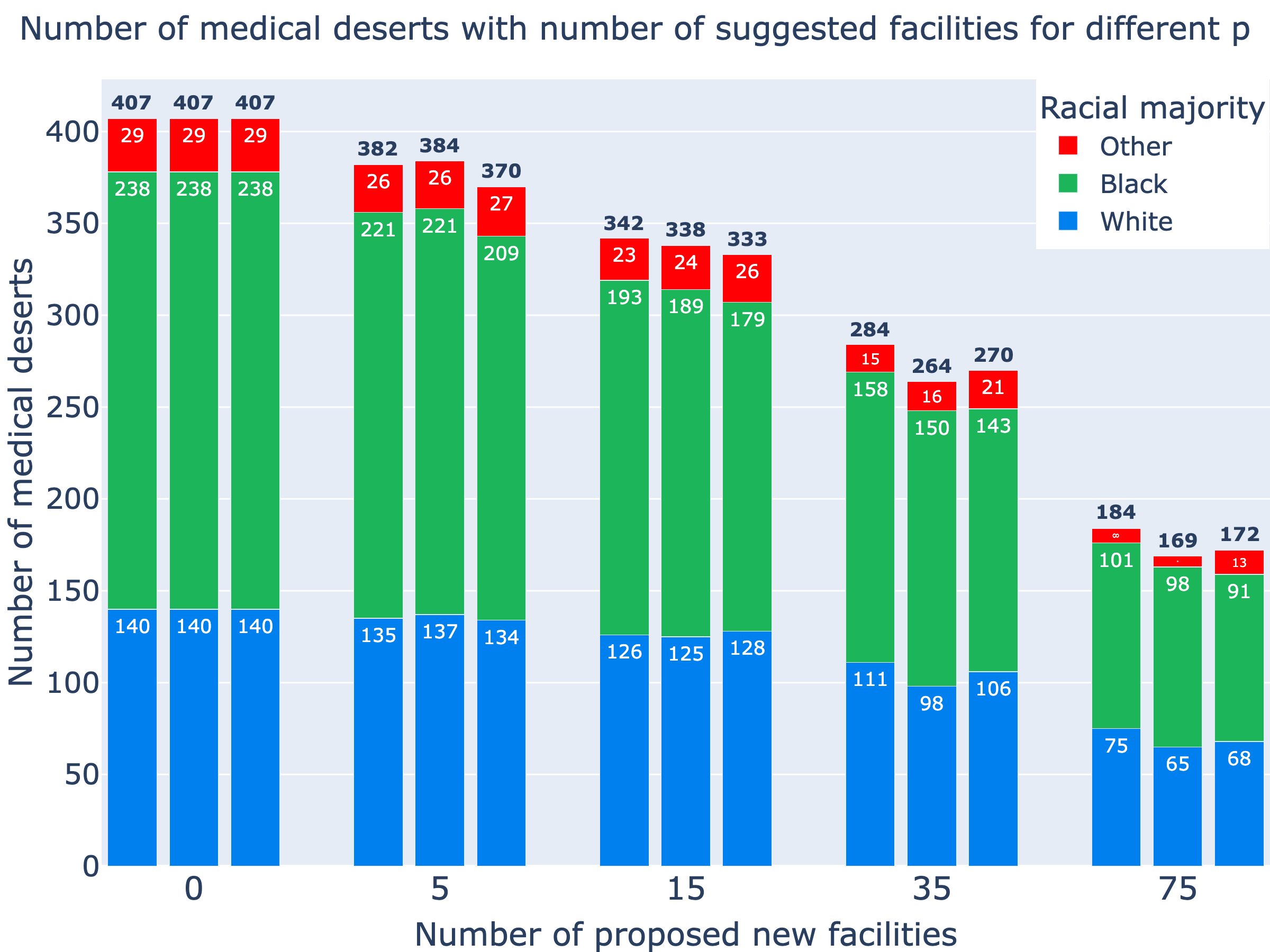}
            \caption{Reduction in the number of medical deserts for $L_p, p\in \{1, 2, \infty\}$ norm solutions with $k = 5, 15, 25$ or $75$ newly open facilities. Each bar represents the number of medical deserts for a $(p, k)$ pair. We observe a consistent reduction in the number of medical deserts for each $p$.}
            \label{fig: experiment-portfolio-bar-chart}
        \end{minipage}
        \hfill
        \begin{minipage}{0.35\textwidth}
            \centering
            \includegraphics[width=0.7\textwidth]{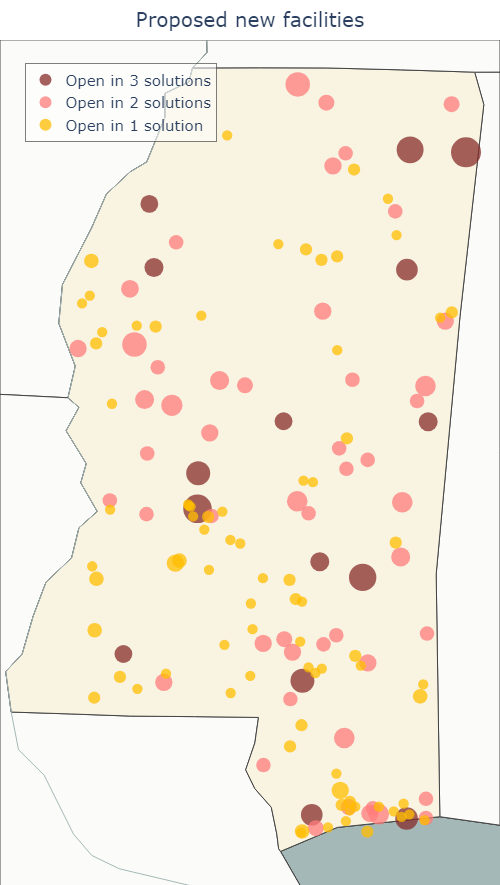}
            \caption{A heatmap for all facilities suggested by the three solutions. Facilities opened earlier are drawn larger. The color denotes whether the facility is in only one, in two, or in all three solutions.}
            \label{fig: mississippi-heatmap}
        \end{minipage}
    \end{figure}

    In this section, we return to medical deserts in Mississippi, USA (see Figure \ref{fig: online-tool}) and present a portfolio of recommendations for new pharmacy placements in addition to existing CVS, Walgreens, and Walmart pharmacies. Our recommendations satisfy the facility condition for refinements and therefore can be opened with rolling budgets.
    Our portfolio has three solutions corresponding to three $L_p$ norm objectives; these are obtained using the ideas in Sections \ref{sec: portfolio-upper-bound} and \ref{sec: refinements}, suitably modified for real data.
    Each solution suggests a sequence of $75$ facilities to be opened over time; however, the solutions are flexible, and opening just a few of these $75$ facilities can potentially drastically reduce the number of medical deserts identified by our tool (see Figure \ref{fig: experiment-portfolio-bar-chart}).

\begin{figure}[t]
        \begin{minipage}{0.67\textwidth}
            \centering
            \includegraphics[width=\textwidth]{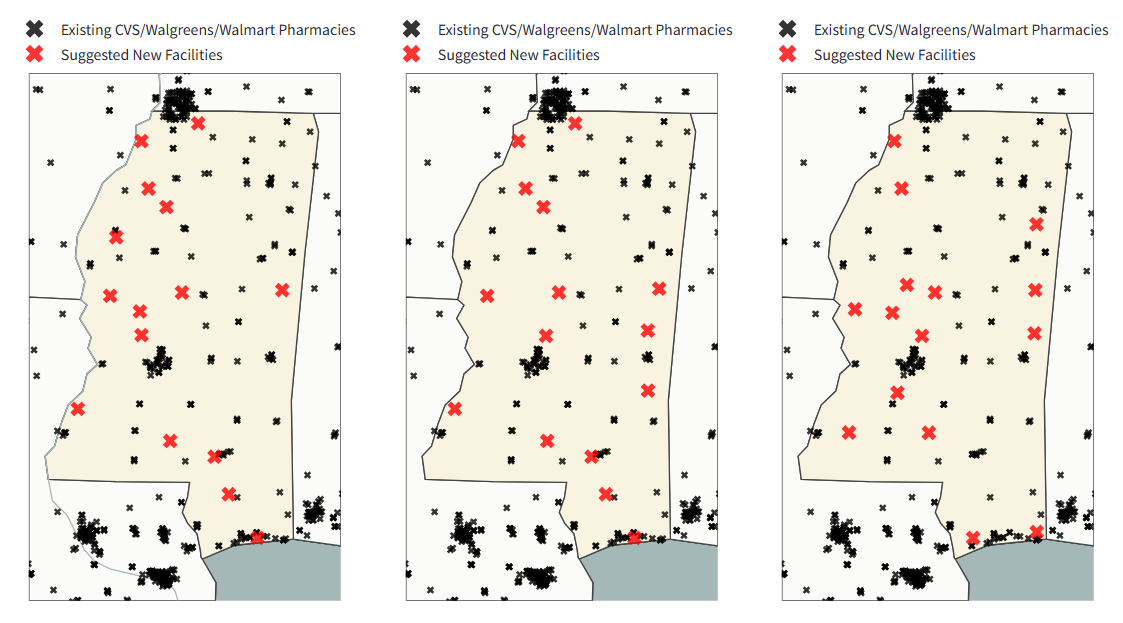}
        \end{minipage}
        \hfill
        \begin{minipage}{0.32\textwidth}
            \caption{(In red) The first $15$ sites for new pharmacies suggested by our tool corresponding to (left) $L_1$, (center) $L_2$, and (right) $L_\infty$ objectives respectively in Mississippi, USA. These facilities are suggested in addition to existing CVS, Walgreens, and Walmart pharmacies (which are in black).}
            \label{fig: experiment-suggested-facilities-portfolio}
        \end{minipage}

    \end{figure}

    Recall that our tool defines a medical desert as a US census blockgroup with over $p\%$ people living below the poverty line and over $n$ miles away from the closest CVS/Walgreens/Walmart pharmacies. Parameters $p$ and $n$ can be chosen by the user, and $n$ can be different for rural and urban areas.
    Following a definition similar to US Department of Agriculture's (USDA) food deserts \cite{DPF12}, the tool identifies $407$ medical deserts in Mississippi, USA when $n_{\text{urban}} = 2$, $n_{\text{rural}} = 10$ and $p = 20$.
    Majority Black or African American blockgroups form a disproportionately high $58.5\%$ of medical deserts despite being only $35.0\%$ of all blockgroups.
    Pharmacy data, sourced from the HIFLD Open Database (2009)\footnote{We could not find official pharmacy data after 2009. While we recognize that pharmacy locations may have changed since, the general patterns in our tool's findings -- the existence of facility deserts and racial disparity -- are also true for other facilities such as hospitals, private schools, etc for which newer data is available.}, includes 174 CVS, Walgreens, and Walmart pharmacies in Mississippi, and some others across state borders (see Figure \ref{fig: experiment-suggested-facilities-portfolio}).

    \textbf{Choice of clients, client groups, and norms.} To suggest opening new facilities, we model each census blockgroup in Mississippi as a client and assume that it is located at its geographic center. We classify each census blockgroup
    \begin{enumerate}
        \item by its racial/ethnic majority into one of (i) White alone (labeled `White' in figures), (ii) Black or African American alone (`Black'), (iii) other or no single racial/ethnic majority (`Other')
        \item based on whether or not over $20\%$ of the blockgroup's population is below the poverty line
        \item based on whether (i) over $25\%$ of the population has no health insurance (ii) over $25\%$ of the population has two health insurances, (iii) neither of the two.
    \end{enumerate}
    This results in $3 \times 2 \times 3 = 18$ client groups (some of which are empty and therefore discarded). We choose the $L_1, L_2$, and $L_\infty$ norms for the three objective functions in our portfolio. As before, to account for access differences between rural and urban areas, we modify the objective so that the distance for each urban blockgroup has weight five times the distance for each rural blockgroup.

    \begin{figure}
        \begin{subfigure}{0.24\textwidth}
            \includegraphics[width=\columnwidth]{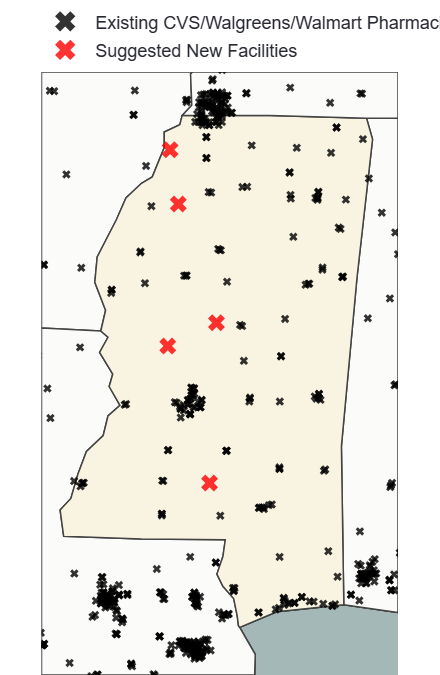}
            \caption*{$k = 5$}
        \end{subfigure}
        \hfill
        \begin{subfigure}{0.24\textwidth}
            \includegraphics[width=\columnwidth]{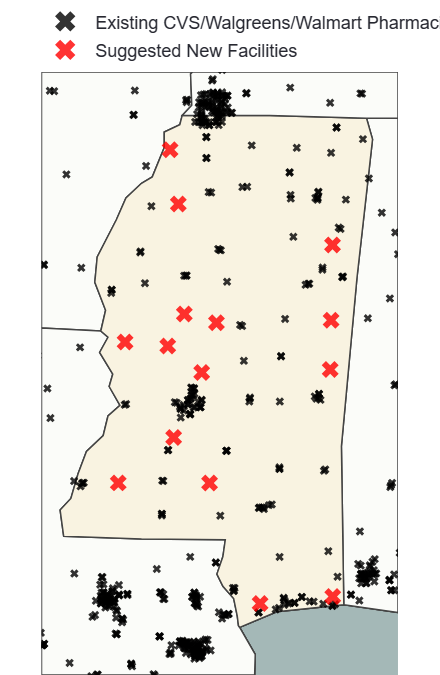}
            \caption*{$k = 15$}
        \end{subfigure}
        \hfill
        \begin{subfigure}{0.24\textwidth}
            \includegraphics[width=\columnwidth]{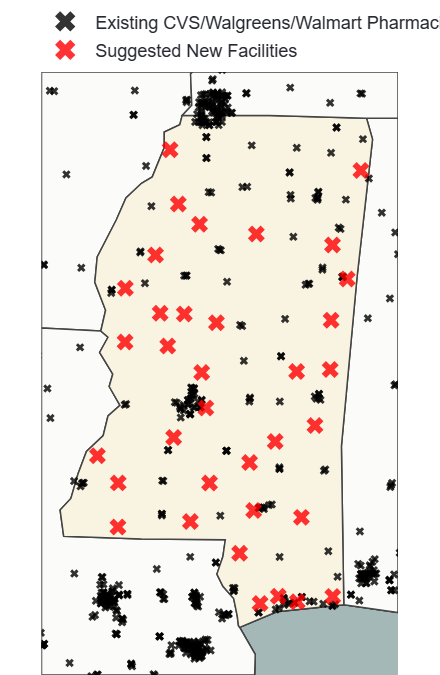}
            \caption*{$k = 35$}
        \end{subfigure}
        \hfill
        \begin{subfigure}{0.24\textwidth}
            \includegraphics[width=\columnwidth]{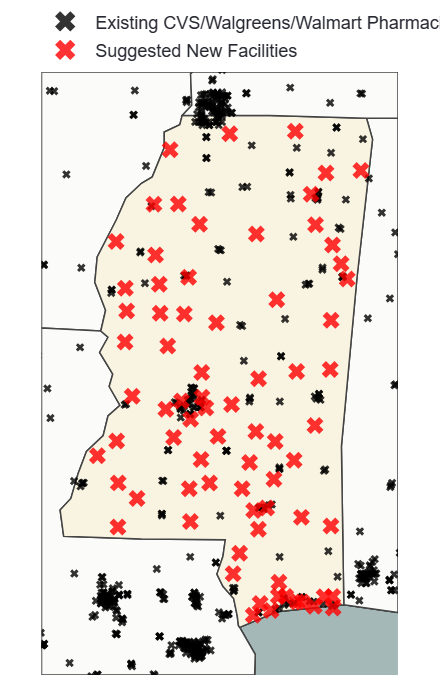}
            \caption*{$k = 75$}
        \end{subfigure}
        \caption{Refinement of suggested facilities for the $L_\infty$ solution with $k = 5, 15, 35, 75$ new open facilities from left to right.}
        \label{fig: experiment-refinement}
    \end{figure}

    \textbf{Choice of $k$ and refinements.} We suitably modify \cite{STA97}'s algorithm (see Theorem \ref{thm: facility-location-solvability-theorem} and Appendix \ref{app: relaxation-and-rounding}) to take in account existing $174$ CVS, Walgreens, Walmart pharmacies. We allow a new facility to open at the location of any blockgroup. To recommend iterative opening of facilities, we first run the algorithm for $k$-clustering suggesting $k = 5$ new facilities for each of the three $L_p$ norm objectives. Next, we run the algorithm to suggest $k = 10$ facilities in addition to the $174 + 5$ facilities, followed by $k = 20$ and then $k = 40$ more facilities. This is based on the doubling idea employed to obtain facility sets for refinements in Section \ref{sec: refinements}, suitably modified to avoid bicriteria solutions and take into account already open facilities. This results in a total of $5 + 10 + 20 + 40 = 75$ new facilities for each $L_p$ norm.

    \textbf{Results.} Figure \ref{fig: experiment-suggested-facilities-portfolio} shows the first $15$ facilities for each of the three solutions in the portfolio; we observe that while some facilities are open in all three solutions, there are also significant differences. Figure \ref{fig: experiment-refinement} shows the sequence of open facilities for the $L_\infty$ norm solution.

    Figure \ref{fig: experiment-portfolio-bar-chart} shows reduction in the number of medical deserts for each of the three solutions with $k = 5, 15, 25, 35, 75$ facilities open. For comparison, the existing state ($k = 0$) is also plotted. We observe that opening a small number of new facilities can significantly reduce the number of medical deserts. E.g., opening just the first $15$ facilities in the $L_\infty$ norm solution leads to $74$ fewer medical deserts. When considering the average population in a blockgroup ($600$ to $3000$ people), this potentially impacts tens of thousands of people. This is true despite the optimization objective not explicitly aiming to reduce the number of medical deserts.
    Next, the optimization partially addresses the disproportionate impact of medical deserts on majority Black population. We see that out of the $239 - 179 = 74$ medical deserts newly served by the first $15$ facilities in the $L_\infty$ norm solution, $59$ ($80\%$) are majority Black. Finally, while all three solutions behave similarly, we find that $L_2, L_\infty$ norms seem better at reducing the number of medical deserts.

    \section{Conclusion}\label{sec: conclusion}

    Significant evidence from the US and from around the world indicates that critical infrastructure is less accessible to certain socioeconomic groups.
    Motivated by this, we studied two challenges faced by the research community in understanding fairness: developing a theoretical framework for fairness in facility location and taking into account changing budgets.
    We introduced the notion of \emph{portfolios} -- a small set of solutions representing all possible fairness criteria -- to help policymakers make choices about fairness while being on a theoretically sound footing.
    We established portfolios formally and studied the trade-off between portfolio quality (measured using approximation ratios) and size (number of solutions in the portfolio) for facility location problems.
    We also studied the refinement model to accommodate rolling budgets and gave several approximation algorithms for it. Finally, we showcased our tool to recommend new facilities that may be useful for policymakers.

    Access costs for various groups can be balanced differently using different $L_p$ norms, making them a useful class of objectives to model fairness. However, a different trade-off -- that between facility opening cost and access cost -- makes the uncapacitated facility location interesting. This latter trade-off results in an unintended consequence for fair facility location portfolios: since $L_p$ norms are non-increasing with $p$, changing $p$ from $1$ to $\infty$ makes distances more balanced but leads to \emph{fewer} open facilities. This can be remedied by introducing a \emph{balancing factor} $\lambda_p$ in the objective: $\text{(facility opening cost)} + \lambda_p (L_p \text{ norm access cost)}$. This is discussed in more detail in Appendix \ref{app: facility-access-cost-tradeoff}.

    We close with a list of open questions that arise from this work. First, we studied portfolios for monotonically interpolating classes of norms (that includes as special case $L_p$ norms, top-$l$ norms, etc) to display a proof-of-concept for the idea of portfolios. However, it is open to determine how portfolio size varies with the approximation factor for other classes of norms such as ordered norms and symmetric monotonic norms \cite{chakrabarty2022approximation}. In particular, what is the size of the smallest $O(1)$-approximate portfolios for ordered and symmetric monotonic norms for facility location problems?

    Next, we showed a bicriteria $(O(1), O(1))$-approximation for fair $k$-clustering that potentially violates the budget constraint and opens $O(k)$ facilities. It is open whether there is a unicriteria $O(1)$-approximation for fair $k$-clustering that opens exactly $k$ facilities.

    Given sets of facilities $F_1 \subseteq \ldots \subseteq F_l$, our algorithm \textsc{DiscountedLookahead} gives an $\alpha$-reassignment for $\alpha = \poly(e^{\sqrt{l}})$, i.e. given assignments that form a refinement and assign clients to within factor $\alpha$ of their distance to the closest facility in each facility set $F_t$. It is open if there even exists a better than $\alpha$-reassignment (polynomial-time or not). Similarly, is there an algorithm that achieves $o(l)$-reassignments on line metrics?

    While our work on the facility location problem leads to many such open questions that are interesting from a theoretical viewpoint and impactful from a practical perspective, our approach to fairness with portfolios goes beyond this specific problem and potentially gives rise to other exciting questions while being useful for policymaking.

    \paragraph{Acknowledgements.} This work was supported by NSF Grants CCF-2106444, CCF-1910423, 2112533 and NSF CAREER Grant 2239824.


    \appendix

    \section{Proof of Theorem \ref{thm: facility-location-solvability-theorem}}\label{app: relaxation-and-rounding}

    In this section, we use a generalization of the linear programming technique used in \cite{STA97} for uncapacitated facility location and $k$-median to give approximation guarantees for fair facility location and fair $k$-clustering. Given a fixed norm $g$ on $\R^r$, we state and prove the $4$-approximation algorithm for FFL$_g$; the proof for FC$^{(k)}_g$ is analogous and omitted.
    We start with an integer formulation: for each point $i \in X$, variable $y_i \in \{0, 1\}$ indicates whether facility at $i$ is opened and for each client $j \in X$, variable $x_{ij} \in \{0, 1\}$ indicates whether client $j$ is assigned to facility $i$.
    The distance traveled by client $j$ to its assigned facility is then $z_j = \sum_{i \in X} x_{ij} d(i, j)$, and the total (weighted by group memberships) distance traveled by clients in group $s \in [r]$ is $w_s = \sum_{j \in X} \mu_{j, s} z_j$.
    Then $g(w)$ denotes the access cost with group distance vector $w$ and norm $g$. It is easy to check that an integral solution (with $x_{i, j}, y_i \in \{0, 1\}$ for all $i, j \in X$) to the above convex program corresponds to a feasible solution to FFL$_g$ and vice-versa. Consider the convex relaxation

    \vspace{-2em}
    \begin{align}\label{eqn: convex-relaxation}
    \min_{x, y} &\sum_{i \in X} c_i y_i + g\left(w\right) &
    \text{s.t.} \tag{CP} \\
    w_s &= \sum_{j \in X} \mu_{j, s} z_j & \forall \: s \in [r], \\
    z_j &= \sum_{i \in X} x_{ij} d(i, j) & \forall \: j \in X, \\
    \sum_{i \in X} x_{ij} &\ge 1, & \forall \: j \in X, \label{constraint_1}\\
    x_{ij} &\le y_i, & \forall \: i \in X, j \in X, \label{constraint_2} \\
    0 &\le x, y. \label{constraint_3}
    \end{align}

    \begin{algorithm}[!t]
        \caption{\textsc{Filter}$(x, y)$}\label{alg: round-1}
        \nonl \textbf{input}: fractional optimal solution $(x, y)$ to \ref{eqn: convex-relaxation} \\
        \nonl \textbf{output}: filtered fractional solution $(\ov{x}, \ov{y})$ \\
        $\ov{y_i} = \frac{1}{\alpha} y_i$ for each $i \in X$ \\
        \For{$j \in X$}
        {
            define $S_j = \{i \in X: x_{ij} > 0 \}$ \\
        let $\sigma$ be the permutation arranging facilities in $S_j$ such that $d(\sigma(1), j) \le d(\sigma(2), j) \le \ldots $ \\
        define $i_j = {\arg\min}_{k} \sum_{k' \in [k]} x_{\sigma(k')j}\ge \alpha$, and $d_j(\alpha) = d(\sigma(i_j), j)$ \\
        \For{$i \in X$}
        {\label{step: rounding-ov-x-condition}
        set $\ov{x}_{ij} = \frac{1}{\alpha} x_{ij}$ if $d(i, j) \le d_j(\alpha)$ and $\ov{x}_{ij} = 0$ otherwise
        }
        }
        \Return{$(\ov{x}, \ov{y})$} \\
    \end{algorithm}

    Since $g$ is convex in $w$ (and hence in $z$ and $x$), we can get an optimal fractional solution to (\ref{eqn: convex-relaxation}) in a polynomial number of oracle calls to $g$.
    Further, since $g$ is a norm, increasing all client distances $x$ by a factor $\alpha$ increases the cost $g(w)$ by a factor at most $\alpha$. Therefore, to get a $4$-approximation for FFL$_g$, it is sufficient to give a rounding algorithm that
    \begin{enumerate}
        \item takes in a fractional solution $(x, y, z, w)$ to (\ref{eqn: convex-relaxation}), and
        \item outputs a solution $(x^*, y^*, z^*, w^*)$ such that each $x_{i, j}^*, y_i^* \in \{0, 1\}$ (i.e., this is an integer solution) satisfying (a) $z^*_j \le 4 z_j$ and (b) $\sum_{i \in X} c_i y_i^* \le 4 \sum_{i \in X} c_i y_i$.
    \end{enumerate}

    We show that the rounding algorithm of \cite{STA97} achieves this. Since $w, z$ are determined completely by $x, y$, we omit them in the algorithm description. The algorithm works in two stages: a filtering stage (Algorithm \ref{alg: round-1}) that takes in the fractional optimal $(x, y)$ and removes `far away' facilities for each client, and a rounding stage where this modified solution is rounded to an integral solution (Algorithm \ref{alg: round-2}). We proceed to prove that these two algorithms together satisfy the desired conditions.

    \begin{algorithm}[!t]
        \caption{\textsc{Round}$(\ov{x}, \ov{y})$}\label{alg: round-2}
        \nonl \textbf{input}: fractional solution $(\ov{x}, \ov{y})$ output by Algorithm \ref{alg: round-1} \\
        \nonl \textbf{output}: integral solution $(x^*, y^*)$ \\
        form a bipartite graph $G$ with two copies of $X$, denoted $X_1, X_2$ \\
        for all $i \in X_1, j \in X_2$, $ij \in E(G)$ iff $\ov{x}_{ij} > 0$ \\
        for $j \in X_2$, denote by $N_G(j) = \{i \in X_1: \ov{x}_{ij} > 0\} \subseteq X_1$ the neighborhood of $j$ in $G$ and by $N^2_G(j) \subseteq X_2$ the set of all $j' \in X_2$ that share a neighbor with $j$ in $G$ \\
        set iteration counter $k = 0$ and $X' = X_2$ \\
        \While{$X' \neq \emptyset$ \label{step: rounding-loop}}
        {
            choose $j_k = {\arg\min}_{j \in X'} \max_{i: ij \in E(G)} d(i, j)$ \label{step: choose-rounding-client} \\
        choose $i_k = {\arg\min}_{i \in N_G(j_k)} c_i$ \label{step: choose-rounding-facility} \\
        open $i_k$, i.e., set $y^*_{i_k} = 1$ \\
        assign $\Pi(j) = i_k$ for all $j \in N^2_G(j_k)$ i.e., $x^*_{i_k, j} = 1$ for all such $j$ \\
            $X' \gets X' \setminus (N^2_G(j_k) \cup \{j_k\})$ and update $G$ by deleting all vertices in $(N^2_G(j_k) \cup N_G(j_k) \cup \{j_k\})$; set all corresponding $x^*, y^*$ variables to $0$ \\
            $k \gets k + 1$ \\
        }
        \Return{$(x^*, y^*)$}
    \end{algorithm}

    Consider the solution $(\ov{x}, \ov{y})$ output by \textsc{Filter}. Since $\ov{x} \le \frac{1}{\alpha} x$ and $\ov{y} = \frac{1}{\alpha} y$, it satisfies feasibility constraints (\ref{constraint_2}) and (\ref{constraint_3}). Further, by definition of $d_j(\alpha)$, we get $\sum_{i \in X} \ov{x}_{ij} = \sum_{i: d(i, j) \le d_j(\alpha)} \frac{x_{ij}}{\alpha}  \ge \frac{1}{\alpha} \cdot \alpha = 1$. Therefore, $\ov{x}, \ov{y}$ is feasible for (\ref{eqn: convex-relaxation}). Next, for each client $j \in X$,
    $$
    \sum_{i \in X} d(i, j) x_{ij} \ge \sum_{\substack{i: d(i, j) \ge d_j(\alpha)}} d(i, j) x_{ij} \ge d_j(\alpha) \sum_{\substack{i: d(i, j) \ge d_j(\alpha)}} x_{ij} \ge d_j(\alpha)(1 - \alpha),
    $$ so we get
    \begin{align}\label{eqn: d_j_alpha_bound}
    d_j(\alpha) \le \frac{1}{1 - \alpha} \sum_{i \in X} d(i, j) x_{ij} \quad \forall \; j \in X.
    \end{align}

    Now consider the solution $(x^*, y^*)$ output by Algorithm \ref{alg: round-2}. For facility $i_k$ and client $j_k$ chosen in iteration $k$ of loop \ref{step: rounding-loop}, by step \ref{step: choose-rounding-facility} and since $\sum_{i \in X}\ov{x}_{ij_k} \ge 1$,
    $$
    c_{i_k} \le c_{i_k} \sum_{i \in X} \ov{x}_{i j_k} \le \sum_{i \in N_G(j_k)} c_i \ov{x}_{i j_k} \le \sum_{i \in N_G(j_k)} c_i \ov{y}_i.
    $$

    Furthermore, since each client in set $N_G^2(j_k)$ is assigned to $i_k$ and then removed from $G$, the sets $N_G(j_1), N_G(j_2), \ldots$ are disjoint. Along with the above, this implies
    $$
    \sum_{k} c_{i_k} \le \sum_k \sum_{i \in N_G(j_k)} c_i \ov{y}_i \le \sum_{i \in X} c_i \ov{y}_i = \frac{1}{\alpha} \sum_{i \in X} c_i y_i.
    $$
    That is, the facility cost of integral soltution is $\sum_{i \in X} c_i y_i^* = \sum_{k} c_{i_k} \le \frac{1}{\alpha} \sum_{i \in X} c_i y_i$.

    Suppose client $j$ was assigned to $i_k$. Then $j \in N_G^2(j_k)$ and by step \ref{step: choose-rounding-client}, $d_j(\alpha) \ge d_{j_k}(\alpha)$. Since $j \in N_G^2(j_k)$, there exists some $i \in N_G(j) \cap N_G(j_k)$. By metric property of $d$, we have $d(i_k, j) \le d(i_k, j_k) + d(i, j_k) + d(i, j)$.
    Further, by step \ref{step: rounding-ov-x-condition} and since $\ov{x}_{i j}, \ov{x}_{i j_k}, \ov{x}_{i_k j_k} > 0$, we must have $d(i, j) \le d_j(\alpha)$ and both $d(i, j_k), d(i_k, j_k) \le d_{j_k}(\alpha)$, so that $d(i_k, j) \le 2 d_{j_k}(\alpha) + d_j(\alpha) \le 3 d_j(\alpha)$. Eqn. (\ref{eqn: d_j_alpha_bound}) then implies that $d(i_k, j) \le 3 d_j(\alpha) \le \frac{3}{1 - \alpha} \sum_{i \in X} x_{ij} d(i, j)$,
    or that $\sum_{i \in X} x^*_{ij} d(i, j) \le \frac{3}{1 - \alpha} \sum_{i \in X} x_{ij} d(i, j)$.
    Choosing $\alpha = \frac{1}{4}$ implies $z^*_{j} \le 4 z_j$ and $\sum_i c_i y_i^* \le 4 \sum_i c_i y_i$.
    This completes the proof of Theorem \ref{thm: facility-location-solvability-theorem} for FFL.

    For FC$^{(k)}$, the relaxation has constraint $\sum_i y_i \le k$ instead of the term $\sum_i c_i y_i$ in the objective. The rest of the proof is similar and omitted.

    \section{Omitted Proofs from Section \ref{sec: refinements}}\label{app: missing-refinements-proofs}

    This section contains proofs omitted from Section \ref{sec: discounted-lookahead} for Algorithm \textsc{DiscountedLookahead}. The results are restated for convenience.
    We begin with a technical claim on assignments output by \textsc{DiscountedLookahead}.

    \begin{claim}\label{claim: metric-refinement-claim-1}
    Given $k, t \in [l + 1]$ with $t \le k$, we have for any $f \in F_t \subseteq F_k$ that $\Pi_{t}^{(k)}(f) = f,$ where $\Pi_t^{(k)}: F_k \to F_t$ is the assignment in Algorithm \ref{alg: metric-partition-refinement}.
    \end{claim}

    \begin{proof}
        We use induction on $k - t$. When $k - t = 0$, i.e., $t = k$, by definition, $\Pi_{t}^{(k)}(f) = \Pi_{k}^{(k)}(f) = f$. Suppose $k - t > 0$. Then $F_t \subseteq F_{k - 1} \subseteq F_{k}$. Therefore, $h_{k - 1}(f) = f$ and $d(f, h_{k - 1}(f)) = 0$. By step \ref{step: metric-partition-index-selection} and the tie-break condition, we must have $s^* = k - 1$. Therefore, for $t \in [1, k]$, by step \ref{step: metric-partition-assignment},
        $$
        \Pi_t^{(k)}(f) = \begin{cases}
                             h_{k - 1}(f) = f & \text{if} \: t = k - 1, \\
                             \Pi_t^{(k - 1)}(h_{k - 1}(f)) = \Pi_t^{(k - 1)}(f) & \text{if} \: t \in [k - 2].
        \end{cases}
        $$
        By the induction hypothesis, $\Pi_t^{(k - 1)}(f) = f$.
    \end{proof}

    Next, we prove Claim \ref{claim: metric-refinement-claim-2}, or the convolution property for assignments:

    \ConvolutionProperty*

    \begin{proof}
        We induct on $k$. When $k = 1$, $t = s = 1$. By definition, $\Pi_{1}^{(1)}(f) = f$ and the claim holds.

        Suppose $k > 1$. If $s = k$ then $\Pi_{t}^{(s)}\left(\Pi_{s}^{(k)}\left( f\right)\right) = \Pi_{t}^{(s)}\left(f\right) = \Pi_{t}^{(k)}\left(f\right)$. Therefore, assume $t \le s < k$.

        As in step \ref{step: metric-partition-index-selection} of the algorithm, let $s^* = {\arg\min}_{t \in [k - 1]} \gamma^t \cdot d(f, h_t(f))$. Also denote $h = h_{s^*}(f)$. There are three cases:

        \textbf{Case I}: $t, s \ge s^*$. Since $t \le s \le k - 1$, by step \ref{step: metric-partition-assignment}, we have
        $$
        \Pi^{(k)}_t(f) = \Pi^{(k)}_s(f) = h.
        $$
        Further, $h \in F_{s^*} \subseteq F_t \subseteq F_s$. By Claim \ref{claim: metric-refinement-claim-1}, $\Pi_t^{(s)}(h) = h$. Therefore,
        $$
        \Pi^{(k)}_t(f) = h = \Pi_t^{(s)}(h) = \Pi_t^{(s)}\left(\Pi^{(k)}_s(f)\right).
        $$

        \textbf{Case II}: $t < s^* \le s$. As above, by step \ref{step: metric-partition-assignment}, we have $\Pi^{(k)}_s(f) = h$.

        Since $h \in F_{s^*} \subseteq F_s$, by Claim \ref{claim: metric-refinement-claim-1}, $\Pi^{(s)}_{s^*}(h) = h$. Since $s < k$, by the induction hypothesis,
        $$
        \Pi_t^{(s)}(h) = \Pi_t^{(s^*)}\left(\Pi_{s^*}^{(s)}(h)\right) = \Pi_t^{(s^*)}\left(h\right).
        $$

        Finally, by step \ref{step: metric-partition-assignment} again, $\Pi_t^{(k)}(f) = \Pi_t^{(s^*)}(h)$. Putting these together,
        $$
        \Pi_t^{(k)}(f) = \Pi_t^{(s^*)}(h) = \Pi_t^{(s)}(h) = \Pi_t^{(s)} \left(\Pi^{(k)}_s(f)\right).
        $$

        \textbf{Case III}: $t, s < s^*$. By step \ref{step: metric-partition-assignment}, $\Pi_t^{(k)}(f) = \Pi_t^{(s^*)}(h)$. and $\Pi_s^{(k)}(f) = \Pi_s^{(s^*)}(h)$. Further, since $t \le s < s^* < k$, by the induction hypothesis on $h \in F_{s^*}$, we get
        $$
        \Pi_t^{(s^*)}(h) = \Pi_t^{(s)}\left(\Pi_s^{(s^*)}(h)\right)
        $$
        Therefore, $\Pi_t^{(k)} = \Pi_t^{(s^*)}(h) = \Pi_t^{(s)}\left(\Pi_s^{(s^*)}(h)\right) = \Pi_t^{(s)}\left(\Pi_s^{(k)}(f)\right)$. This completes the proof.
    \end{proof}

    Finally, we prove Lemma \ref{lem: general-metric-recurrence-relation} used to upper bound the algorithm approximation ratio:

    \RecurrenceRelation*

    \begin{proof}
        Recall that for $t < k$, $u_{k, t} = \max\{\gamma^{t - 1}, \max_{s \in [t, k - 1]} \left(\gamma^{t - s} + u_{s, t} (1 + \gamma^{t - s})\right)\}$. Therefore, we get $u_{k, t} = \max\{\gamma^{t - 1}, \gamma^{t - k + 1} + u_{k - 1, t} \left(1 + \gamma^{t -k +1}\right), \max_{s \in [t, k - 2]} \left(\gamma^{t - s} + u_{s, t} (1 + \gamma^{t - s})\right)\}$. But this latter term equals $\max\{u_{k - 1, t}, \gamma^{t - k + 1} + u_{k - 1, t} \left(1 + \gamma^{t -k +1}\right)\}$, which is just $\gamma^{t - k + 1} + u_{k - 1, t} \left(1 + \gamma^{t -k +1}\right)$. Therefore $u_{k, t} = \gamma^{t - k + 1} + u_{k - 1, t} \left(1 + \gamma^{t -k +1}\right)$. Since $\gamma > 1$, $u_{k -1 , t} \ge 1$, and so
        $$
        \frac{u_{k, t}}{u_{k - 1, t}} \le \frac{\gamma^{t - k + 1}}{u_{k - 1, t}} + (1 + \gamma^{t + 1 - k}) \le 1 + 2 \gamma^{t - k + 1}.
        $$
        Taking an alternating product,
        $$
        \frac{u_{k, t}}{\gamma^{t - 1}} = \frac{u_{k, t}}{u_{t, t}} \le \prod_{s \in [t + 1, k]} \left(1 + 2\gamma^{t - s + 1}\right).
        $$
        Using $1 + x \le \exp(x)$, we get
        $$
        \prod_{s \in [t + 1, k]} \left(1 + 2\gamma^{t - s + 1}\right) \le  \exp\left(\sum_{s \in [t + 1, k]} 2\gamma^{t -  s + 1}\right) \le \exp\left(\sum_{s \ge t + 1}  2 \gamma^{t - s + 1}\right) = \exp\left(\frac{2\gamma}{\gamma - 1}\right).
        $$
        Therefore, $u_{k, t} \le \gamma^{t - 1} \prod_{s \in [t + 1, k]} \left(1 + 2\gamma^{t - s + 1}\right) \le \gamma^{l} \exp\left(\frac{2\gamma}{\gamma - 1}\right)$.

        We prove the second part of the theorem by choosing $\gamma = 1 + \frac{1}{\sqrt{l}}$. Then $e^\frac{2\gamma}{\gamma - 1} = e^{2\sqrt{l} \left(1 + \frac{1}{\sqrt{l}}\right)} = e^{2 \sqrt{l}} \cdot e^2$. Also, $\gamma^l = \left(1 + \frac{1}{\sqrt{l}}\right)^l \le \left( e^{\frac{1}{\sqrt{l}}} \right)^l = e^{\sqrt{l}}$, implying that $e^\frac{2\gamma}{\gamma - 1} \gamma^{l} \le e^2 \cdot e^{3 \sqrt{l}}$.
    \end{proof}

    \section{Omitted Proofs from Section \ref{sec: line-metric}}\label{sec: line-missing-proofs}

    We fill in the proofs omitted from Section \ref{sec: line-metric}. We prove Claim \ref{claim: line-interval-tree} that characterizes refinements as hierarchy trees, and then we proceed to prove various lemmas for Algorithm \textsc{ExpandIntervals}. The results are restated for convenience.

    \LineHierarchyTreesAreRefinements*

    \begin{proof}
        We begin by showing that the assignments are well-defined. Fix $k \in [0, l]$. By the immediate parent property, $(f, k)$ and $(f', k')$ are siblings iff $k = k'$. Then, since $A(f, k) \cap A(f', k) = \emptyset$ for all siblings, by the completeness property, intervals $\{A(f, k): f \in F_k\}$ partition $\R$. Therefore, each $\Pi_k(j)$ is uniquely defined.

        \begin{enumerate}
            \item Fix $k \in [l]$ and facility $f \in F_k$. Since $H$ is an interval tree, there is a unique edge $((f, k), (h, k')) \in E$. and $A(f, k) \subseteq A(h, k')$. By the immediate parent property, $k' = k - 1$. By definition of $\Pi_k, \Pi_{k - 1}$, we have that $\{j \in X: \Pi_{k}(j) = f\} = A(f, k) \subseteq A(h, k - 1) = \{j \in X: \Pi_{k - 1}(j) = h\}$.

            \item Fix $k \in [0, l]$. Let $f, f'$ be two consecutive facilities in $F_k$ with $x(f) < x(f')$. Consider the interval $[x(f), x(f')]$. Since $A(f, k) \supseteq N_\alpha(f, k)$ and $A(f', k) \supseteq N_\alpha(f', k)$, assignment $\Pi_k$ assigns each point in $[x(f), x(f) + \frac{x(f') - x(f)}{1 + \alpha}] = N_\alpha(f, k) \cap [x(f), x(f')]$ to $f$. Since $\Pi_k(j) = f$ for all such points $j$,
            $$
            d\left(j, \Pi_k(j)\right) \le \alpha \cdot \min_{h \in F_k} d(j, h).
            $$
            Similar result holds for for points in $[x(f') - \frac{x(f') - x(f)}{1 + \alpha}, x(f')]$. The cost upper bound for the rest of the points is satisfied irrespective of whether they are assigned to $f$ or $f'$ since their distances to $f$ and $f'$ are within factor $\alpha$ of each other.
        \end{enumerate}
    \end{proof}

    We show that $H$ is an interval tree at the end of step 1 (Lemma \ref{lem: disjoint-interval-trees-A}). We need the following structural lemma first:

    \begin{lemma}\label{lem: distance-between-A-intervals}
    For all levels $k \in [l]$ and facilities $f \in F_k$, let $x(f'')$ and $x(f')$ be the location of facilities in $F_k$ to the immediate left and right of facility $f$ respectively (if $f$ is the leftmost facility, denote $x(f'') = -\infty$ and $f$ is the rightmost facility, denote $x(f') = \infty$). Then, at the end of step 1,
    $$
    A(f, k) \subseteq \left[x(f) - \frac{(l - k + 1)(x(f) - x(f''))}{\alpha + 1}, x(f) + \frac{(l - k + 1)(x(f') - x(f))}{\alpha + 1}\right].
    $$
    \end{lemma}

    \begin{proof} We use induction on $l - k$. When $k = l$, $A(f, k) = N_\alpha(f, k)$ since $A(f, k)$ is initialized to $N_\alpha(f, k) = \left[x(f) - \frac{(x(f) - x(f''))}{\alpha + 1}, x(f) + \frac{(x(f') - x(f))}{\alpha + 1}\right]$ and never updated in step 1.

    Suppose $k < l$. Note that $A(f, k)$ is only updated in iteration $k$ of the for loop in step 1. Let facility $h \in F_s$ at level $s > k$ be such that $N_\alpha(f, k) \cap A(h, s) \neq \emptyset$. It is sufficient to show that $N_\alpha(f, k) \cup A(h, s) \subseteq \left[x(f) - \frac{(l - k + 1)(x(f) - x(f''))}{\alpha + 1}, x(f) + \frac{(l - k + 1)(x(f') - x(f))}{\alpha + 1}\right]$ for all such $(h, s)$.

    Suppose $h'', h'$ are the facilities in $F_s$ to the immediate left and right of $h$, and suppose that $h$ itself is to the right of $f$ (the proof is similar when $h$ is to the left of $f$); then since $s > k$, $F_s \supseteq F_k$, and so $x(h'') \ge x(f)$ and $x(h') \le x(f')$. By the induction hypothesis, the length of $A(h, s)$ is at most
    $$
    \frac{(l - s + 1)(x(h') - x(h''))}{\alpha + 1} \le \frac{(l - k)(x(f') - x(f))}{\alpha + 1}.
    $$
    Then, since $N_\alpha(f, k) \cap A(h, s) \neq \emptyset$, the right endpoint of $N_\alpha(f, k) \cup A(h, s)$ is at most
    $$
    \left(x(f) + \frac{x(f') - x(f)}{\alpha + 1} \right) + (l - k) \frac{x(f') - x(f)}{\alpha + 1} = x(f) + (l - k + 1)\frac{x(f') - x(f)}{\alpha + 1}.
    $$
    Similarly, the left endpoint of $N_\alpha(f, k) \cup A(h, s)$ is at least $\frac{(l - k + 1)(x(f) - x(f''))}{\alpha + 1}$.
    \end{proof}

    We proceed to prove Lemma \ref{lem: disjoint-interval-trees-A}.

    \LineStepOneIntervalTree*

    \begin{proof}
        For each $k \in [0, l]$, let $H_k$ denote the subgraph of $H$ induced by the vertex set $\bigcup_{k' \ge k} \ov{F}_{k'}$, i.e., facilities at level $k$ or higher. We induct on $l - k$ to prove the following (stronger) statement: each component of $H_k$ is an interval tree with respect to intervals $A$, and furthermore, for two components $C_1, C_2$ of $H_k$ with roots $r_1, r_2$, $A(r_1)$ and $A(r_2)$ are internally disjoint.

        For $k = l$, each component of $H_l$ is an isolated vertex and further, $A(f, l) = N_\alpha(f, l)$ and $A_\alpha(g, l) = N_\alpha(g, l)$ are disjoint intervals for $f \neq g$, implying the claim.

        Assume now that $k < l$ and that the claim is true for all $k' > k$. Let $C$ be a component of $H_k$ rooted at some facility $(f, k')$, $k' \ge k$. We first prove that $C$ is an interval tree. If $k' > k$, then $C$ is also a component of $H_{k + 1}$ and therefore $C$ is an interval tree by the induction hypothesis. Suppose $k' = k$. The only parent-child pairs present in $H_k \cap C$ but not in $H_{k + 1} \cap C$ are $(f, k)$ and one of its children. The only sibling pairs present in $H_k \cap C$ but not in $H_{k + 1} \cap C$ are two children of $(f, k)$.

        Let $(g_1, k_1)$ be some child of $(f, k)$ in $C$. By definition of $H$, interval $A(g_1, k_1) \subseteq A(f, k)$ by line \ref{step: A_update} in the algorithm. Let $(g_2, k_2)$ be some other child of $(f, k)$ in $C$. Since $(f, k) \not\in V(H_{k + 1})$, both $(g_1, k_1), (g_2, k_2)$ are the roots of their corresponding components in $H_{k - 1}$. By the induction hypothesis, $A(g_1, k_1) \cap A(g_2, k_2) = \emptyset$. This proves that $C$ is an interval tree.

        Let $C_1, C_2$ be two components in $H_k$, rooted at $(f_1, k_1), (f_2, k_2)$ respectively. we claim that $A(f_1, k_1)$ and $A(f_2, k_2)$ are disjoint. If both $k_1, k_2 > k$, then $A(f_1, k_1)$ and $A(f_2, k_2)$ are internally disjoint by the induction hypothesis. If $k_1 = k_2 = k$, then Lemma \ref{lem: distance-between-A-intervals} implies that $A(f_1, k_1)$ and $A(f_2, k_2)$ are internally disjoint. Suppose $k_1 = k$ and $k_2 > k$. Since $(f_2, k_2)$ is a root of a component in $H_{k}$, $(f_2, k_2)$ must have been unassigned at the end of iteration $k$ in loop \ref{step: A_assignment_loop} in the algorithm. Therefore, since $((f_2, k_2), (f_1, k_1)) \not\in H$,  we must have $A(f_1, k_1) \cap A(f_2, k_2) = \emptyset$ in line \ref{step: A_update}. This proves the inductive statement.

        Finally, since $A(f_0, 0) = N_\alpha(f_0, 0) = \R$, by line \ref{step: A_update}, $H = H_{0}$ is a tree with root $(f_0, 0)$, and therefore an interval tree by our claim.
    \end{proof}

    \LineStepTwo*

    \begin{proof}
        We first claim that $H$ stays an interval tree at all times in step 2. By Lemma \ref{lem: disjoint-interval-trees-A}, each $A(f, k)$ is an interval at the start of step 2. The only updates to intervals $A$ occur on line \ref{step: combined-convex-hull}. Since convex hulls on $\R$ are intervals, sets $A$ are still intervals after step 2. Denote by $\overline{F}_k = \{(f, k): f \in F_k\}$ for any level $k \in [0, l]$.

        Fix an iteration of the loops in step 2, and let $(f, k), (h, k')$ and $(g, k + 1)$ be the corresponding facilities in this iteration. Before the update in line \ref{step: combined-convex-hull}, $A(g, k + 1), A(h, k') \subseteq A(f, k)$, and therefore $\mathrm{conv}(A(g, k + 1), A(h, k')) \subseteq A(f, k)$. Therefore, after the update, $A(g, k + 1)$ is still a subset of $A(f, k)$. Further, by construction, $A(h, k')$ is a subset of $A(g, k + 1)$ after the update, so that $H$ is an interval tree.

        We now show that interval tree condition is also satisfied. Since $H$ is an interval tree before the update, the intersection of $A(g, k + 1)$ and $A(h, k')$ is empty before the update. Therefore, after the update, the subtree of $H$ rooted at $(g, k + 1)$ is an interval tree. Further, for any other child $(g', k + 1)$ of $(f, k)$, $A(g', k + 1), A(g, k + 1)$ and $A(h, k')$ are disjoint sets before the update. Since $(g, k + 1)$ is the child of $(f, k)$ closest to $(h, k')$, this means that $A(g', k + 1)$ and $\mathrm{conv}(A(g, k + 1) \cup A(h, k'))$ are disjoint before the update, or that $A(g', k + 1)$ and $A(g, k + 1)$ are disjoint after the update. Since the only children of $(f, k)$ are of the form $(g', k + 1)$ after the update, this implies that $H$ is still an interval tree after the update.

        Finally, we show that the immediate parent condition is satisfied by $H$. We show the following stronger statement by induction on $k$: after the $k$th iteration of the outer loop in step 2 (on line \ref{step: step-2-outer-loop}), the subtree of $H$ induced by $\ov{F}_{k + 1} \cup \ov{F}_{k} \cup \ldots \cup \ov{F}_{0}$ satisfies the immediate parent condition. In iteration $k$, all edges of the form $\left((f, k), (h, k')\right)$, $k' > k + 1$ are removed from $H$, and so the only children of $(f, k)$ remaining are of the form $(g, k + 1)$. Since so edges containing any vertex in $\ov{F}_{k - 1} \cup \ldots \cup \ov{F}_{0}$ are modified in iteration $k$, this implies the claim for level $k$ using the induction hypothesis.
    \end{proof}

    \section{Algorithm \textsc{BranchAndLinearize} for Tree Metric}\label{app: tree-metric}

    In this section, we discuss the $\alpha$-reassignment problem (Definition \ref{def: reassignment-problem}) on tree metrics, i.e., the client set $X$ is the vertex set of a tree $T = (X, E, w)$ with edges $E$ and nonnegative edge lengths $w: E \to \R$; distance $d(j, j')$ between clients $j, j' \in X$ is the sum of edge weights on the unique path from $j$ to $j'$. We prove Theorem \ref{thm: reassignment-problem} for tree metrics by giving a polynomial-time algorithm (called \textsc{BranchAndLinearize}) that given facility sets $F_1 \subseteq \ldots \subseteq F_l \subseteq X$, obtains $O(l)$-reassignments $\Pi_t: X \to F'_t$ for $t \in [l]$ for modified facility sets $F_1' \subseteq \ldots \subseteq F_l' \subseteq X$ such that $F_t \subseteq F_t'$ and $|F_t'| \le 2 |F_t|$. That is, the algorithm obtains $O(l)$-reassignments on tree metrics with facility sets at most double the size of the input facility sets.

    First, we introduce some notation to describe the algorithm, followed by its description and analysis. Given any tree $T'$, call a vertex $v \in V(T')$ a \emph{branch vertex} if $\deg_{T'}(v) \ge 3$. Given a set of vertices $S$ of $T'$, let $T'_S$ denote the subtree of $T'$ induced by all vertices in $S$ and those in the paths connecting any two vertices in $S$. That is, $T'_s$ is the minimal induced subtree of $T'$ that contains all vertices in $S$. Two vertices $u, v \in S$ are called \emph{consecutive vertices of $S$} if the unique path in $T'$ (and in $T'_S$) joining $u, v$ does not contain any other vertex of $S$. If $S$ contains all the branch vertices of $T'_S$, then the edges of $T'_S$ can be uniquely partitioned into paths between pairs of consecutive vertices of $S$. We will also use the following observation that follows from the handshaking lemma:

    \begin{fact}\label{fact}
    The number of branch vertices is at most the number of leaves in any tree.
    \end{fact}

    Given tree $T$ on client set $X$, \textsc{branchAndLinearize} (Algorithm \ref{alg: tree-partition-refinement}) first adds all branch vertices of $T_{F_s}$ to $F_s$ for each level $s \in [l]$ to get facility sets $F_{s}'$. By Fact \ref{fact}, $|F_{s}'| \le 2 |F_s|$. Additionally, since $F_s \subseteq F_{s + 1}$, every branch vertex of $T_{F_s}$ is a branch vertex of $T_{F_{s + 1}}$, so that

    \begin{lemma}\label{lem: tree-chain-after-adding-branch-vertices}
    Facility sets $F'_1, \ldots, F'_l$ in Algorithm \ref{alg: tree-partition-refinement} satisfy (1) $F_1' \subseteq \ldots \subseteq F'_l$, (2) $F_t \subseteq F'_t$ and (3) $|F_t'| \le 2 |F_t|$ for each level $t \in [l]$.
    \end{lemma}

    Next, consider the subtree $T_{F_1'}$ of $T$. By our observation above, $T_{F_1'}$ can be decomposed into edge-disjoint paths between facilities in $F_1'$. All such paths $p$ are isomorphic to the line metric, and the algorithm calls subroutine \textsc{ExapndIntervals} each such $p$. This  gives a $2l$-reassignment on $T_{F_1'}$, and it remains to give assignments on $T \setminus T_{F_1'}$. This is done in the second loop (step \ref{step: tree-paths-loop}) of the algorithm.

    Each component $C$ of $T \setminus T_{F_1'}$ is a tree, and connected to $T_{F_1'}$ by a unique vertex $f \in T_{F_1'}$. Let $t$ be the highest level such that $V(C) \cap F_t' = \emptyset$. The main idea of the algorithm is the following: since $F_1' \subseteq \ldots \subseteq F_t'$, for each level $s \in [t]$, $V(C) \cap F_{s'} = \emptyset$. To assign $\Pi_s(j)$ for vertices $j \in V(C)$ we note that $f \in T_{F_{1}'}$, and therefore $\Pi_s(f)$ has been assigned in the second for loop (step \ref{step: tree-paths-loop}). The natural choice is to assign $\Pi_s(j) = \Pi_s(f)$ for each $j \in V(C)$, $s \in [t]$. For $s \in [t + 1, l]$, we recurse on $V(C) \cup \{f\}$. This is done in the third loop (step \ref{step: tree-components-loop}).

    \begin{algorithm}[t]
        \caption{\textsc{BranchAndLinearize}($T, F_1, \ldots, F_l$)}\label{alg: tree-partition-refinement}
        \nonl \textbf{input}: tree $T = (X, E, w)$, facility sets $F_1 \subseteq \ldots \subseteq F_l \subseteq X$ \\
        \nonl \textbf{output}: assignments $\Pi_s: V(T) \to F_s$ for each $s \in [l]$ \\
        initialize empty mappings $\Pi_1, \ldots, \Pi_l = \emptyset$ \\
        \For{$s \in [l]$}
        {
            let $B_s$ be the set of all branch vertices of $T_{F_s}$ \\
        set $F_s' = F_s \cup B_s$
            \label{step: tree-add-branch-vertices}
        }
        \For{each path $p$ between two consecutive vertices of $F_1'$ in $T_{F'_1}$
            \label{step: tree-paths-loop}}
        {
            \label{step: tree-assignments-on-path} treat $p$ as a line with lengths determined by edge weights $w$, and obtain assignments
            \[
                \Pi'_1, \ldots, \Pi'_l \gets \mathrm{\textsc{ExpandIntervals}}(F'_1, \ldots, F'_l),
            \]
            where assignments $\Pi'_s: V(p) \to V(p) \cap F'_{s}$, $s \in [l]$ \\
        update assignments $\Pi_s \gets \Pi_s \cup \Pi'_s$ for $s \in [l]$ \label{step: assign-vertices-on-paths-between-facilities} \\
        }
        \For{each component $C$ of $T \setminus T_{F'_1}$
            \label{step: tree-components-loop}}
        {
            let $f$ be the vertex that connects $C$ to $T_{F'_1}$ \\
        let $t \in [l]$ be the maximum level such that $V(C) \cap F'_t = \emptyset$ \label{step: tree-choose-threshold} \\
        \For{$s \in [t]$}
        {
            assign $\Pi_{s}(j) = \Pi_s(f)$ for all $j \in V(C)$ \label{step: tree-assign-all-in-component-to-same-facility} \\
        }
            $\Pi'_{t + 1}, \ldots, \Pi'_{l} = \textsc{BranchAndLinearize}(C \cup \{f\}, F'_{t + 1}, \ldots, F'_{l})$ \label{step: tree-recursive-assignments} \\
        update $\Pi_s \gets \Pi_s \cup \Pi'_s$ for all $s \in [t + 1, l]$ \\
        }
        \textbf{return} assignments $\Pi_1, \ldots, \Pi_l$ \\
    \end{algorithm}

    We prove in Lemma \ref{lem: tree-metric-assignment-refinement} that the output assignments $\Pi_s, s \in [l]$ satisfy the assignment condition, and in Lemma \ref{lem: tree-metric-cost-bounds} that they satisfy the cost upper bound. First, we prove that they are well-defined:

    \begin{lemma}\label{lem: tree-metric-well-defined-assignments}
    Algorithm \ref{alg: tree-partition-refinement} outputs well-defined assignments $\Pi_1, \ldots, \Pi_l$, i.e., $\Pi_s(j)$ is uniquely defined for each client $j \in X$ and level $s \in [l]$.
    \end{lemma}

    \begin{proof}
        Each vertex is covered by each $\Pi_s, s \in [l]$ at least once since each $j \in T_{F'_1}$ is covered by some path $p$ in the loop at line \ref{step: tree-paths-loop}, and each $j \in T \setminus T_{F'_1}$ is covered in the loop at line \ref{step: tree-components-loop}. We show by induction that when some assignment $\Pi_s(j)$ is defined multiple times for some $j \in X$, then $j \in F_s'$ and the algorithm always assigns $\Pi_s(j) = j$. There are two cases in which $\Pi_s(j)$ is defined multiple times:

        \textbf{Case 1}: $j$ is the endpoint of multiple paths connecting consecutive vertices of $F_1'$; $\Pi_s(j)$ is then defined multiple times in the first loop (step \ref{step: tree-paths-loop}). But each endpoint of such paths is in $F_1'$, so $j \in F_s'$ since $F_1' \subseteq F'_s$.

        \textbf{Case 2}: Some component $C$ of $T \setminus T_{F_1'}$ is connected to $T_{F_1'}$ at $j$. In this case, $\Pi_s(j)$ is first defined for some path(s) in the loop at step \ref{step: tree-paths-loop}. Note that $\Pi_s(j) = j$ in this case since \textsc{ExpandIntervals} always assigns an open facility at a level to itself. It is defined potentially again in step \ref{step: tree-recursive-assignments} if $s > t$ for $t$ chosen in step \ref{step: tree-choose-threshold}. But $t$ is chosen such that $F_{s}' \cap C \neq \emptyset$. If $j \not\in F_1'$, then $\deg_{T_{F_1'}}(j) = 2$ and so $j$ is a branch vertex in $T_{F_s'}$, implying that $j \in F_s'$ and so $\Pi_s(j) = j$, consistent with the previous assignment. If $j \in F_1'$, then $j \in F_s'$ anyway since $F_1' \subseteq F_s'$. This completes the proof.
    \end{proof}

    \begin{lemma}\label{lem: tree-metric-assignment-refinement}
    Assignments $\Pi_s: X \to F'_s$, $s \in [l]$ output by Algorithm \ref{alg: tree-partition-refinement} satisfy the assignment condition and therefore form a refinement.
    \end{lemma}

    \begin{proof}
        Any vertex $v$ in a tree induces two subtrees $T_1, T_2$ on either side that only intersect at $v$ and satisfy $T_1 \cup T_2 = T$. It is sufficient to prove that the assignments $\Pi_1, \ldots, \Pi_l$ satisfy the assignment condition when restricted to each of the two subtrees induced by $v$. By generalizing this argument, it is sufficient to show that these assignments satisfy the assignment condition on each of the paths that decompose $F'_{1}$, and for each component of $T \setminus T_{F'_1}$.

        We prove this using induction on $l$, the number of facility sets. For $l = 1$, the claim is trivially true. Suppose $l > 1$, and that the result is true for all $k \le l - 1$ on all trees.

        If $p$ is one of the paths that decompose $T_{F_1'}$, the assignments on $p$ are defined in line \ref{step: tree-assignments-on-path} in the algorithm, using \textsc{ExpandIntervals}. By the corresponding guarantee on the line metric (Theorem \ref{thm: reassignment-problem}), $\Pi_1, \ldots, \Pi_l$ form an $O(l)$-reassignment when restricted to $p$.

        Fix a component $C$ of $T \setminus T_{F_1'}$, connected to $T_{F_1'}$ at $f$. Let $t$ be the level chosen in line \ref{step: tree-choose-threshold}. Note that $t$ exists because $V(C) \cap F_1' = \emptyset$ since $C$ is a subtree of $T \setminus T_{F_1'}$. Assignments $\Pi_1, \ldots, \Pi_t$ are defined on $C$ in line \ref{step: tree-assign-all-in-component-to-same-facility}, and assignments $\Pi_{t + 1}, \ldots, \Pi_{l}$ are defined on $C \cup \{f\}$ recursively in line \ref{step: tree-recursive-assignments}.

        By the induction hypothesis, $\Pi_{t + 1}, \ldots, \Pi_{l}$ satisfy the assignment condition on $C \cup \{f\}$. Further, $\Pi_s(j) = \Pi_s(f)$ for all $s \in [t]$, and therefore assignments $\Pi_1, \ldots, \Pi_t$ also satisfy the assignment condition. Therefore, it remains to show that for each facility $h \in F'_{t + 1}$, there is some facility $g \in F'_{t}$ such that all clients in $V(C)$ assigned to $h$ under $\Pi_{t + 1}$ are assigned to $g$ under $\Pi_{t}$. By step \ref{step: tree-assign-all-in-component-to-same-facility} $\Pi_{t}(j) = \Pi_{t}(f)$ for $j \in V(C) \cup \{f\}$ under $\Pi_{k + 1}$,the choice $g = \Pi_t(f)$ works for all $h$.
    \end{proof}

    \begin{lemma}\label{lem: tree-metric-cost-bounds}
    Assignments $\Pi_s: X \to F'_s$, $s \in [l]$ output by Algorithm \ref{alg: tree-partition-refinement} satisfy the cost guarantee, i.e., for all $j \in V(T)$ and each $s \in [l]$,
    $$
    d(j, \Pi_s(j)) \le 2l \cdot \min_{f \in F'_s} d(j, f) \le 2l \cdot \min_{f \in F_s} d(j, f).
    $$
    \end{lemma}

    \begin{proof}
        We use induction on $l$, the number of facility sets input. As noted above, we can decompose $T$ into $T_{F'_1}$ and components of $T \setminus T_{F_1'}$, and $T_{F'_1}$ can be further decomposed into edge-disjoint paths between pairs of consecutive vertices in $F_1'$.

        Suppose a client $j$ lies on one such path $p$. The endpoints of $p$ are both in $F'_1$, and therefore in $F'_s$ for all $s \in [l]$. Therefore, for each client $j \in V(p)$, the nearest facility at each of the levels $s \in [l]$ lies on this path. Since $\Pi_1(j), \ldots, \Pi_l(j)$ are assigned on line \ref{step: tree-assignments-on-path} using \textsc{ExpandIntervals}, the result then holds for $j$ by the corresponding claim for line metric.

        If $j$ is not on such a path, let $C$ be the component of $T \setminus T_{G_k}$ containing $j$, and let $C$ be connected to $T_{F_1'}$ at vertex $f$. Let $t$ be the level chosen in line \ref{step: tree-choose-threshold}, i.e., $t$ is the maximum level in $[l]$ such that $V(C) \cap F_{t}' = \emptyset$.

        If $s > t$, then $V(C) \cap F_s' \neq \emptyset$, and so $f$ is either in $F_1'$ or is a branch vertex in $F_s'$ and therefore itself in $F_s'$. Therefore, the two nearest facilities of $j$ in $F_s'$ both lie in $C \cup \{f\}$. In this case, the result holds by the induction hypothesis applied on the recursion in step \ref{step: tree-recursive-assignments}.

        If $s \in [t]$, then $V(C) \cap F_s' = \emptyset$, and therefore the facility in $F_s'$ closest to $j$ is the facility in $F_s'$ closest to $f$; call this facility $g$. Since $f$ lies on some path connecting two consecutive vertices in $F_1'$, we have by our earlier claim that $d\left(f, \Pi_{s}(f)\right) \le 2l \cdot d(f, g)$. Now, since $\Pi_{s}(j) = \Pi_s(f)$ by construction in line \ref{step: tree-assign-all-in-component-to-same-facility}, we get
        $$
        \dfrac{d(j, \Pi_s(j))}{d(j, g)} = \dfrac{d(j, \Pi_s(f))}{d(j, g)} = \dfrac{d(j, f) + d(f, \Pi_s(f))}{d(j, f) + d(f, g)} \le \max\Big( \dfrac{d(j, f)}{d(j, f)}, \dfrac{d(f, \Pi_s(f))}{d(f, g)} \Big) \le 2l.
        $$
    \end{proof}

    \section{Second Experiment: $k$-Clustering Refinement}\label{app: second-experiment}

    \begin{figure}[h]
        \begin{minipage}{0.53\textwidth}
            \centering
            \includegraphics[width=0.9\textwidth]{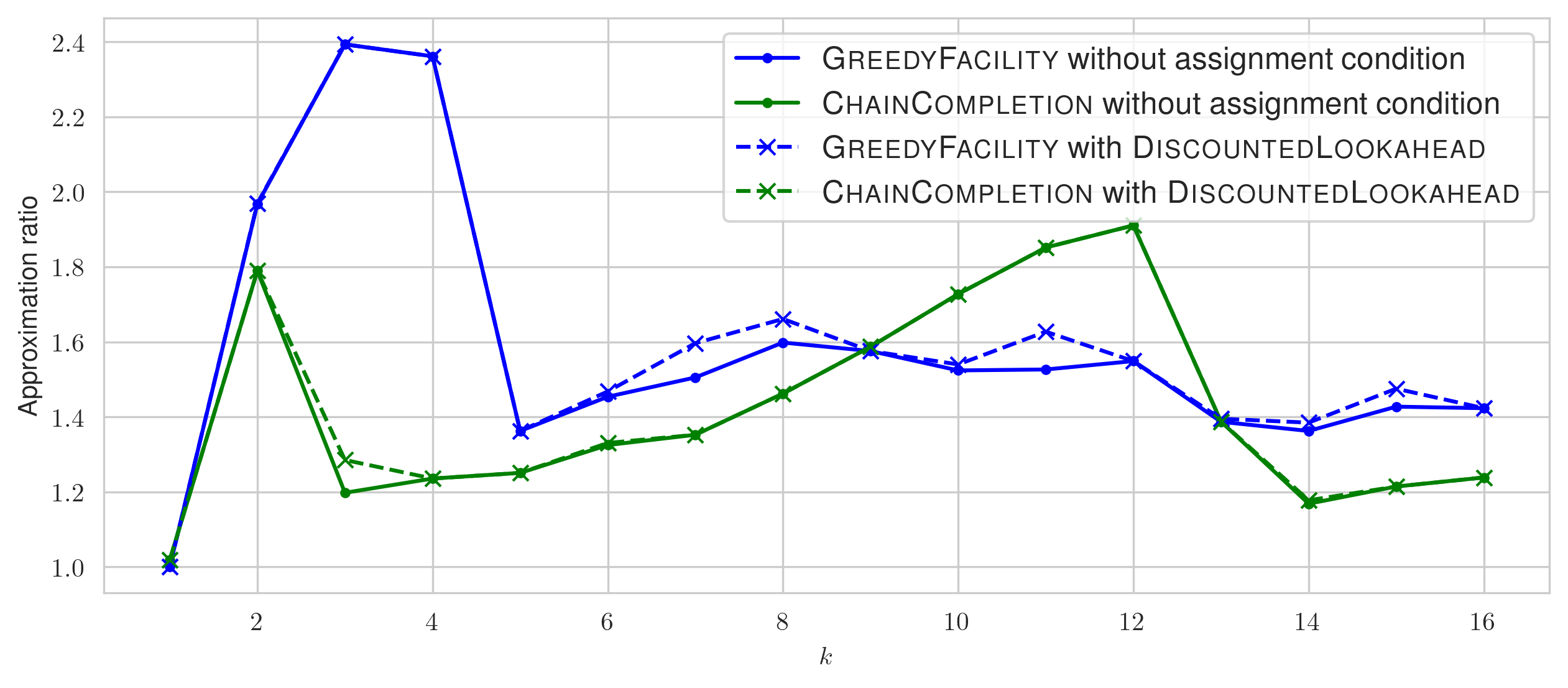}
            \caption{A comparison of approximation ratios for $L_\infty$ norm $k$-clustering refinement between \textsc{GreedyFacility} and our algorithm \textsc{ChainCompletion}, both without and with \textsc{DiscountedLookahead} for Fulton County, Georgia, USA.}
            \label{fig: approximation-comparison}
        \end{minipage}
        \hfill
        \begin{minipage}{0.43\textwidth}
            \centering
            \includegraphics[width=\textwidth]{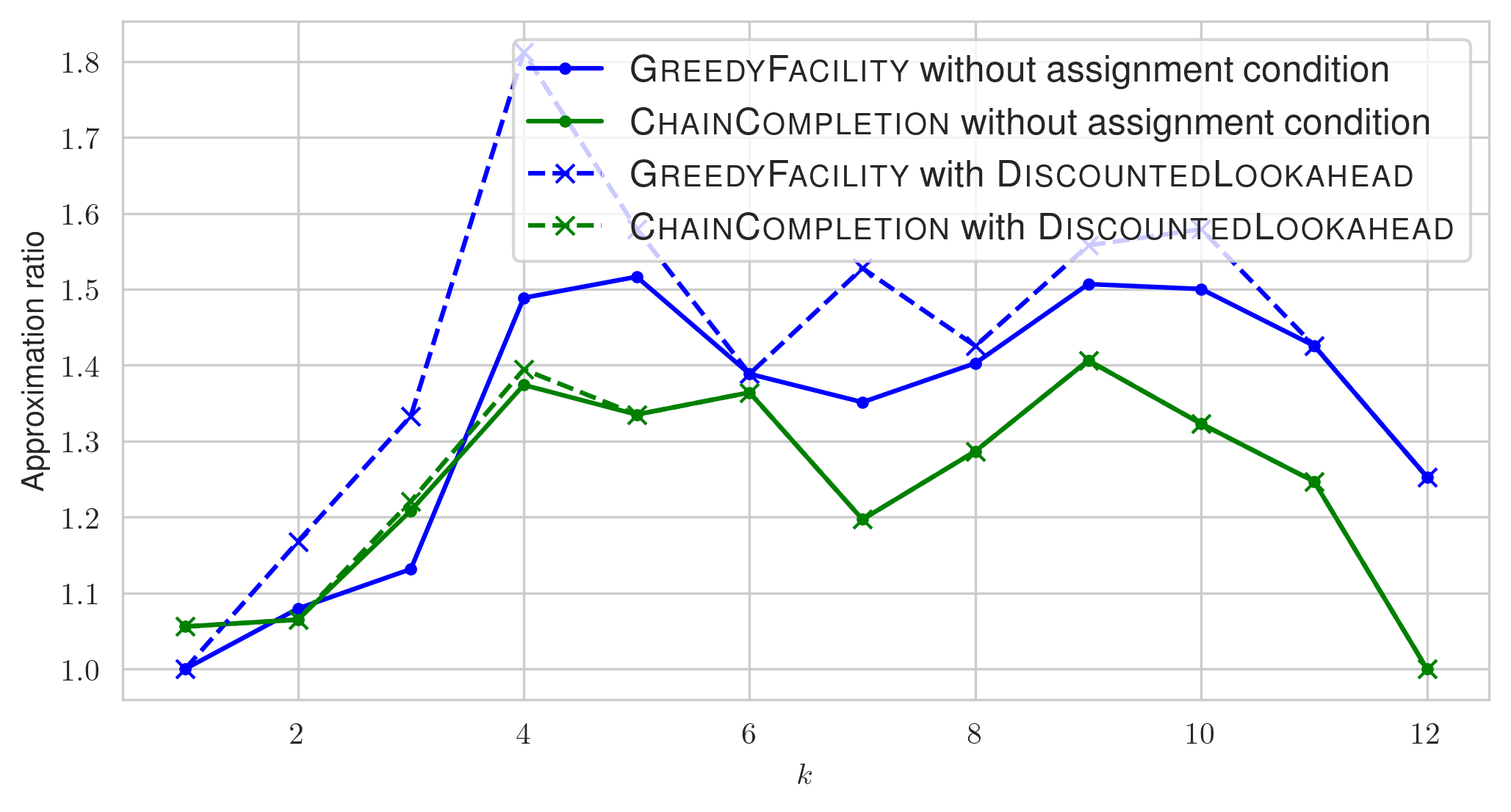}
            \caption{A comparison of the approximation ratios for Chatham County, Georgia, USA. Note that \textsc{ChainCompletion} consistently performs better than \textsc{GreedyFacility}.}
            \label{fig: approximation-comparison-chatham}
        \end{minipage}
    \end{figure}

    \begin{figure}[h]
        \begin{minipage}{0.4\textwidth}
            \centering
            \includegraphics[width=0.65\textwidth]{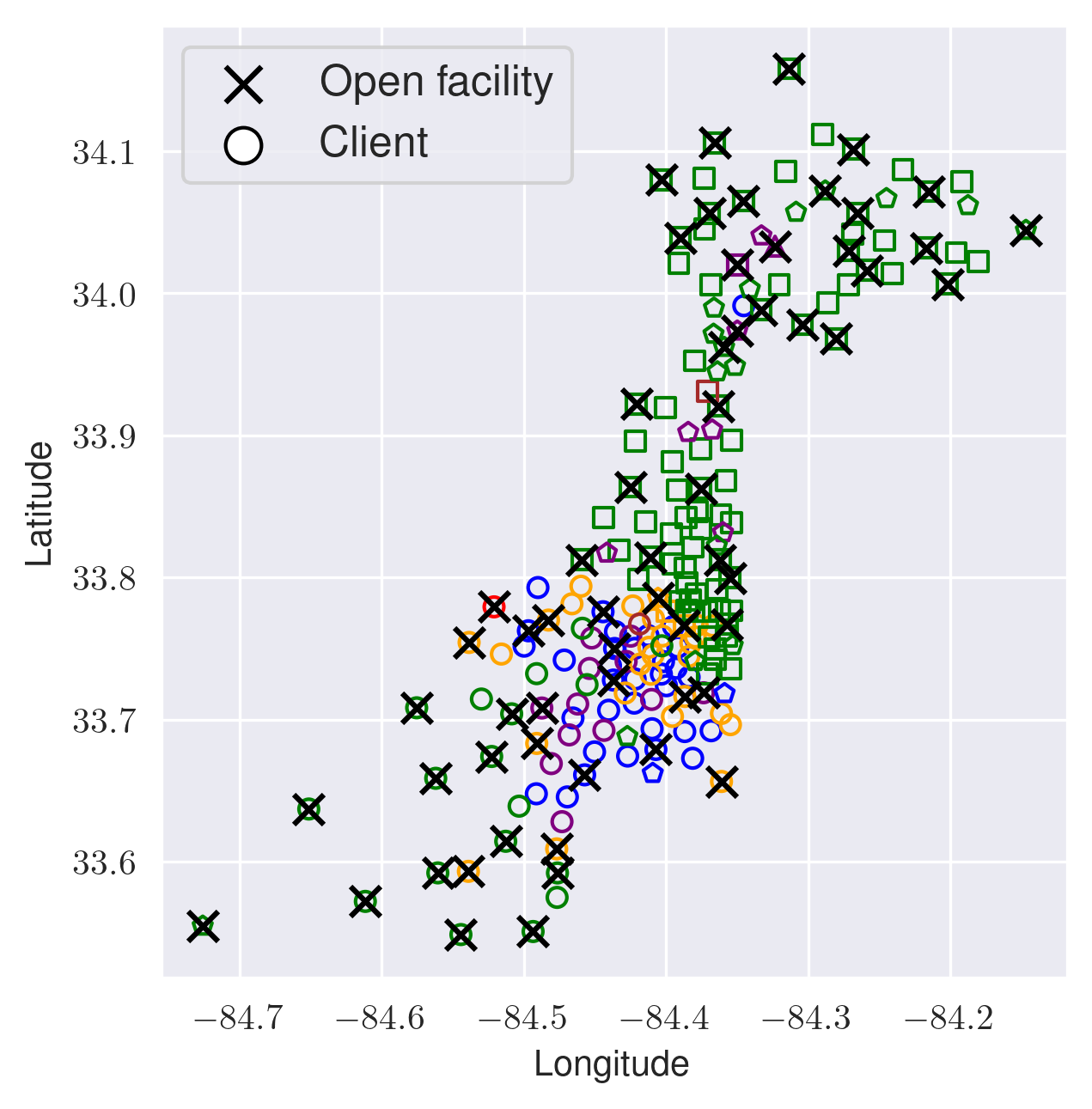}
        \end{minipage}
        \begin{minipage}{0.58\textwidth}
            \centering
            \includegraphics[width=0.7\textwidth]{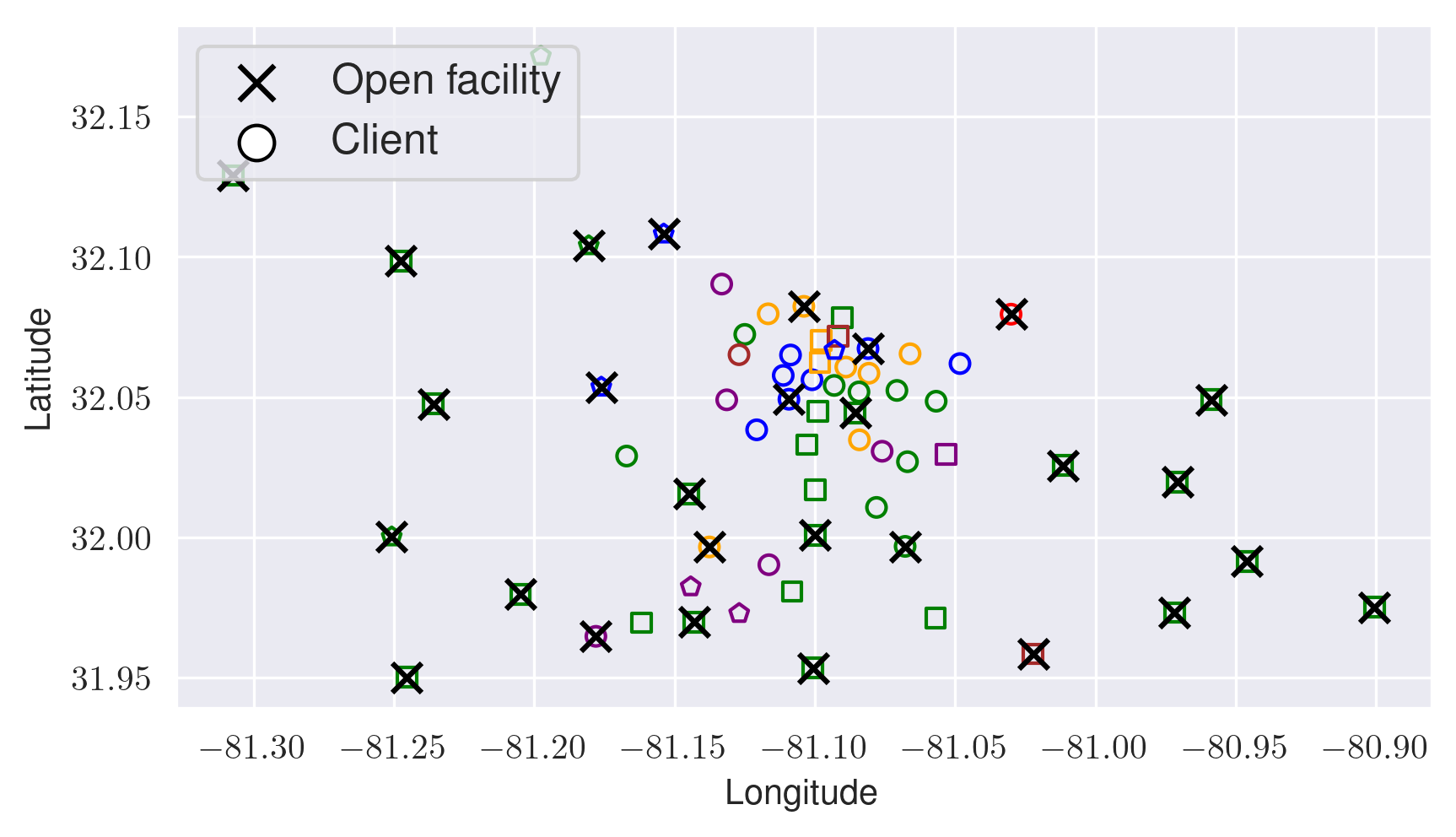}
        \end{minipage}
        \caption{Potential open facilities for (left) Fulton County, and (right) Chatham County.}
        \label{fig: potential-open-facilities}
    \end{figure}

    In this section, we discuss experiments for refinement for $k$-clustering (Definition \ref{def: refinements}) on Fulton County and Chatham County in Georgia, USA. For each county, consider opening several facilities (e.g. hospitals) one by one with the aim to provide equitable access to each demographic/client group in the county.
    We model this using refinement for $k$-clustering by choosing an appropriate set of clients and using straight-line distances as proxies for access costs.
    Our client groups are based on indicators for race, health insurance, and poverty.
    In the first stage of the experiment, we only seek the list $f_1, \ldots, f_l$ of facilities across $l$ iterations and the order in which to open them. All clients are assumed to be assigned to their nearest open facility at this stage. That is, the facility condition is satisfied but not the assignment condition.
    In the second stage, we also enforce the assignment condition by applying \textsc{DiscountedLookahead} to the facility sets $\{f_1\}, \{f_1, f_2\}, \ldots$ produced in the first stage.

    We compare two candidate algorithms for the first stage. The first (called \textsc{ChainCompletion}) is based on techniques in this work, and the second, (called \textsc{GreedyFacility}) greedily opens the best facility in each of the $l$ iterations. We observe (see Figures \ref{fig: approximation-comparison}, \ref{fig: approximation-comparison-chatham}) that \textsc{ChainCompletion} achieves a significantly better approximation ratio than \textsc{GreedyFacility}. Further, approximation ratios for \textsc{ChainCompletion} in the second stage (with \textsc{DiscountedLookahead}) are only marginally smaller as compared to the first stage, despite the additional structure of the assignment condition. We give more details on the settings of the experiments and describe algorithms next, before presenting additional results.

    \begin{algorithm}[t]
        \caption{\textsc{GreedyFacility}}\label{alg: greedy facility}
        \nonl {\textbf{input}:} client set $X$, partition $(X_1, \ldots, X_r)$ into client groups, set of potential open facilities $G \subseteq X$, integer $l \ge 1$, and some norm $g: \R^r \to \R$ on client group distances \\
        \nonl \textbf{output}: facility sets $F_1 \subseteq \ldots \subseteq F_l \subseteq G$ with $|F_k| = k$ for all $k \in [l]$ \\
        initialize empty ordering $F \gets ()$ \\
        \For{$k = 1$ to $l$}
        {
            find facility $f_k$ with least objective value when added to $F$:
            \[
                f_k = {\arg\min}_{f \not\in F} g(F \cup \{f\})
            \]
            append $f_k$ to $F$ \\
        }
        for each $k \in [l]$, set $F_k$ to be the first $k$ facilities in $F$ \\
        \textbf{return} $(F_1, \ldots, F_l)$ \\
    \end{algorithm}

    \begin{algorithm}[!t]
        \caption{\textsc{ChainCompletion}}\label{alg: enhanced-greedy}
        \nonl \textbf{input}: client set $X$, partition $(X_1, \ldots, X_r)$ into client groups, set of potential open facilities $G \subseteq X$, integer $l \ge 1$, and some norm $g: \R^r \to \R$ on client group distances \\
        \nonl \textbf{output}: facility sets $F_1 \subseteq \ldots \subseteq F_l \subseteq G$ with $|F_k| = k$ for all $k \in [l]$ \\
        initialize empty ordering $F \gets ()$
        \For{$k = 2^0, 2^1, \ldots, 2^{\lceil \log_2 l \rceil}$}
        {
            let $G_{k}$ be the solution to FC$^{(k)}_g$ with norm $g$ obtained using Theorem \ref{thm: facility-location-solvability-theorem} \label{step: chain-completion-solutions} \\
        prune $G_k$ to keep only the first $k$ facilities \\
        order facilities in $G_{k} \setminus F$ greedily for objective $g$, and append them to $F$ in this order \\
        }
        for each $k \in [l]$, set $F_k$ to be the first $k$ facilities in $F$ \\
        \textbf{return} $(F_1, \ldots, F_l)$
    \end{algorithm}

    \textbf{Choice of clients, demographic groups, and distances.} The set $X$ of clients is the set of all census tracts in the given county. Each census tract is classified into one of $15$ demographic groups as follows: it is either (1) majority White alone (not including Hispanic), (2) majority Black or African American alone (not including Hispanic), (3) majority Hispanic, or (4) mixed (i.e. no racial group is in the majority). Tracts are also placed into three categories based on health insurance levels: (a) at least a quarter of the people in the tract have two health insurances, (b) at least a quarter of the people in the tract have no insurance, and (c) neither of the above two. Finally, tracts are placed into two categories based on whether or not over 35 percent of the people in the Census tract are below the poverty line. This results in $4 \times 3 \times 2 = 24$ groups that partition $X$. $9$ of these are empty for both counties so we end up with $r = 15$ groups, denoted $X_1, \ldots, X_r$.

    \begin{figure}
        \begin{minipage}{0.32\textwidth}
            \centering
            Optimal solution for $k$-clustering \\
            \phantom{Phantom text}
            \includegraphics[width=0.8\textwidth]{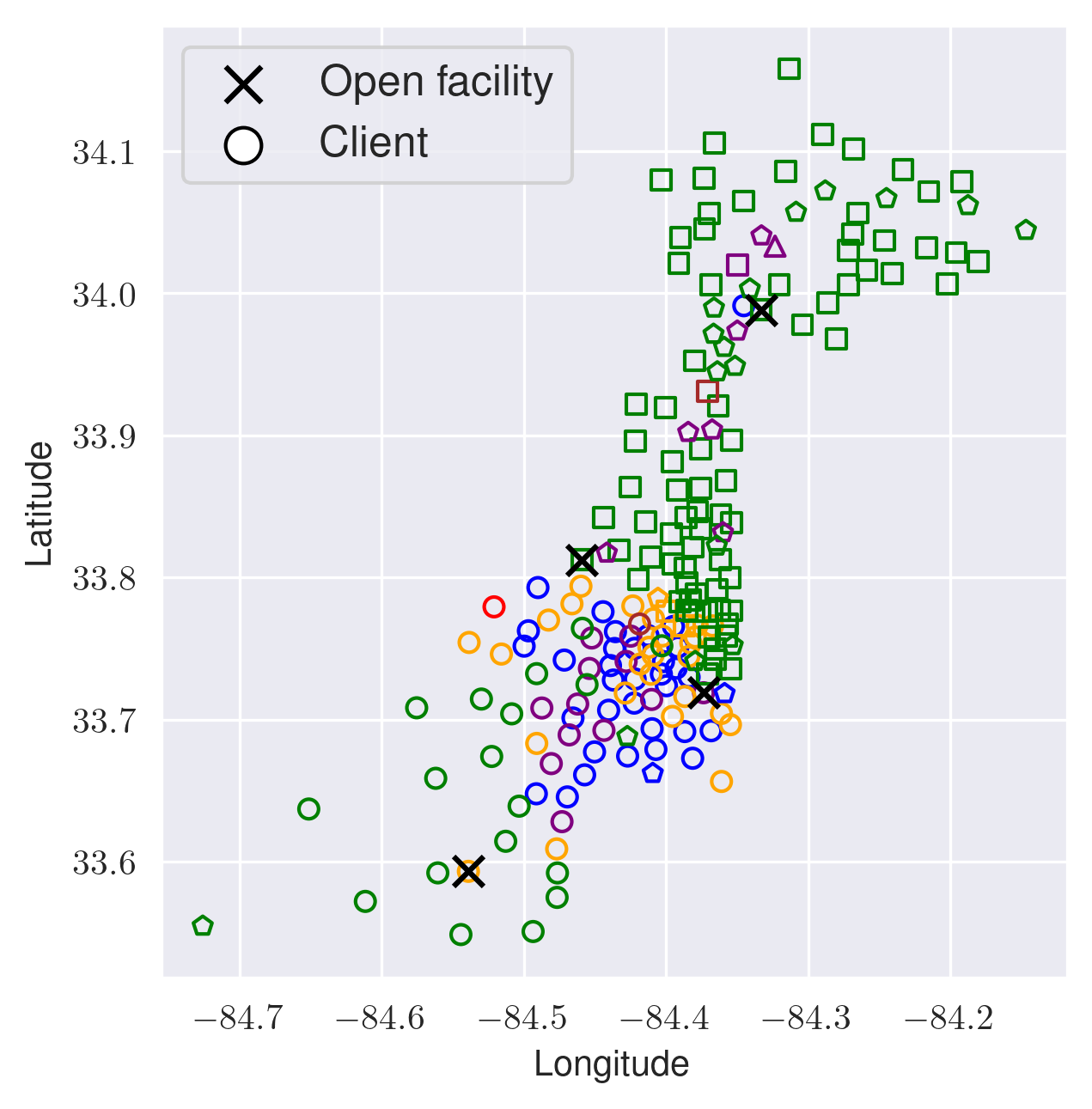}
            \phantom{} \\
        \end{minipage}
        \hfill
        \begin{minipage}{0.32\textwidth}
            \centering
            \textsc{GreedyFacility} \\
            {\phantom{A} $k = 4$ \phantom{A}}
            \includegraphics[width=0.8\textwidth]{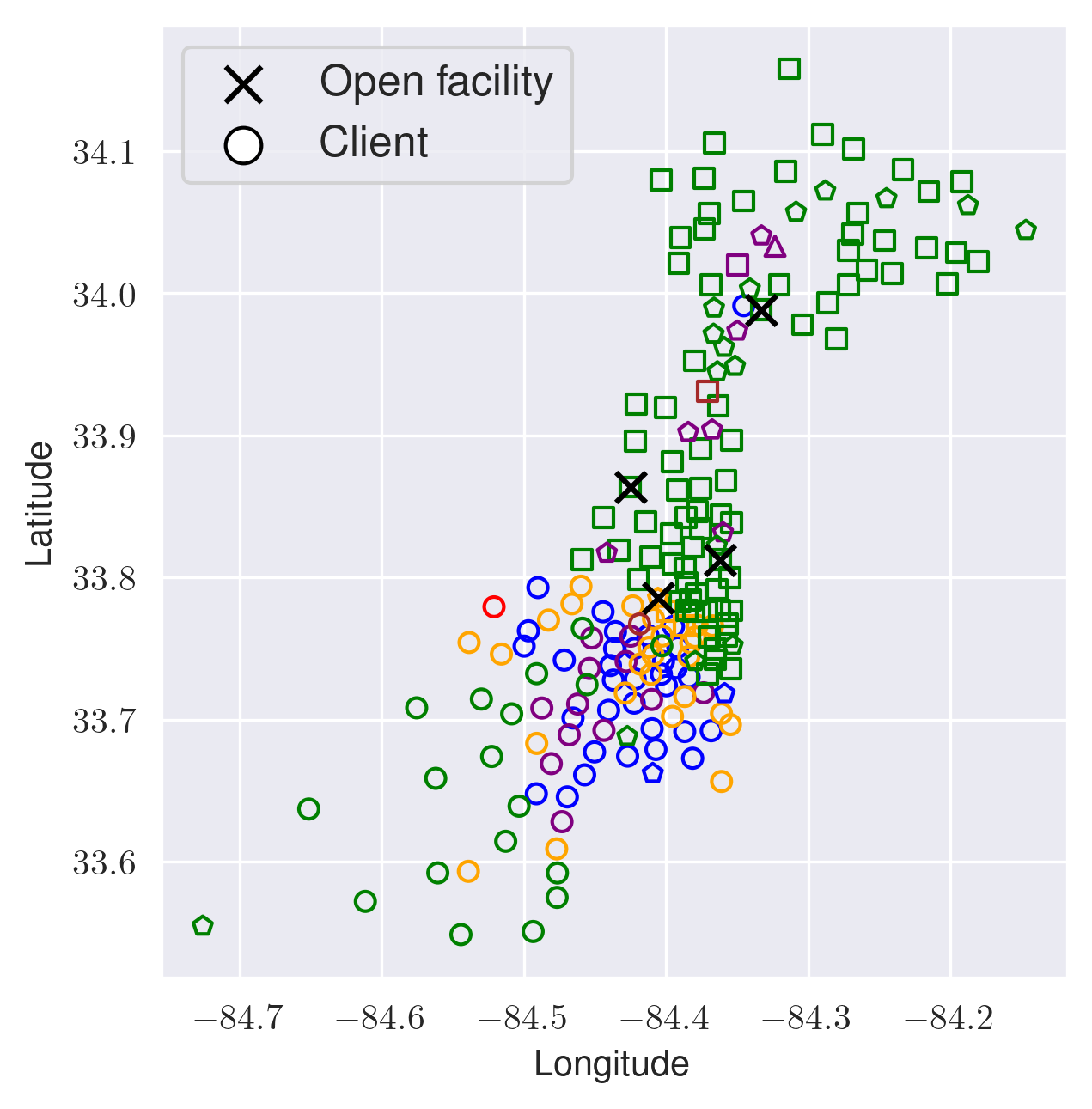}
            \phantom{} \\
        \end{minipage}
        \hfill
        \begin{minipage}{0.32\textwidth}
            \centering
            \textsc{ChainCompletion} \\
            \phantom{} \\
            \includegraphics[width=0.8\textwidth]{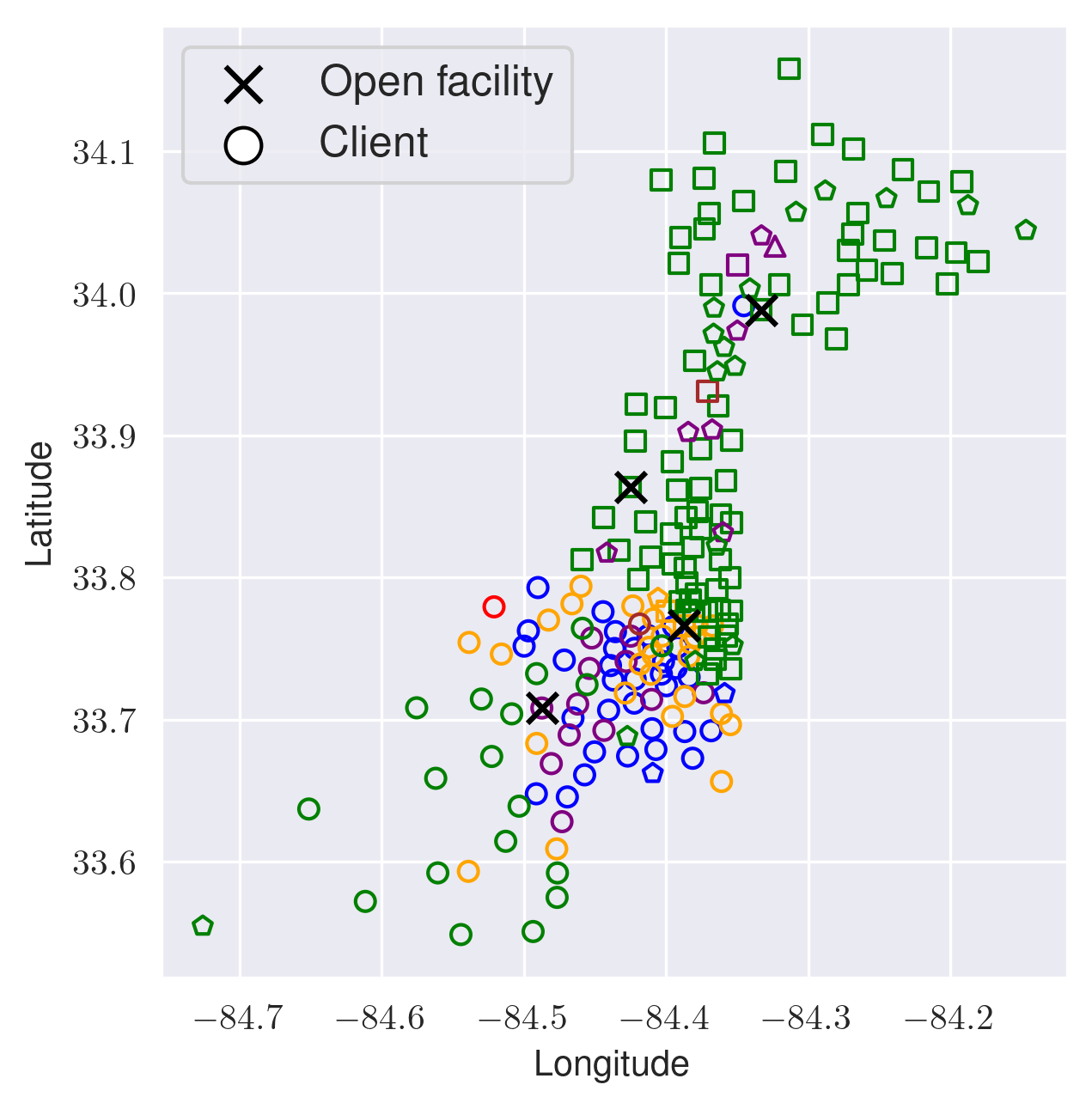}
            \phantom{} \\
        \end{minipage}
        \centering $k = 6$ \\
        \begin{minipage}{0.32\textwidth}
            \centering
            \includegraphics[width=0.8\textwidth]{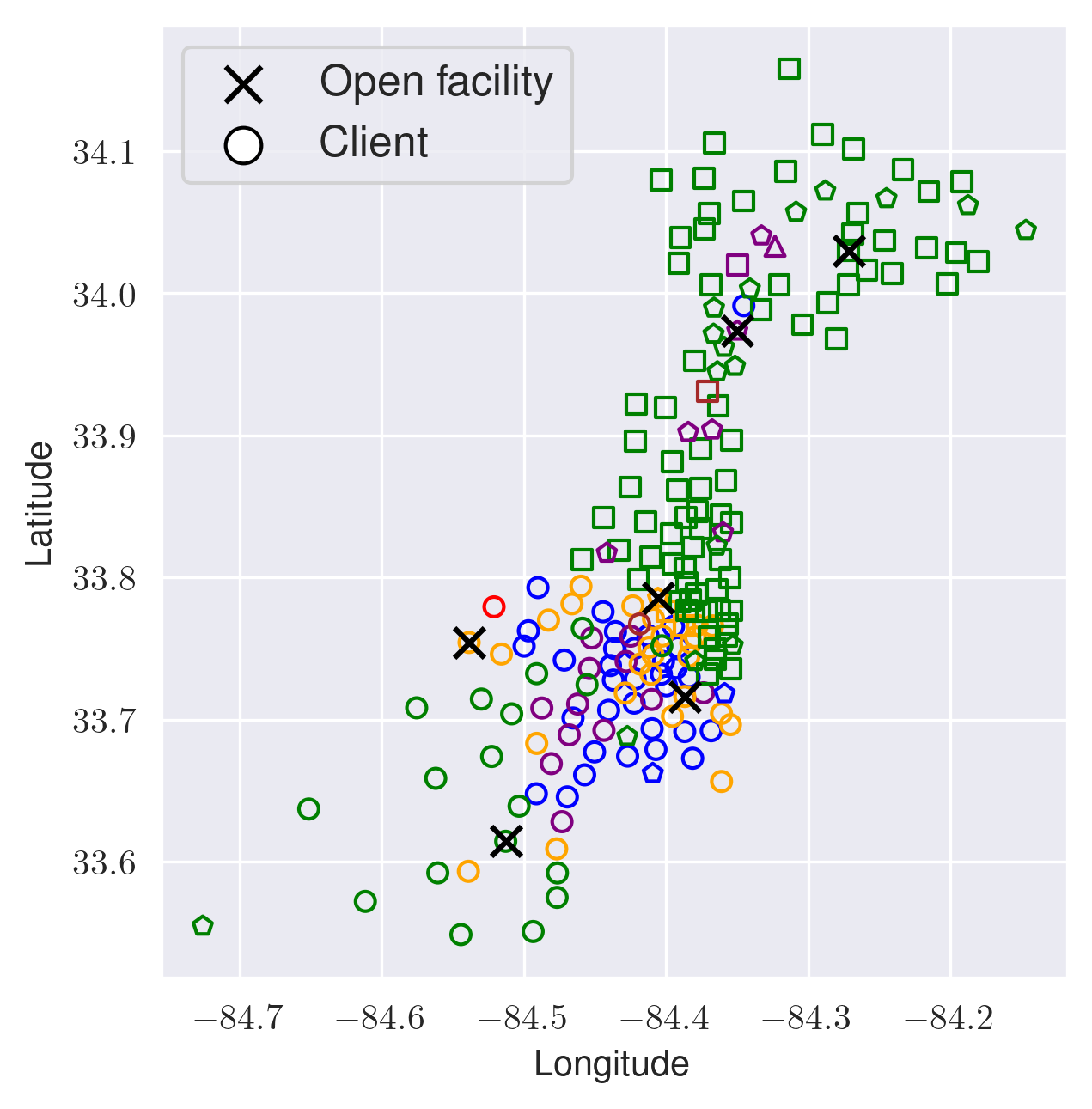}
            \phantom{} \\
        \end{minipage}
        \hfill
        \begin{minipage}{0.32\textwidth}
            \centering
            \includegraphics[width=0.8\textwidth]{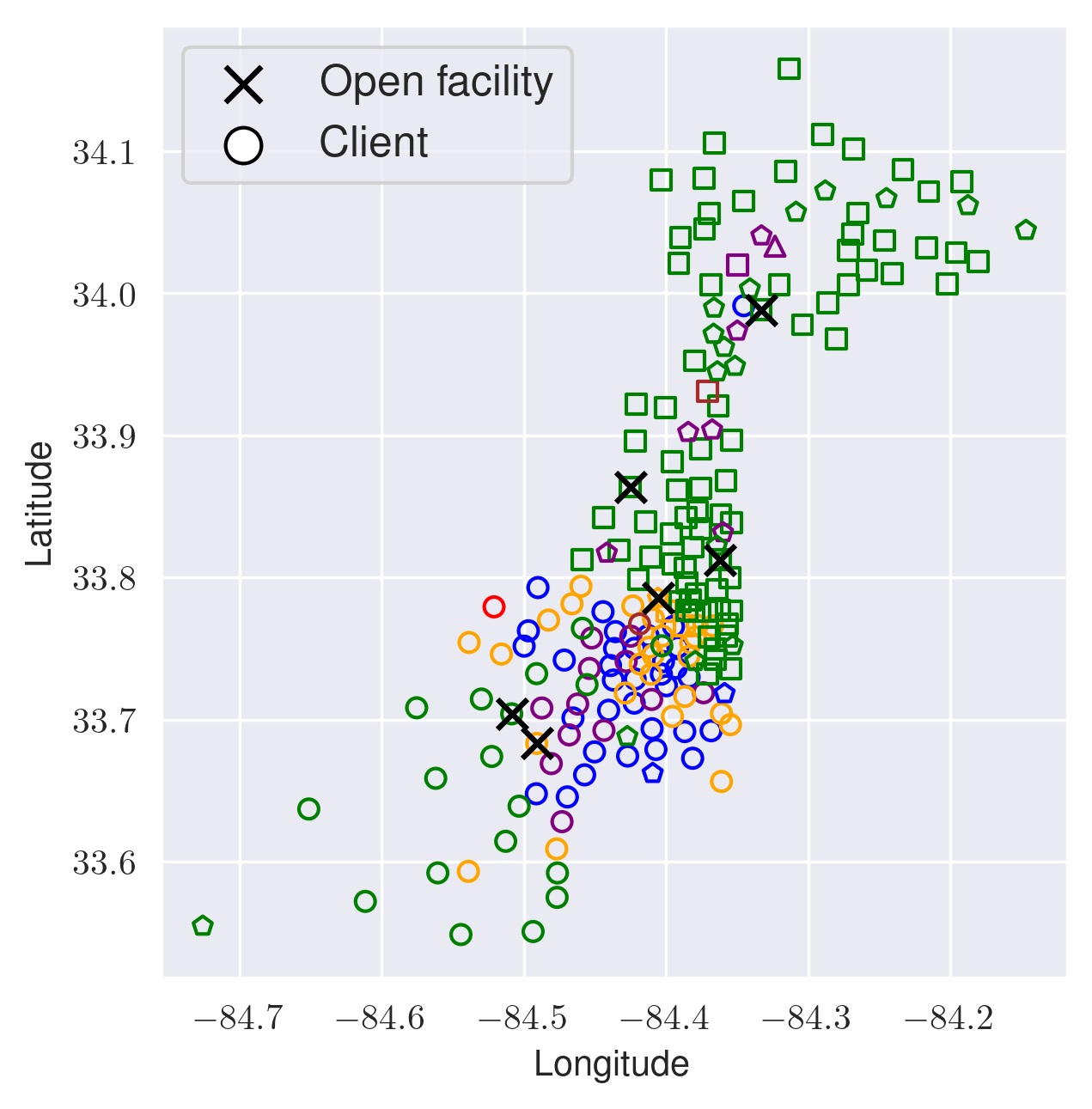}
            \phantom{} \\
        \end{minipage}
        \hfill
        \begin{minipage}{0.32\textwidth}
            \centering
            \includegraphics[width=0.8\textwidth]{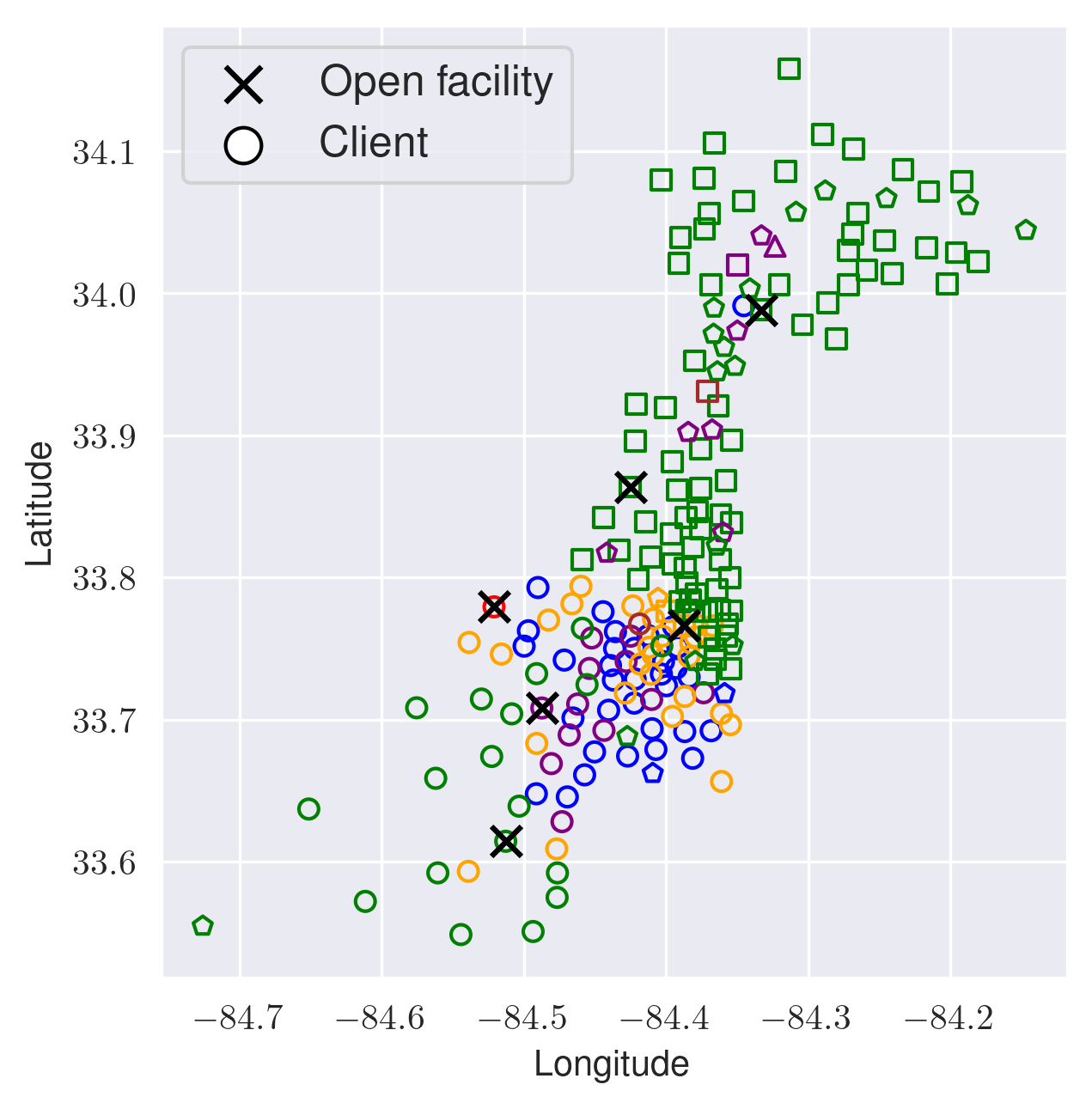}
            \phantom{} \\
        \end{minipage}
        \centering $k = 9$ \\
        \begin{minipage}{0.32\textwidth}
            \centering
            \includegraphics[width=0.8\textwidth]{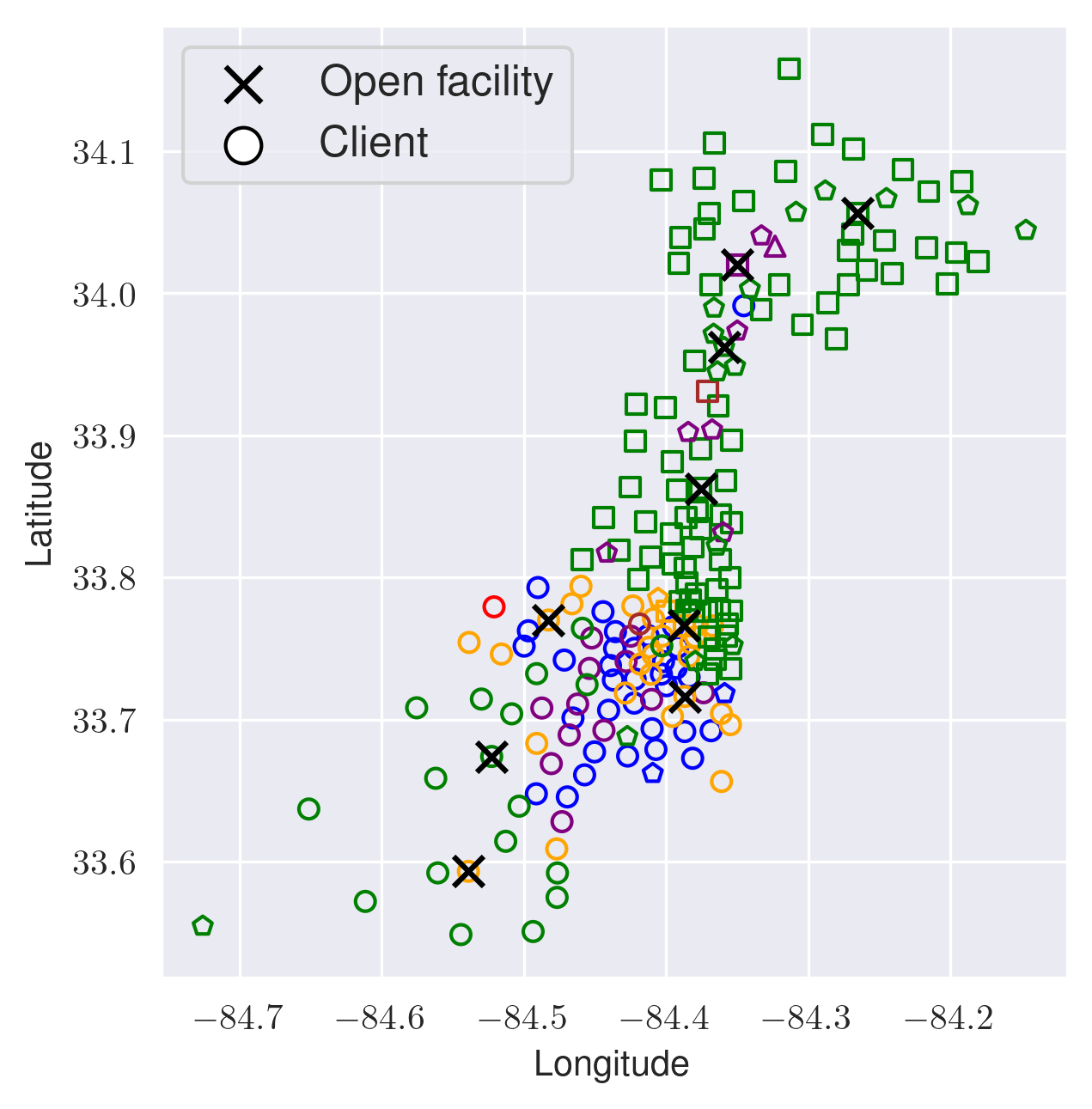}
            \phantom{} \\
        \end{minipage}
        \hfill
        \begin{minipage}{0.32\textwidth}
            \centering
            \includegraphics[width=0.8\textwidth]{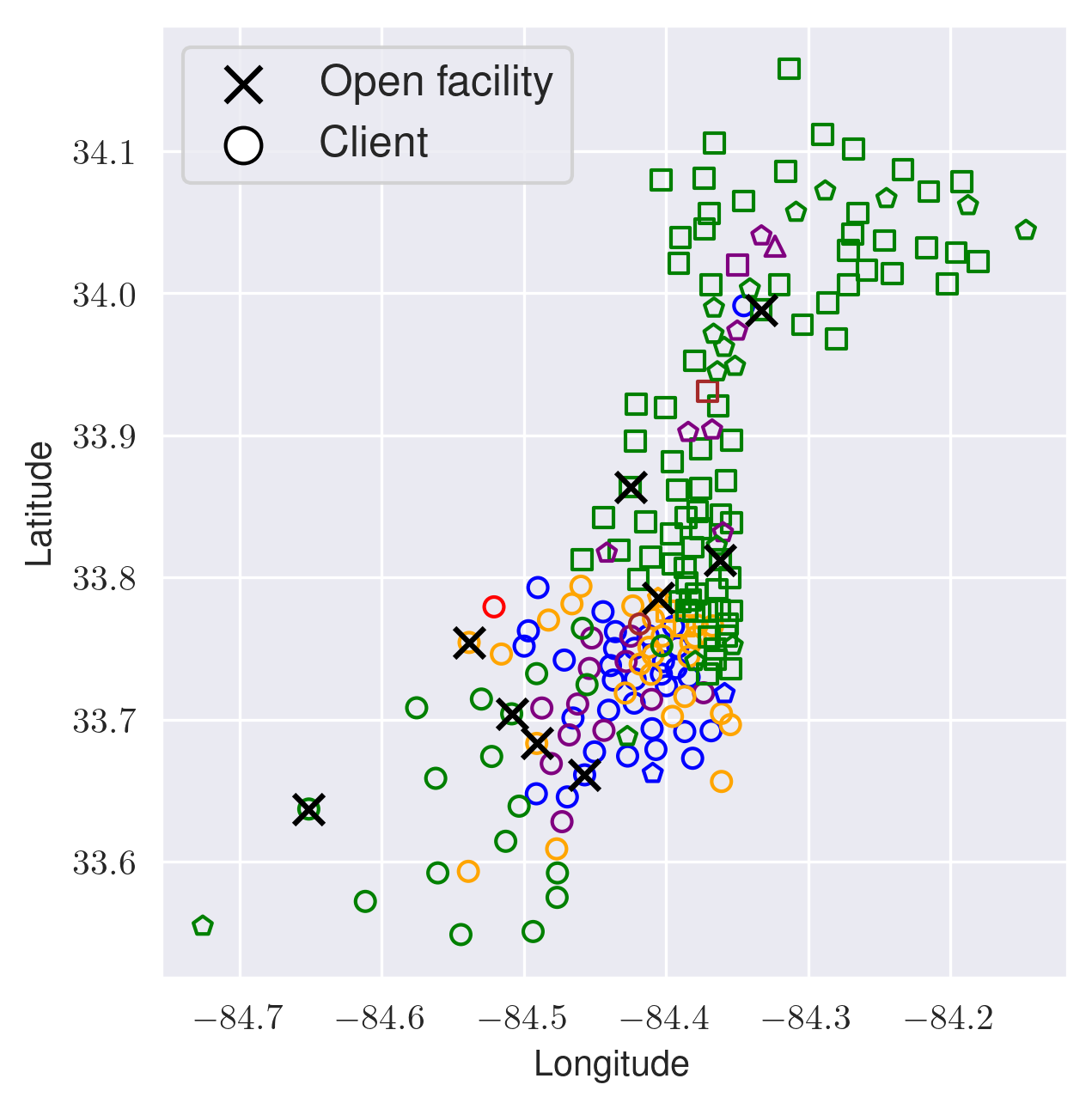}
            \phantom{} \\
        \end{minipage}
        \hfill
        \begin{minipage}{0.32\textwidth}
            \centering
            \includegraphics[width=0.8\textwidth]{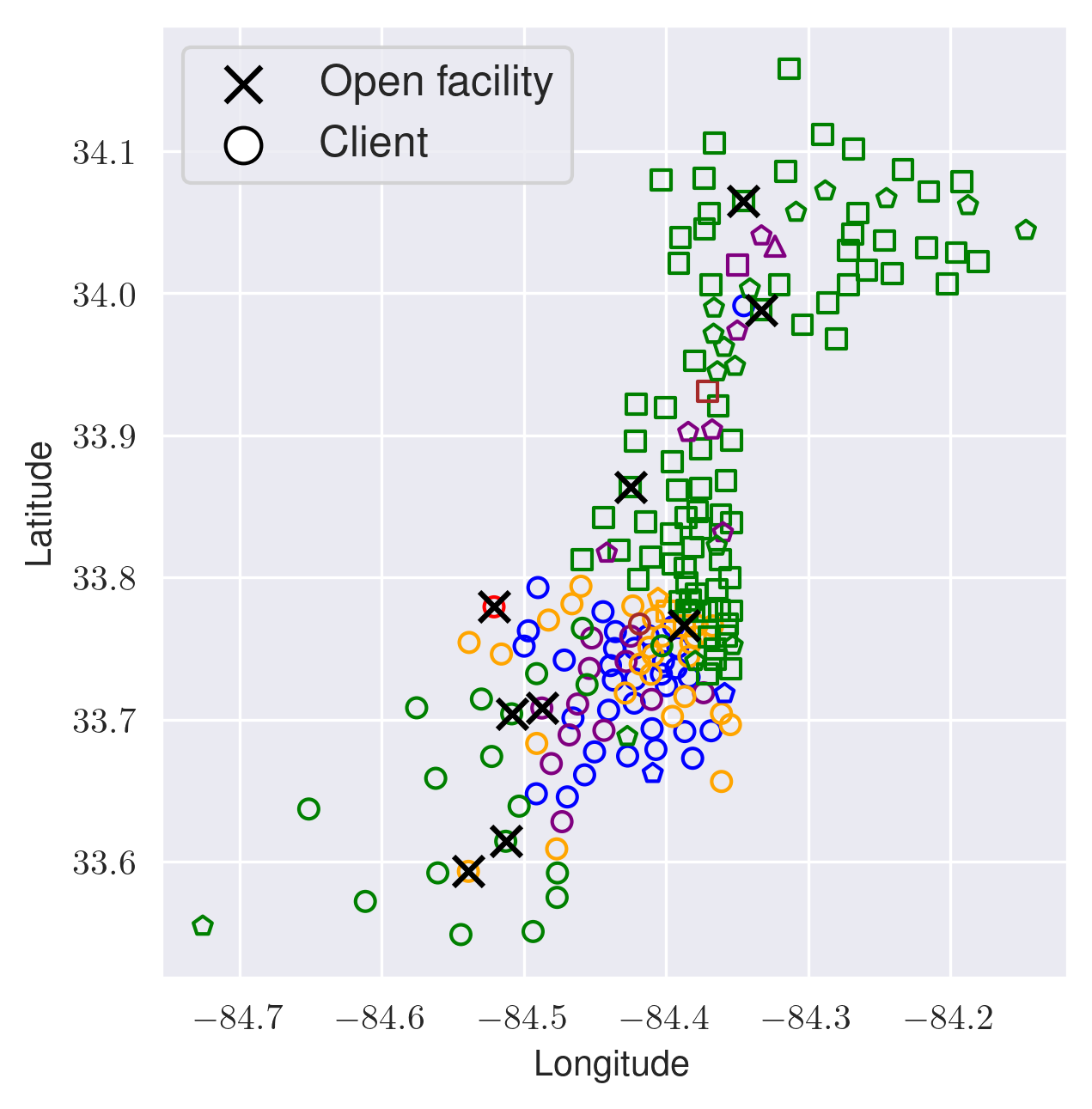}
            \phantom{} \\
        \end{minipage}
        \caption{Open facility sets for FC$^{(k)}_\infty$ optimal, \textsc{GreedyFacility}, and \textsc{ChainCompletion} for $k \in \{4, 6, 9\}$ for Fulton county, Georgia, USA. For each census tract (i.e. each client), different colors indicate different poverty and insurance levels in the census tract, while different shapes indicate different racial majorities.}
        \label{fig: census-experiment}
    \end{figure}

    We choose fractional group memberships $\mu_{j, s} = \frac{1}{|X_s|}$ if tract $j$ is in group $s$ and $\mu_{j, s} = 0$ otherwise, so that the group distance vector contains the average distance to assigned facilities for clients in the group for each group. To facilitate computation, we (1) restrict the locations of open facilities; Figure \ref{fig: potential-open-facilities} shows these locations, (2) assume that census tracts are point masses at their geometric center, and (3) compute straight-line distances between clients using their latitude and longitude values. The number of iterations $l$ is set to be $16$ for Fulton County and $12$ for Chatham County as reasonably large values for this experiment; however we remark that other values of $l$ also lead to similar performance outcomes.

    \begin{figure}[t]
        \centering
        \includegraphics[width=0.8\textwidth]{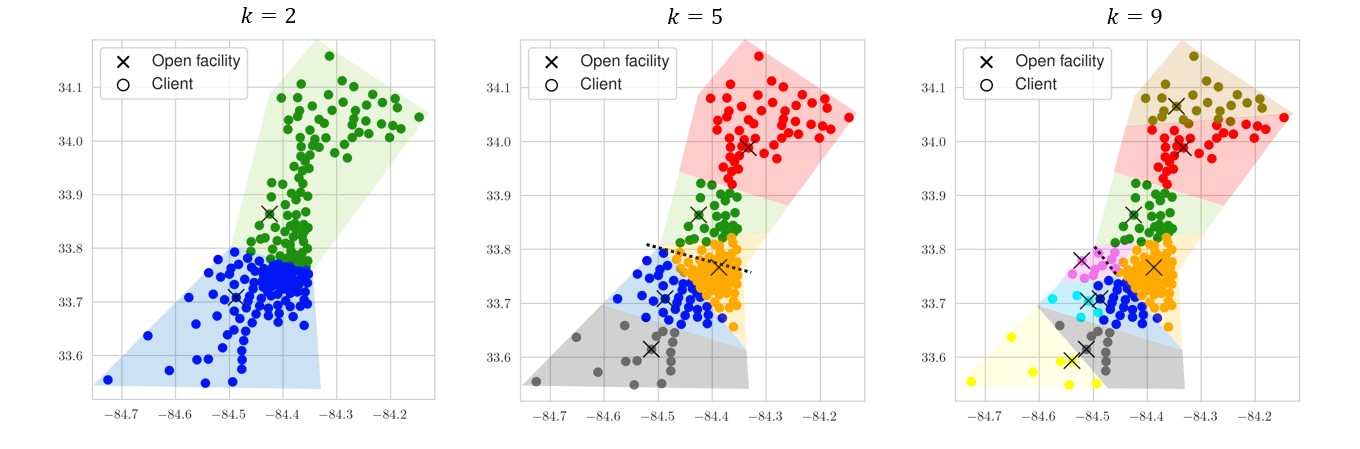}
        \includegraphics[width=0.8\textwidth]{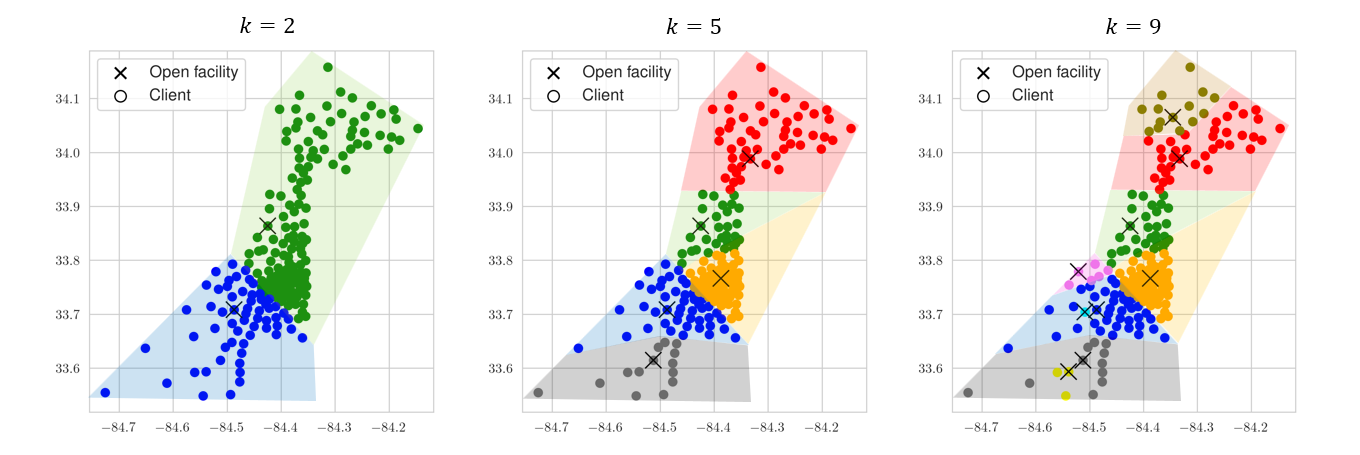}
        \caption{Open facilities and assignments for $k$-clustering refinement  using \textsc{ChainCompletion} with and without \textsc{DiscountedLookahead}. The top row shows assignments of clients (circles) to their respective closest facilities (crosses), indicated by the color of the client. These assignments do not satisfy the assignment condition. Dashed lines indicate where the assignment condition is not satisfied. The bottom row shows assignments produced by \textsc{DiscountedLookahead}; notice that these satisfy the assignment condition.}
        \label{fig: assignments-experiments}
    \end{figure}

    \textbf{Computation of approximation ratios.} For each $k \in [l]$, for $k$ open facilities, we fix the the $L_\infty$ norm objective $\min_{\substack{(F, \Pi), |F| = k, \\ \Pi: X \to F}} \max_{s \in [r]} d_\Pi(s)$, where $d_\Pi(s) = \frac{1}{|X_s|} \sum_{j \in X_s} d(j, \Pi(j))$ is the average distance travelled by clients in group $X_s$ to their assigned facilities. We seek facility sets $F_1, \ldots, F_l$ and assignments $\Pi_k: X \to F_k$, $k \in [l]$ that satisfy form a Refinement: (1) $|F_k| = k$ for all $k \in [l]$, (2) (facility condition) $F_1 \subset \ldots \subset F_l$, and (3) (assignment condition) for each $f_k \in F_k$, there is some $f_{k - 1} \in F_{k - 1}$ such that
    \[
        \{j \in X: \Pi_k(j) = f_k\} \subseteq \{j \in X: \Pi_{k - 1}(j) = f_{k - 1}\}.
    \]
    Recall that the approximation ratio of $(F_k, \Pi_k)$ is the ratio of its objective value to the optimal solution for $k$ open facilities (i.e., the Fair Clustering FC$^{(k)}_\infty$ objective).

    \textbf{Stage 1 algorithms.} The two algorithms \textsc{GreedyFacility} and \textsc{ChainCompletion} for stage 1 return only facility sets $F_k, k \in [l]$ that satisfy the facility condition.
    \textsc{GreedyFacility} (Algorithm \ref{alg: greedy facility}) was first used by \cite{Hochbaum09} in the context of a different facility location problem: for any given objective function, this algorithm iteratively opens the facility that improves the objective function. Algorithm \ref{alg: greedy facility} formally describes this algorithm for an arbitrary objective function $g$.
    \textsc{ChainCompletion} relies on the strategy in Section \ref{sec: refinements}. However. First, for each $k \in \{2^0, 2^1, \ldots, 2^{\lfloor \log_2 \rfloor}\}$,
    we obtain facility set $F_k$ by using Theorem \ref{thm: facility-location-solvability-theorem} for FC$^{(k)}$ and objective $g = L_\infty$ norm. Since the theorem only gives the bicriteria guarantee that $|F_k| \le 4k$, we prune each $F_k$ so that it satisfies $|F_k| \le k$.
    We next open $l$ facilities in the following order: first, open the only facility in $F_1$. Next, open the facilities in $F_2 \setminus F_1$, in the order suggested by \textsc{GreedyFacility}. Next, open facilities in $F_4 \setminus (F_2 \cup F_1)$ in the greedy order, and so on.

    \textbf{Results}. The locations of facilities in Fulton County for \textsc{GreedyFacility} and \textsc{ChainCompletion} are shown in Figure \ref{fig: census-experiment} alongside optimal facilities for each FC$^{(k)}_\infty$. Visually, \textsc{ChainCompletion} opens facilities much closer to the optimal facilities than \textsc{GreedyFacility}.
    For Fulton County (resp. Chatham County), Figure \ref{fig: approximation-comparison} (resp. Figure \ref{fig: approximation-comparison-chatham}) compares all four combinations (\textsc{GreedyFacility} and \textsc{ChainCompletion} for facilities with or without \textsc{DiscountedLookahead} for assignments) against the optimal for each $k \in [l]$. We note that \textsc{ChainCompletion} (in {\color{ForestGreen} green}) outperforms \textsc{GreedyFacility} (in {\color{blue} blue}) for most values of $k$. Further, \textsc{DiscountedLookahead} (dashed lines in the Figure) is nearly as good as assigning each client to its closest facility despite being additionally constrained. In particular, it significantly outperforms its worst-case theoretical guarantee from Theorem \ref{thm: reassignment-problem}. These assignments for \textsc{ChainCompletion} are shown in Figure \ref{fig: assignments-experiments}.

    \section{Trade-off Between Facility Cost and Access Cost in Fair Facility Location}\label{app: facility-access-cost-tradeoff}

    \begin{figure}[h]
        \begin{minipage}{\textwidth}
            \begin{minipage}{0.32\textwidth}
                \includegraphics[width=\textwidth]{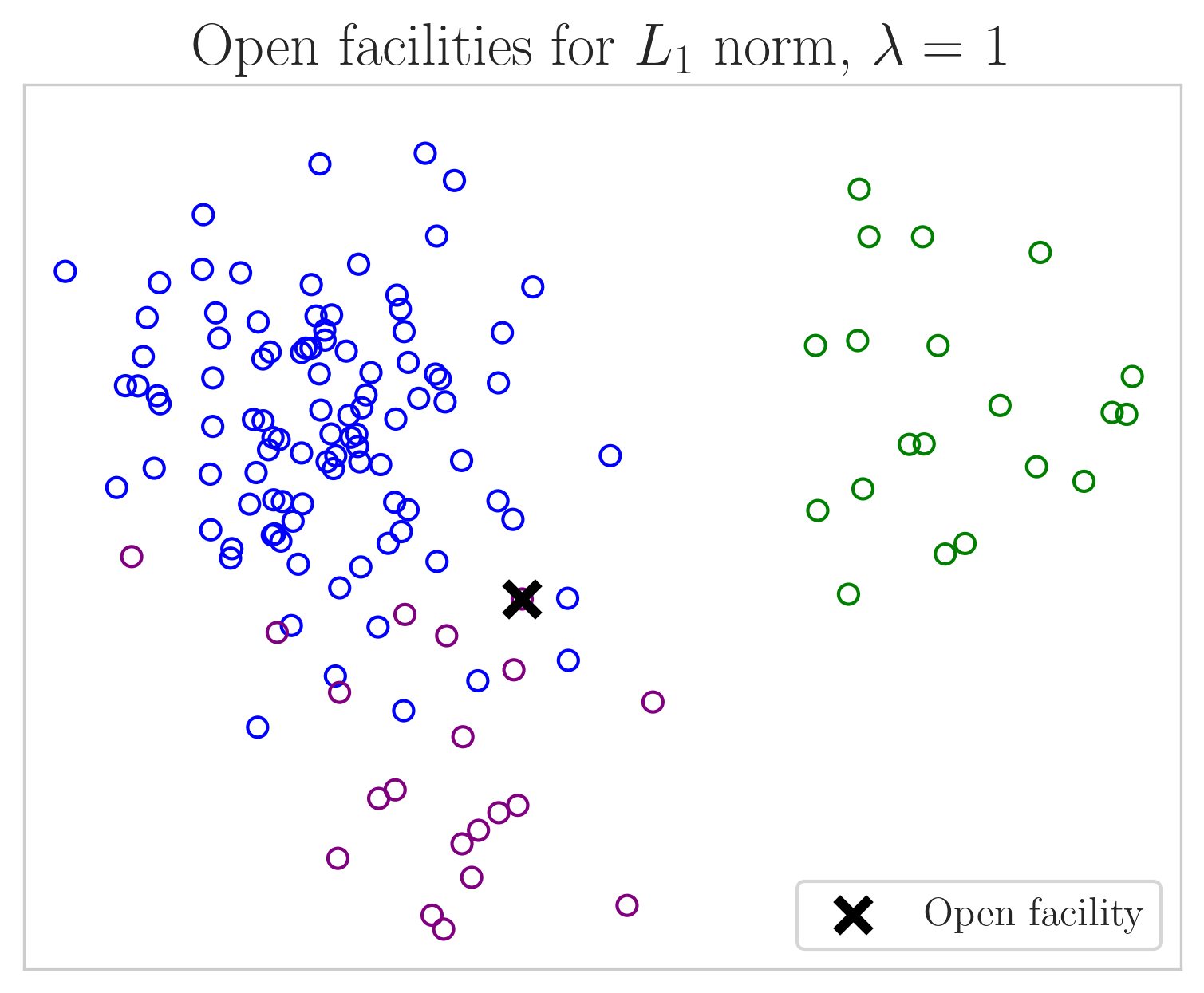}
            \end{minipage}
            \hfill
            \begin{minipage}{0.32\textwidth}
                \includegraphics[width=\textwidth]{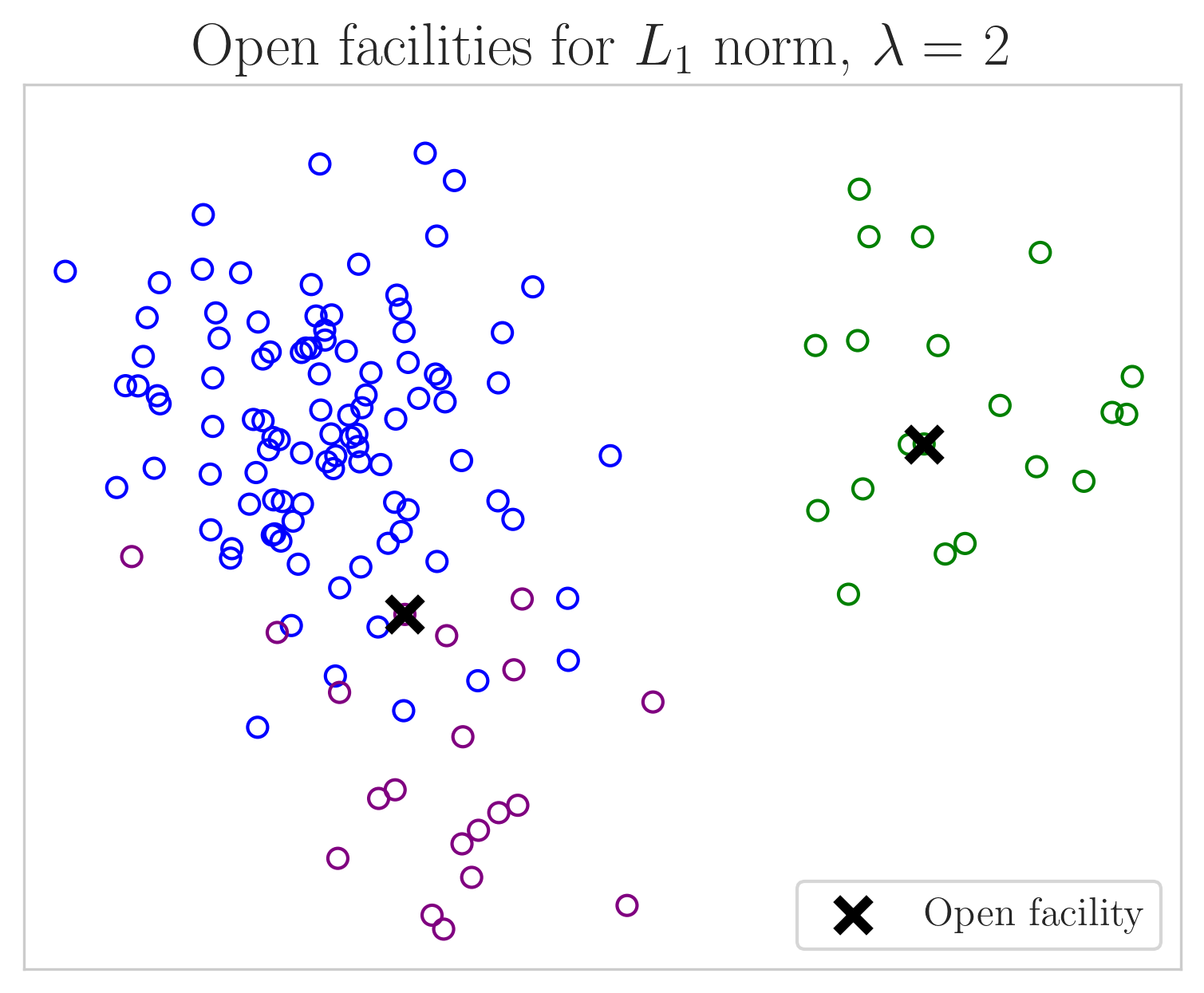}
            \end{minipage}
            \hfill
            \begin{minipage}{0.32\textwidth}
                \includegraphics[width=\textwidth]{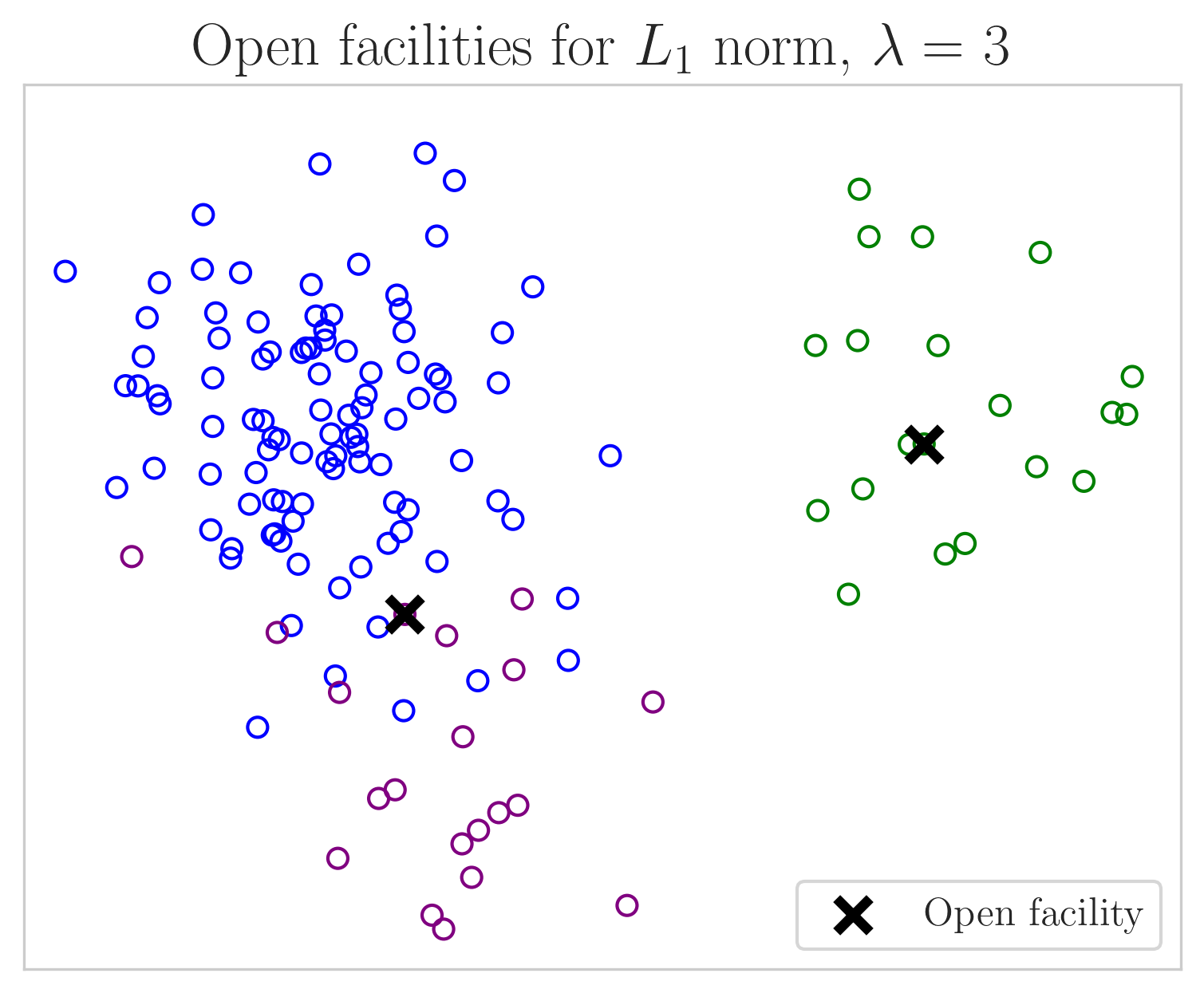}
            \end{minipage}
        \end{minipage}
        \begin{minipage}{\textwidth}
            \begin{minipage}{0.32\textwidth}
                \includegraphics[width=\textwidth]{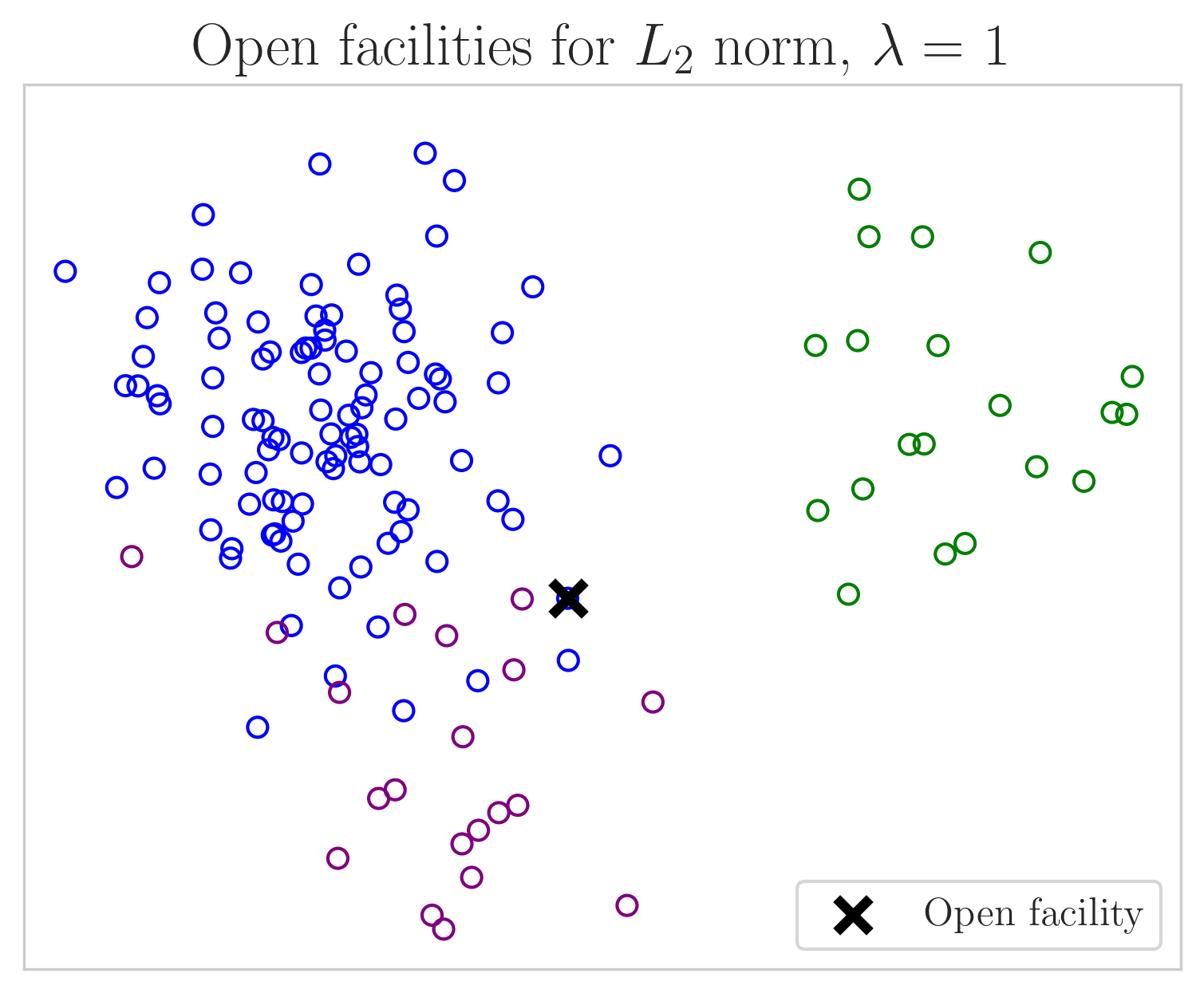}
            \end{minipage}
            \hfill
            \begin{minipage}{0.32\textwidth}
                \includegraphics[width=\textwidth]{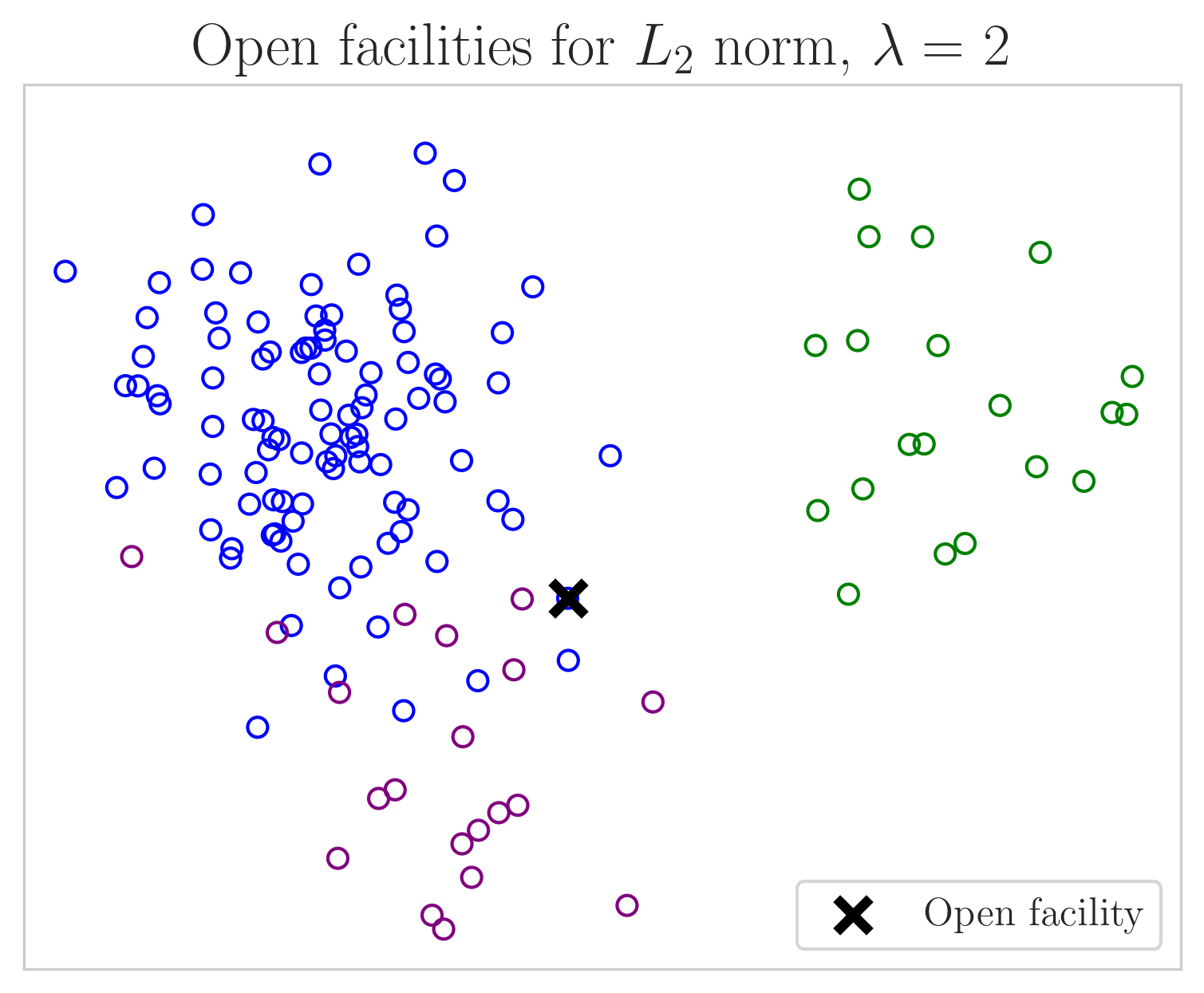}
            \end{minipage}
            \hfill
            \begin{minipage}{0.32\textwidth}
                \includegraphics[width=\textwidth]{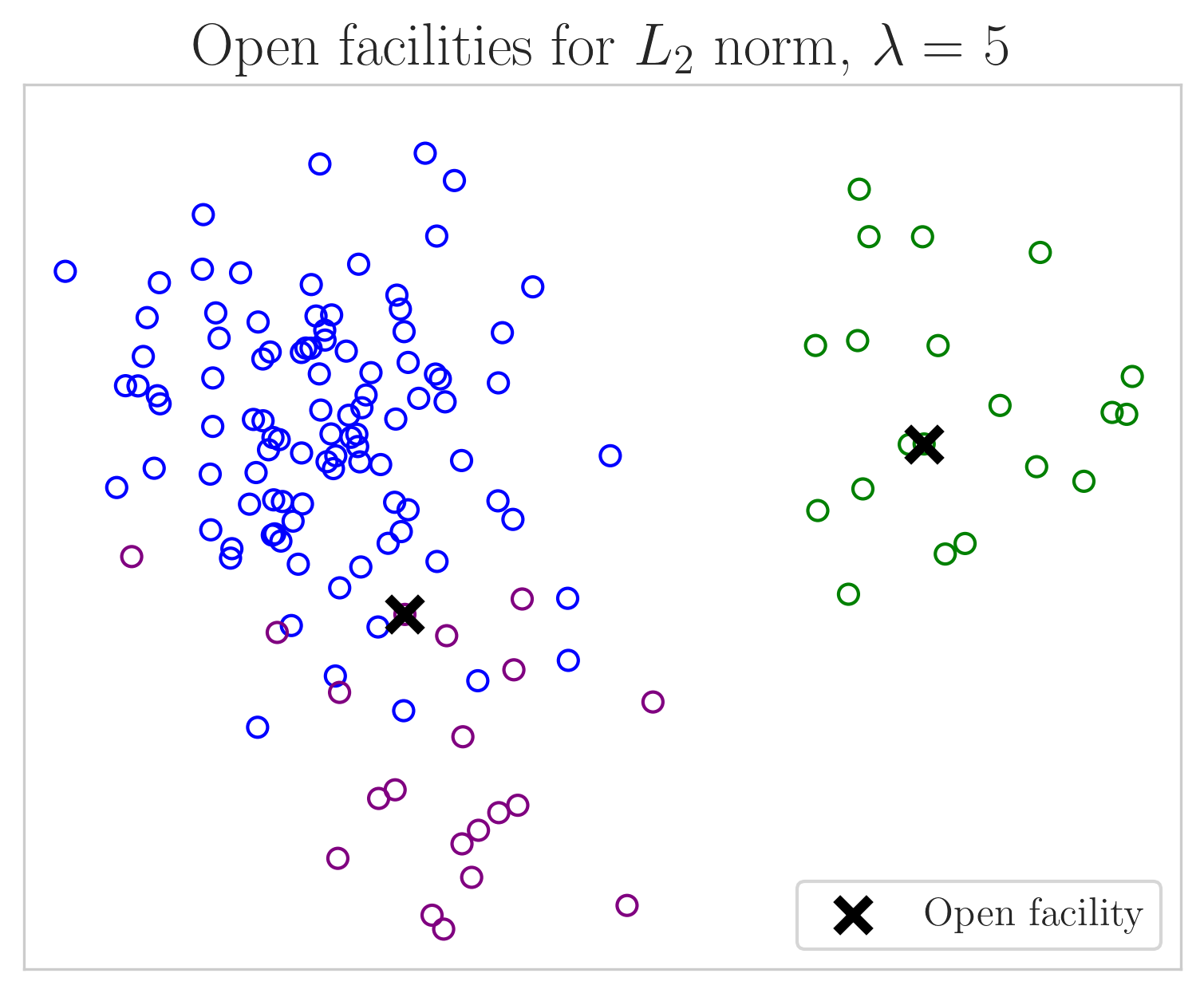}
            \end{minipage}
        \end{minipage}
        \begin{minipage}{\textwidth}
            \begin{minipage}{0.32\textwidth}
                \includegraphics[width=\textwidth]{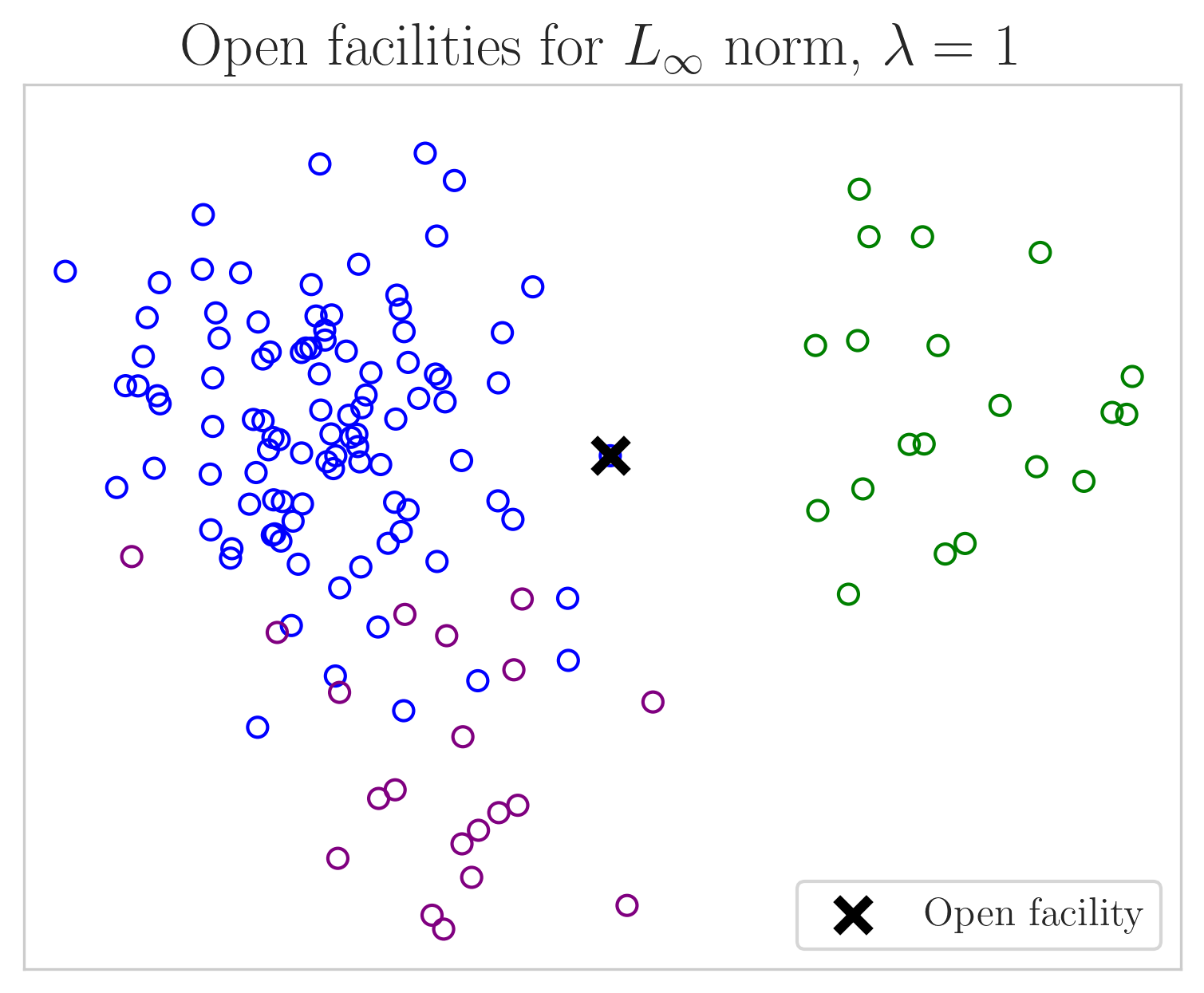}
            \end{minipage}
            \hfill
            \begin{minipage}{0.32\textwidth}
                \includegraphics[width=\textwidth]{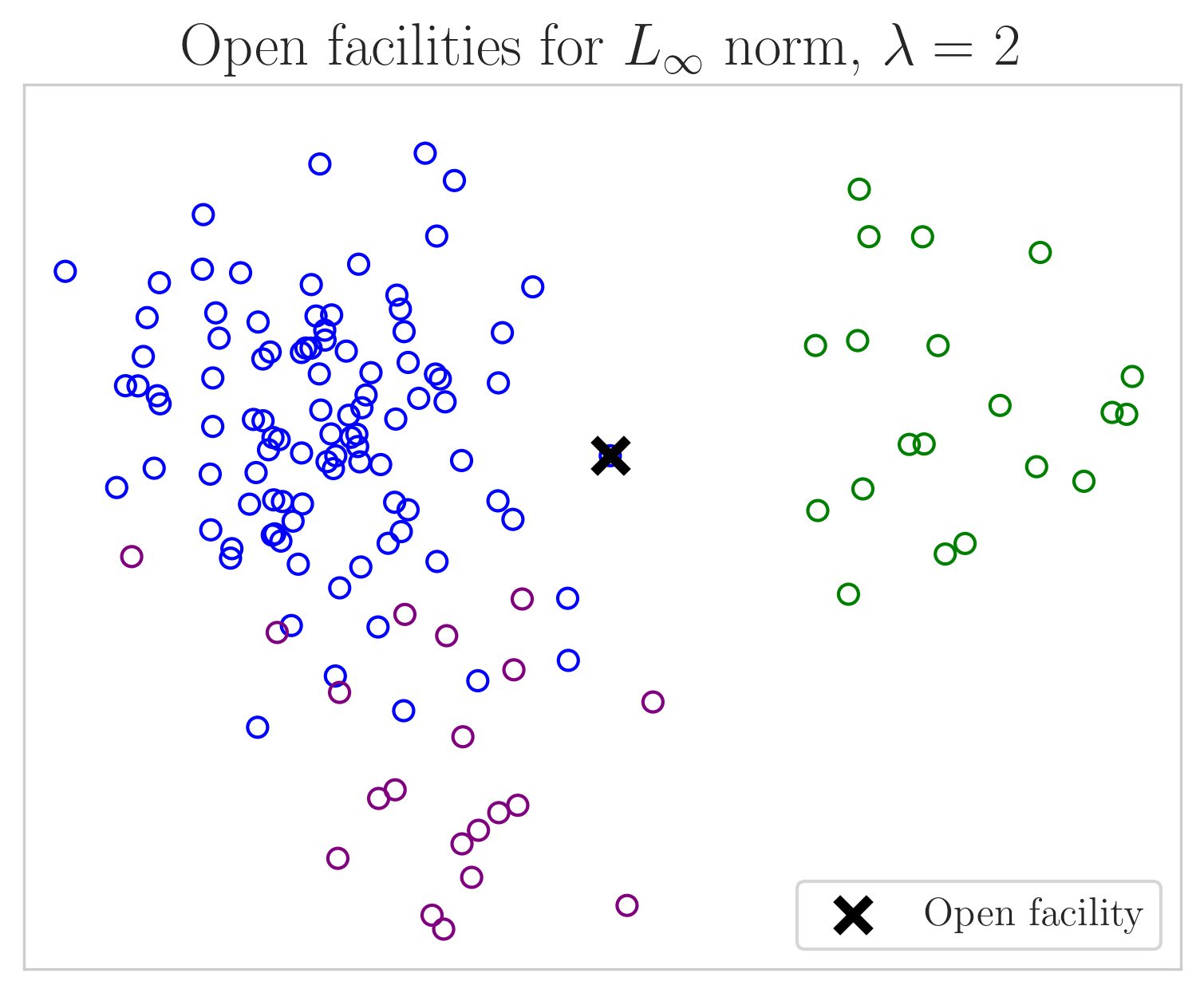}
            \end{minipage}
            \hfill
            \begin{minipage}{0.32\textwidth}
                \includegraphics[width=\textwidth]{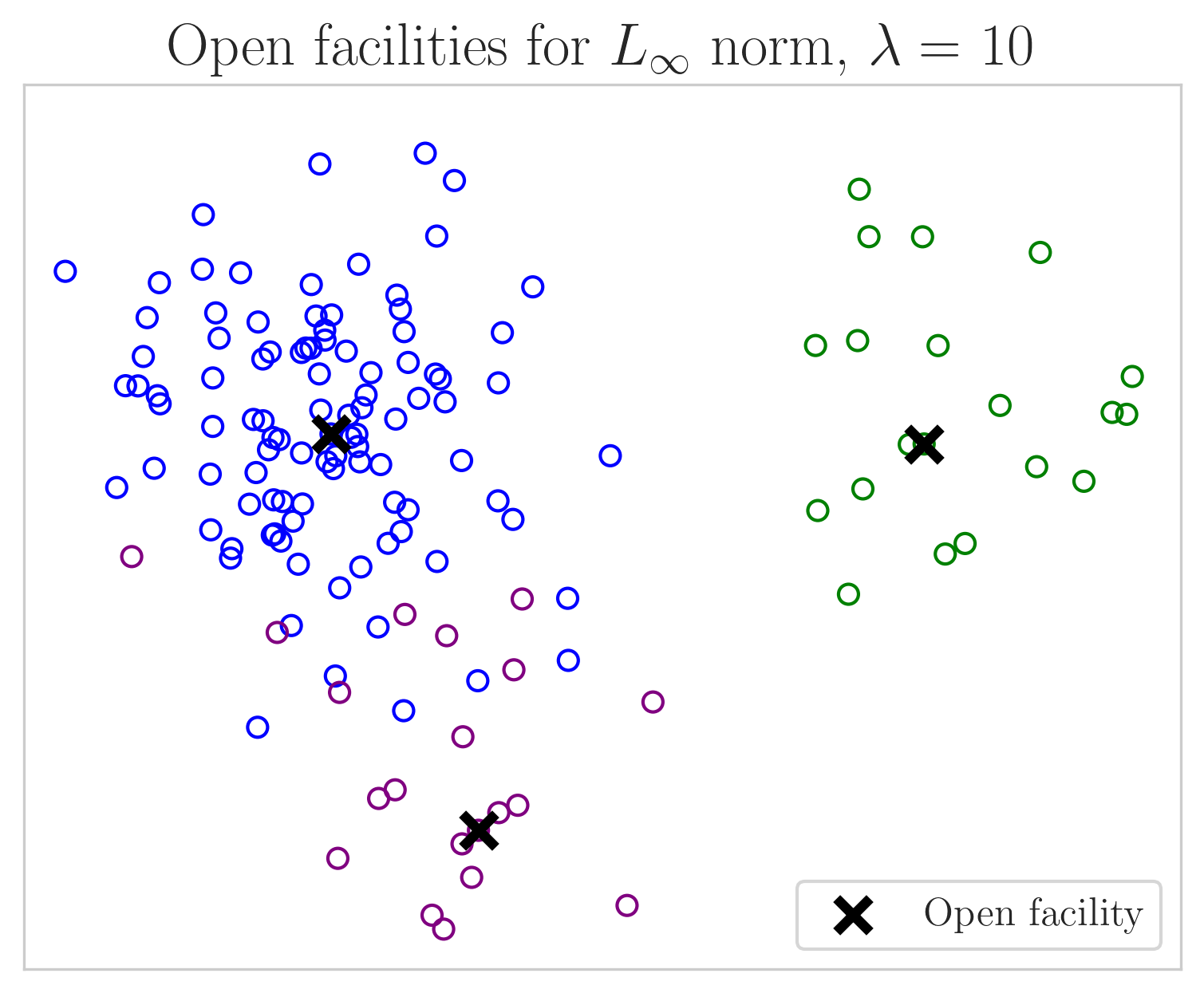}
            \end{minipage}
        \end{minipage}
        \caption{A portfolio of solution for different $L_p$ norms $p \in \{1, 2, \infty\}$ and various values of $\lambda_p$.}
        \label{fig: ffl-portfolio-with-lambda}
    \end{figure}

    Consider the trade-off between facility cost $c(F)$ and the $L_p$ norm access cost $\|d_\Pi\|_p$ for a solution $(F, \Pi)$ in fair facility location as norm $p$ changes from $1$ to $\infty$. Since $\|d_{\Pi}\|_p$ decreases with $p$, the total cost $c(F) + \|d_{\Pi}\|_p$ becomes more dominated by the facility cost $c(F)$. Therefore, optimal solutions to FFL$_p$ for higher $p$ tend to emphasize minimizing the facility cost more and may open fewer facilities as $p$ increases. This leads to an unintended consequence for portfolios: as $p$ increases, distances become more equitable but also larger.

    To amend this, consider the following total cost with an additional parameter $\lambda_p$ that controls the relative weight of access cost with respect to the facility cost: $c(F) + \lambda_p \|d_{\Pi}\|_p$. As $p$ increases, we can increase $\lambda_p$ to `compensate' for the decrease in $L_p$ norm. In particular, for $r$ client groups, it follows using the properties of $L_p$ norms that choosing $\lambda_p = r^{1 - 1/p}$ ensures that the access cost $\lambda_p \|d_{\Pi}\|_p$ is an increasing function of $p$. Therefore, solutions for higher $p$ are now more equitable \emph{and} have smaller group distances. Figure \ref{fig: ffl-portfolio-with-lambda} shows a portfolio of solutions with varying $p$ and $\lambda_p$ values.

    \section{Gap Between Portfolio Sizes for Top-$l$ Norms and $L_p$ Norms}\label{app: portfolios-are-not-tranferable}

    In this section, we discuss further connections between portfolios for different classes of norms and consider whether approximation guarantees for portfolios are transferable from one class of norms to another (see Section \ref{sec: portfolio-upper-bound} for a discussion). This is indeed the case for portfolios of size $1$: We gave an example of fair facility location where the optimal portfolio for top-$l$ norms had size $2$ but the optimal portfolio for $L_p$ norms had size $3$. We give another example of a different problem domain where top-$l$ norms admit an \emph{optimal} portfolio of size $2$ but where any $O(1)$-approximate portfolio for $L_p$ norms must have size $\simeq (\log n)^{1/3}$. This shows that large gaps between portfolio sizes for different norms are possible.

    While we do not have such an example for facility location specifically, consider generally the problem of finding portfolios for sets of vectors. Then we show that for all large enough dimension $n \in \Z$, there exists a set of vectors $\mathcal{V} \subseteq \R_{\ge 0}^n$ such that
    \begin{enumerate}
        \item There is an $O(1)$-approximate portfolio $X$ of size $2$ for all top-$l$ norms, and
        \item Any $O(1)$-approximate portfolio $X'$ for all $L_p$ norms must have size $\Omega\left(\left(\frac{\log n}{\log\log n}\right)^{1/3}\right)$.
    \end{enumerate}

    \begin{proof}
        Let $S = S(n)$ be a super-constant that we will fix later, and $L$ be the largest integer such that $S^{L^2} \le n$. Then $L = \Omega(\sqrt{\log_S n})$. For $s \in [1, L]$, define the vectors $v(s) \in \R_{\ge 0}^d$ as:
        \[
            v(s) = (\underbrace{S^{-2s}, \ldots, S^{-2s}}_{S^{s^2}}, 0, \ldots, 0).
        \]

        Let $\mathcal{V} = \{v(1), \ldots, v(L)\}$. We will prove part (1) of the theorem statement first. Specifically, we will show that either $v(1)$ or $v(L)$ is optimum for all top-$l$ norms, so that $X = \{v(1), v(L)\}$ is an optimal portfolio for all top-$l$ norms. We have
        \[
            \left( \text{top-}l \: \text{norm of}\: v(s)\right) = \begin{cases}
                                                                      l S^{-2s} & \text{if}\: k \le S^{s^2},  \\
                                                                      S^{s^2 - 2s} & \text{if}\: k >  S^{s^2}.
            \end{cases}
        \]
        Fix $l$. For $s$ such that $S^{s^2} < l$, $\left( \text{top-}l \: \text{norm of}\: v(s)\right) = S^{s^2 - 2s}$ increases as $s$ increases since $s \ge 1$. For $s$ such that $S^{s^2} > l$, $\left( \text{top-}l \: \text{norm of}\: v(s)\right) = l S^{-2s}$ decreases as $s$ increases. Therefore, for each top-$l$ norm, either $v(1)$ or $v(L)$ is optimum. This proves part (1) of the theorem statement.

        We move to part (2) of the theorem statement. Consider $L_p$ norms for $p \in [1, L]$. Then we claim that for appropriate choice of $S = S(n)$, (1) for all $p \in [1, L]$, ${\arg\min}_{v(s)} \|v(s)\|_{p} = v(p)$. That is, vector $v(p)$ has the minimum norm $\|\cdot\|_{p}$ among all vectors in $\mathcal{V}$, and (2) for all $p \in [1, L]$ and $l \neq p$, $v(s)$ is not an $O(1)$-approximation for minimizing $\|\cdot\|_{p}$. Together, the two claims imply that any $O(1)$-approximate portfolio for $L_p$ norms, $p \in [1, L]$ must contain each of $v(1), \ldots, v(L)$. Note first
        \begin{equation}
            \|v(s)\|_{p} = \left(S^{s^2} \cdot S^{2ps}\right)^{1/p} = S^{\frac{s^2}{p} - 2s}.
        \end{equation}
        To show that this is minimum at $s = p$, consider $f(x) = \frac{x^2}{p} - 2x$. It attains its minimum at $x = p$. Since $S > 1$, this implies that ${\arg\min}_{v(s)} \|v(s)\|_p = v(p)$, and the minimum is $S^{-p}$.

        Further, for any $s \neq p = t^2$, say $s = t^2 + \Pi$ for $\theta \in [1, L]$, we have $\log_S \|v(s)\|_{p} = \frac{(p + \theta)^2}{p} - 2 (p + \theta) = (p + \theta)\left(1 + \frac{\theta}{p} - 2\right) = \frac{\theta^2 - p^2}{p} = \frac{\theta^2}{p} - p = \frac{\theta^2}{p} + \log_S \|v(p)\|_p$.
        Therefore, $\frac{\|v(s)\|_p}{\|v(p)\|_p} \ge S^{\frac{\theta^2}{p}} \ge S^{\frac{1}{p}} = \exp((\log S)/p)$ Since $p \le L$, this is at least $\exp(\frac{\log S}{L})$.
        Choose $S$ such that $\log S = (\log n)^{1/3} (\log\log n)^{2/3}$. Then $L = \Theta(\sqrt{\log_S n}) = \Theta\left(\sqrt{\frac{\log n}{\log S}}\right) = \Theta\left(\left(\frac{\log n}{\log\log n}\right)^{\frac{1}{3}}\right)$. Therefore, $\log \left(\frac{\|v(s)\|_p}{\|v(p)\|_p}\right) \ge \frac{\log S}{L} = \Theta\left(\log\log n\right)$. That is, $\|v(s)\|_p = \omega(\|v(p)\|_p)$. This proves claim 2. Lastly, note that the size of portfolio $\{v(1), \ldots, v(L)\}$ for $L_p$ norms is $\Theta\left(\left(\frac{\log n}{\log\log n}\right)^{1/3}\right)$.
    \end{proof}

\end{document}